%% file: main.tex
\def \<{\langle}
\def \>{\rangle}

\documentclass[a4paper,11pt]{article}  

\usepackage{jcappub,graphicx,epsfig,natbib,color,times,bm,amsmath,multirow,hyperref}

\usepackage{subfig}
\usepackage[flushleft]{threeparttable}

\title{Efficient ILC analysis on polarization maps after EB leakage correction}

\input{authors.tex}
\begin{document}

\abstract{
\input{abstract.tex}
}

\maketitle

\section{Introduction}\label{sec:intro}
\input{introduction.tex}

\section{Methodology}\label{sec:metho}
\input{methodology.tex}

\section{Data Simulations}\label{sec:simu}
\input{simulation.tex}

\section{Results and discussion}\label{sec:result}
\input{results.tex}

\section{Conclusions}\label{sec:conclusions}
\input{conclusion.tex}

\acknowledgments
We thank Hao Liu, Hua Zhai, Yi-Wen Wu, Zi-Xuan Zhang and those of AliCPT group for useful discussion. This study is supported in part by the National Key R\&D Program of China No.2020YFC
2201600 and by the NSFC No.11653002.

\nocite{*}
\bibliography{main}
\bibliographystyle{JHEP} 

\end{document}

%% file: authors.tex
\author[a,b]{Zirui Zhang}\emailAdd{zhzr@mail.sdu.edu.cn}
\author[b]{Yang Liu\footnote{Corresponding author\label{corresauth}}}\emailAdd{liuy92@ihep.ac.cn}
\author[b]{Si-Yu Li}\emailAdd{lisy@ihep.ac.cn}
\author[b,c]{De-Liang Wu}\emailAdd{wudl@ihep.ac.cn}
\author[a]{Haifeng Li}\emailAdd{lihf@sdu.edu.cn}
\author[b]{Hong Li\textsuperscript{\ref{corresauth}}}\emailAdd{hongli@ihep.ac.cn}

\affiliation[a]{Institute of Frontier and Interdisciplinary Science and Key Laboratory of Particle Physics and Particle Irradiation (MOE), Shandong University, 72 Binhai Road, Jimo,  Qingdao, P.R. China}
\affiliation[b]{Key Laboratory of Particle Astrophysics, Institute of High Energy Physics, Chinese Academy of Sciences, 19B Yuquan Road, Beijing, P.R. China}

\affiliation[c]{University of Chinese Academy of Sciences, 19A Yuquan Road, Beijing, P.R. China}

%% file: abstract.tex
The Internal Linear Combination (ILC) is widely used to extract the cosmic microwave background (CMB) signal from multi-frequency observation maps, especially for Satellite experiments with quasi-full sky coverage. We extend ILC method to CMB polarization map analysis with a small sky patch which is especially typical for ground-based experiments, by combing ILC with a template cleaning method which can give pure $B$ map free from $EB$ leakage caused by partial sky coverage. The feature of our methods is that we do the ILC analysis on pseudo-scalar $B$ maps, and the advantage is that it totally avoids the impact of $EB$ leakage on ILC, so that it can improve the efficiency of component separation dramatically. We demonstrate our methods with mock data of a future ground-based experiment with a deep survey on a clean patch in the northern sky, and the results show that the level of foreground residual can be well controlled, it biases the tensor to scalar ratio ($r$) at the order of $10^{-3}$ which is comparable to the statistical error by noise.

%% file: introduction.tex
The Cosmic Microwave Background (CMB) has made great contribution to modern cosmology. The precise measurements of CMB temperature and polarization fluctuations from the exiting experiments, like the space missions of COBE\cite{1992ApJ...396L...1S}, WMAP (Wilkinson Microwave Anisotropy Probe)\cite{2003ApJS..148....1B} and Planck\cite{2010A&A...520A...1T}, has established the current best-fit $\Lambda$CDM cosmological scenario and measured its six main parameters with precision ranging from fractions of a percent to a few percent\cite{2018arXiv180706209P}.
After the accurate measurement of CMB temperature and $E$-mode polarization, $B$-mode has become one of the next core scientific goals in the field. Compared with temperature signal, the detection of $B$-mode polarization is more challenging, as the signal is very tiny. The upper limit of the measurement given by the current experiment and the predicted theoretical value given by existing inflation models are much smaller than $E$ mode, far lower than temperature fluctuation. The accurate measurement of $B$ mode attracts lots of effort. For now, the joint analysis of BICEP2/\textit{Keck Array}, Planck and WMAP yields the constraint  on tensor-scalar ratio $r$ is $r_{0.05} < 0.036$ at $95\%$ confidence level\cite{BICEP:2021xfz}. Also, in near future, some ground-based microwave telescopes are known to be deployed or planned for the $B$ mode detection in South pole\cite{2013PhRvL.111n1301H,2018PhRvL.121v1301B,2019Univ....5...42M, 2019arXiv191005748S}, Atacama\cite{2014JCAP...10..007N,2018JCAP...09..005K,2019arXiv190800480D,2018JLTP..193.1066N,2019arXiv191002608A,2019JCAP...02..056A}, and Tibet\cite{10.1093/nsr/nwy019}. Some future satellite missions, like LiteBIRD\cite{2022arXiv220202773L} and stratospheric balloon missions \cite{2018ApJS..239....7E, 2018JLTP..193.1112G, 2012SPIE.8452E..3FD, 2018SPIE10708E..06P} are also focusing on $B$ mode science.

The acquisition of CMB $B$-mode is very challenging, as many factors will interfere with the process of the measurement, such as instrument noise and systematic errors. In addition, some extra $B$-mode, for example, the lensing $B$-mode distorted from $E$-mode caused by the existence of large-scale structure in the universe, as well as $EB$ leakage caused by $E$-mode and $B$-mode decomposition in an incomplete $QU$ map, and so on, will affect the observation. 

The foreground radiation is a particularly serious challenge, the diffuse galactic emission in the microwave from several processes contaminates the observational CMB data, and the background of radio and infrared compact sources, galactic or extra-galactic, makes a contribution to the total emission even at high galactic latitudes. Multiband observations show that the foreground radiation, no matter its composition or its physical mechanism, is quite complex, thus, the foreground cleaning strategy remains to be a key step in CMB data processing. On the other hand, with the improvement of experimental sensitivity, foreground analysis technology is more and more emphasized in data analysis, especially in $B$ mode polarization detection, the methods of polarization foreground elimination are in urgent requirement.

There are many component separation methods used in CMB analysis. Generally speaking, they can be divided into three categories:  non-parametric methods, parametric methods,  and template fitting methods. Non-parametric methods make no assumptions about the foreground, such as ILC  (Internal Linear Combination)\cite{2003ApJS..148...97B, 2004ApJ...612..633E,2003PhRvD..68l3523T, 2009A&A...493..835D}, FastICA  (Fast
Independent Component Analysis)\cite{2002MNRAS.334...53M} and SMICA (Spectral Matching Independent Component Analysis) \cite{2003MNRAS.346.1089D, 2008ISTSP...2..735C}. Parametric methods rely on the assumption on the components model, like their respective spectral laws, then fit the model parameters using the observation data, such as Commander\cite{2006ApJ...641..665E}. Template fitting methods construct several foreground templates from real observational data and scale them to other frequency channels by fitting, such as SEVEM (Spectral Estimation Via Expectation Maximisation)\cite{2003MNRAS.345.1101M, 2008A&A...491..597L, 2012MNRAS.420.2162F}. 
In most cases, these methods focus on quasi-fullsky coverage and are mainly used in $T$ map cleaning,  however, when it comes to small partial sky, extracting the right cleaned $B$ mode signal becomes more and more challenging.  

For an incomplete CMB polarization map analysis, $EB$ leakage induced by partial sky appears during the process of $EB$ decomposition in $QU$ map and it generates non-ignorable influence to primordial CMB $B$-mode, this effect is especially serious for ground-based experiments which usually concentrate on a small cleaned patch of sky. Eliminating $B$ mode caused by $EB$ leakage is one of the most important steps in CMB data processing for CMB $B$ mode measurement. Many literature provide effective methods for its correction on harmonic domain \cite{2019MNRAS.484.4127A, 2003PhRvD..67b3501B} or pixel domain \cite{PhysRevD.100.023538, 2019JCAP...01..045R, 2019JCAP...04..046L}. 
It is worth noting that the extra $B$-mode caused by $EB$ leakage will affect the implementation of ILC in polarization maps, because the $B$-mode caused by $EB$ leakage will participate in the trade-off of $B$-modes in ILC.  We note that in previous studies, ILC has not been widely used in the processing of incomplete CMB polarization maps from ground observation experiments, and the issues related to partial sky map analysis have not been discussed in depth.

In this paper, we will provide a new idea to effectively avoid the interference of $EB$ leakage to the performance of ILC in polarization maps. The gist of our method is to avoid the leaked $B$-mode at the first step, then to perform ILC on the pure $B$ map.
The advantage of this method is that it can completely avoid the leaked $B$-mode to participate in the trade-off between $B$-modes from different sources in ILC. We adopt a blind strategy for correcting $EB$ leakage provided in literature\cite{PhysRevD.100.023538}, and this method is argued to be the best blind algorithm for the correction of the leaked $B$-mode for generating pure $B$-map.  Several cases of ILC analyses working in different situations are investigated with properly corrected $EB$ leakage maps. Compared with the normal ILC analysis based on $QU$ maps, it can completely get rid of the influence of $B$-mode contamination caused by $EB$ leakage in ILC analysis, so as to effectively improve the efficiency of ILC component separation. It is worth pointing out that, we, in this paper, mainly focus on the ground-based CMB observations with partial sky coverage. Theoretically, there is no intrinsic limitation on applying our methods on satellite configurations with quasi-fullsky coverage, whose foreground emissions have exhibit variations on large scales, it may require specialized modifications for PixelILC and NILC methods used in this paper on pure $B$ maps , we will leave these in the future study. 

We do a series of map-based simulations for the next generation ground CMB observations, and with the mock data, which includes sky emission of both CMB and realistic Galactic foregrounds signal based on Planck observation, we study the ability of reconstructing primordial $B$-mode. As a working example, two sky patches were considered in the analysis. The first one is a typical trapezoid-shaped sky patch at the Northern hemisphere with $f_{sky}\sim7\%$, and the second sky patch is proposed by CMB-S4 at Southern-hemisphere with $f_{sky}\sim3\%$, which is a circular patch.

The paper is organized as follows: In section \ref{sec:metho} we review some of the ILC methods and discuss how they are applied on $B$ maps. Section \ref{sec:simu} introduces the multi-frequency map simulations. Section \ref{sec:result} shows the results and discussion. Finally, the conclusions are summarized in Section \ref{sec:conclusions}.

%% file: methodology.tex
In this section, we give a brief introduction to the ILC method. Different from previous studies, which mainly focus on $T$ map analysis, we pay more attention to the performance of ILC on polarization maps.

\subsection{Map domain EB leakage correction}\label{sec:bilc}
$EB$ leakage occurs when the spherical harmonic transformation is performed on the partial sky because the orthogonality of spherical harmonics is not satisfied anymore. For a ground-based CMB observation, such leakage is inevitable because of the limited sky coverage and it will introduce the bias on $r$. It's necessary to correct such $EB$ leakage.

Recent studies show that there's a good way called template cleaning method for a blind estimation and correction of $EB$ leakage \cite{PhysRevD.100.023538} comes from the partial sky, and can directly give a pure $B$ map, here we propose to extend the ILC method for scalar
$T$ map to $B$ map, as $B$ is a pseudo-scalar for a strictly speaking.

\begin{figure}[bthp]
    \centering
            \subfloat[]{\includegraphics[width=0.27\textwidth]{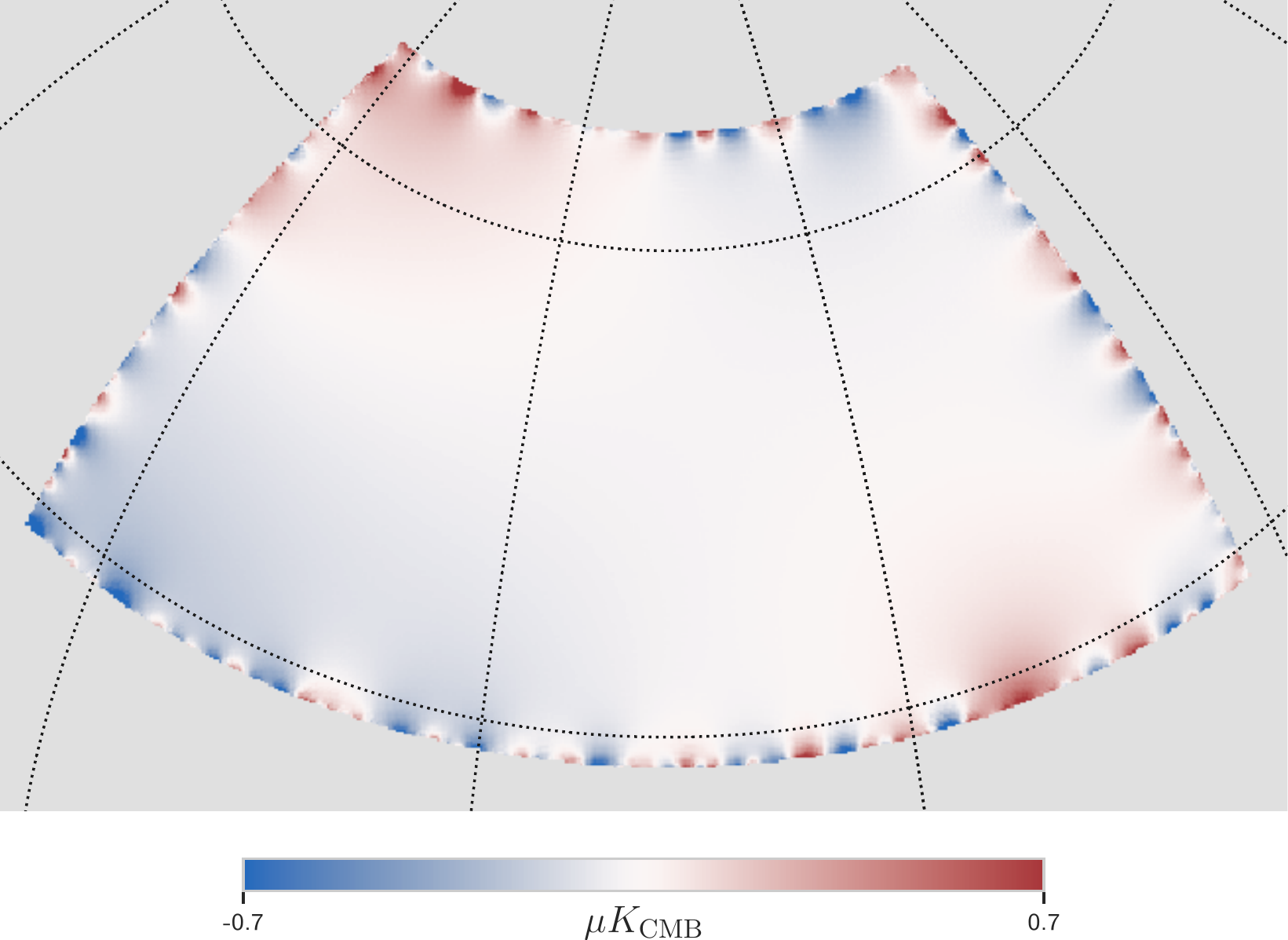}\hskip 1em \includegraphics[width=0.27\textwidth]{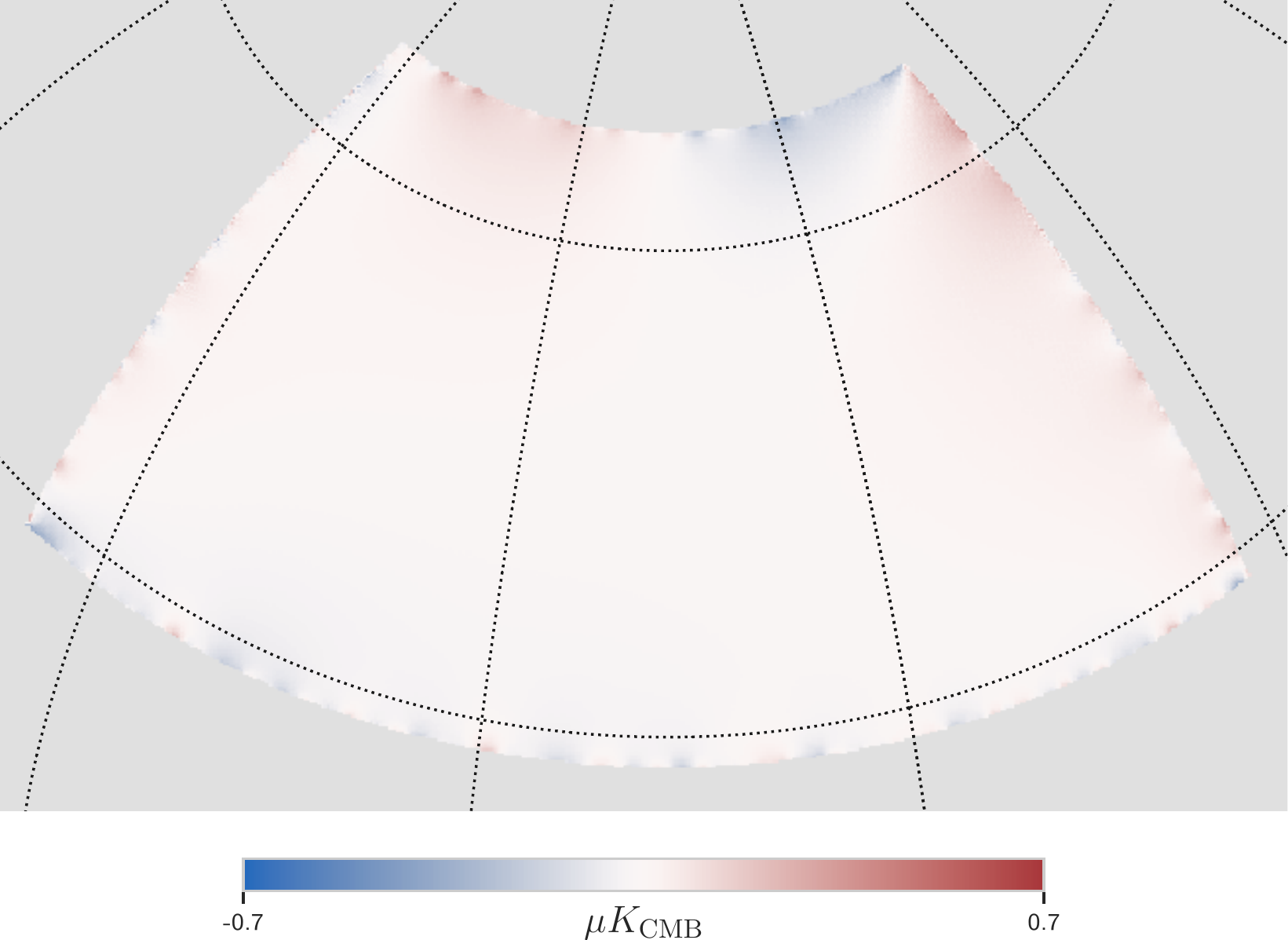}}\\
            \subfloat[]{\includegraphics[width=0.27\textwidth]{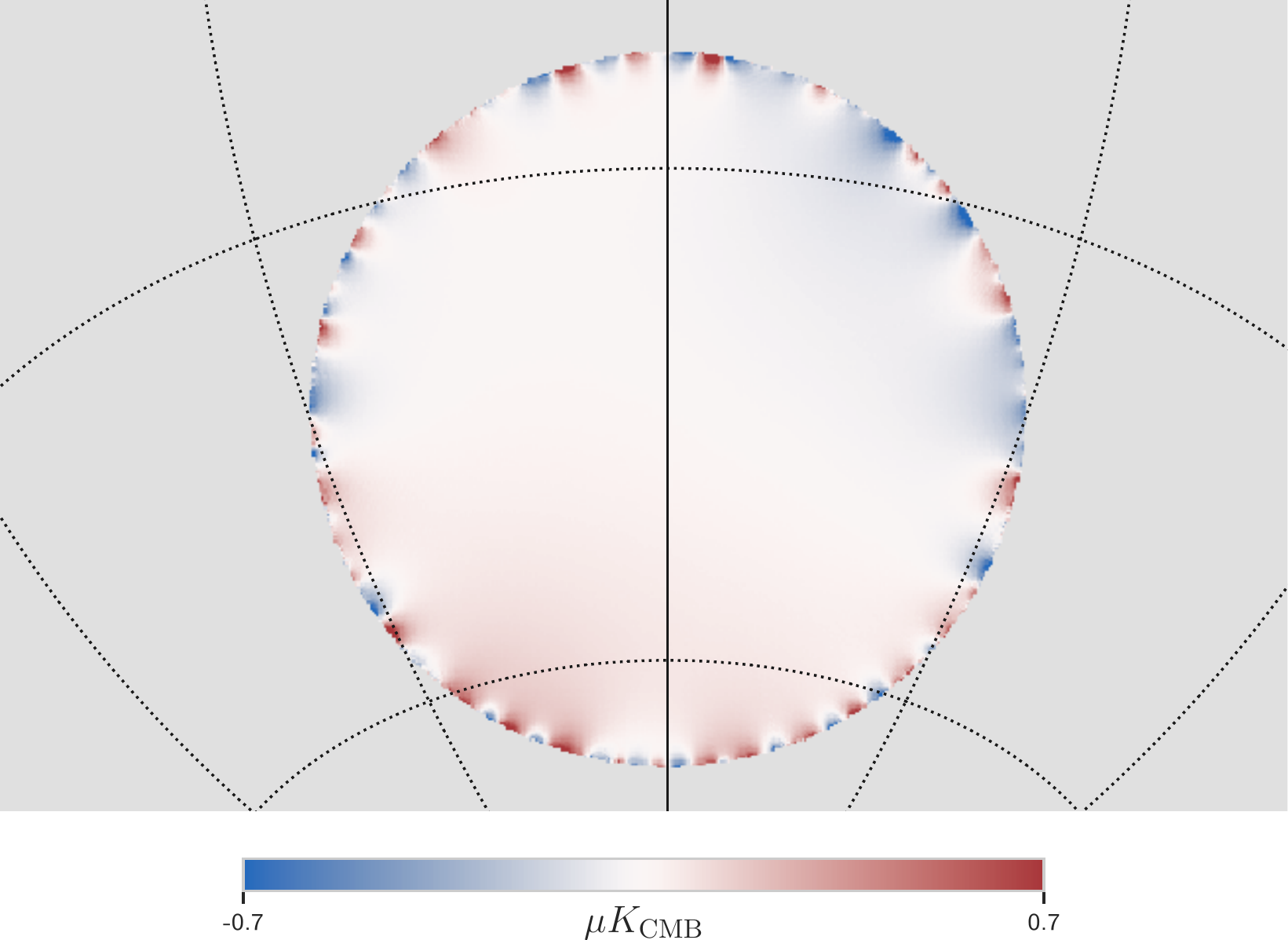}\hskip 1em \includegraphics[width=0.27\textwidth]{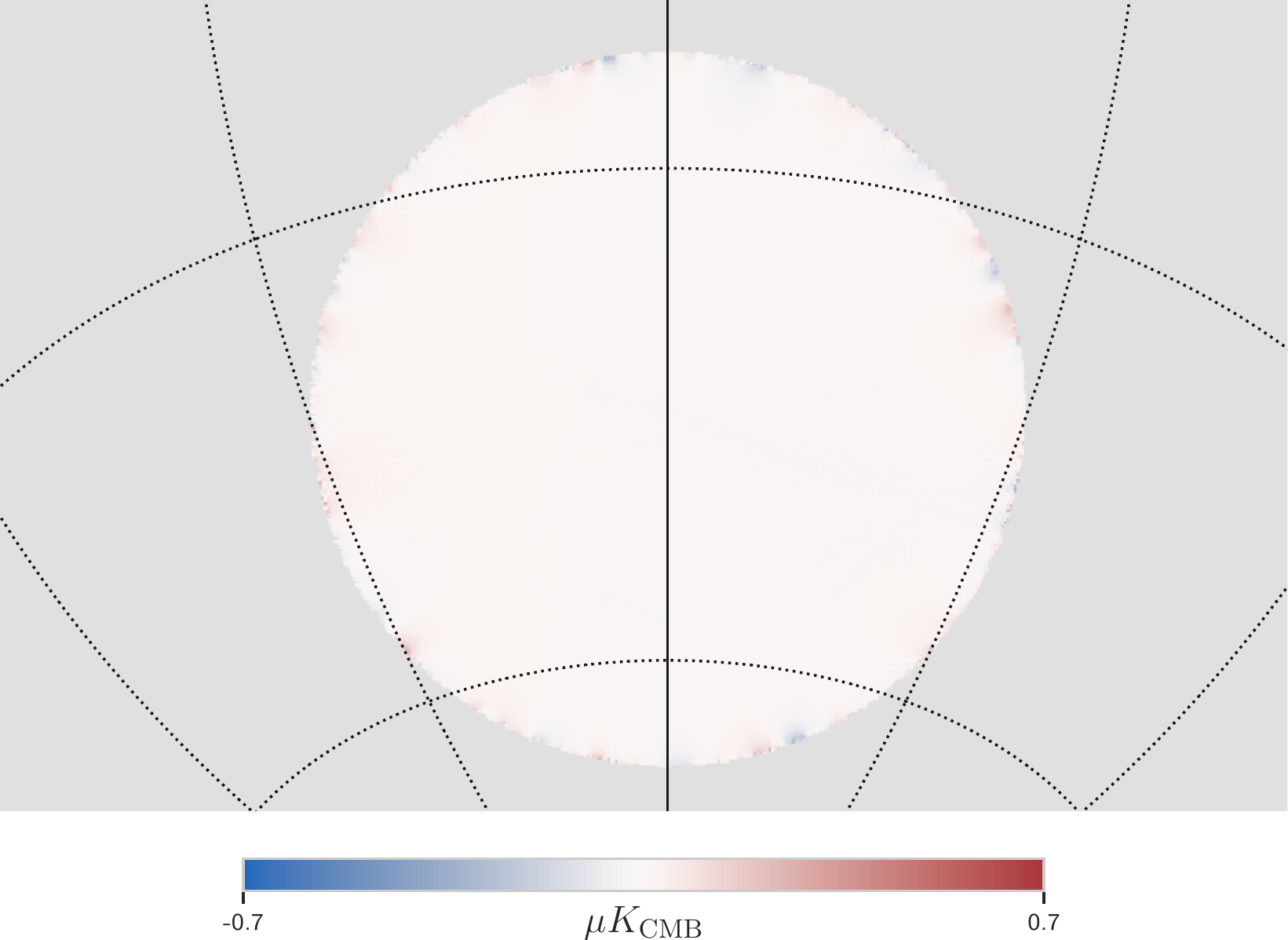}}\\
            \subfloat[]{\includegraphics[width=0.6\textwidth]{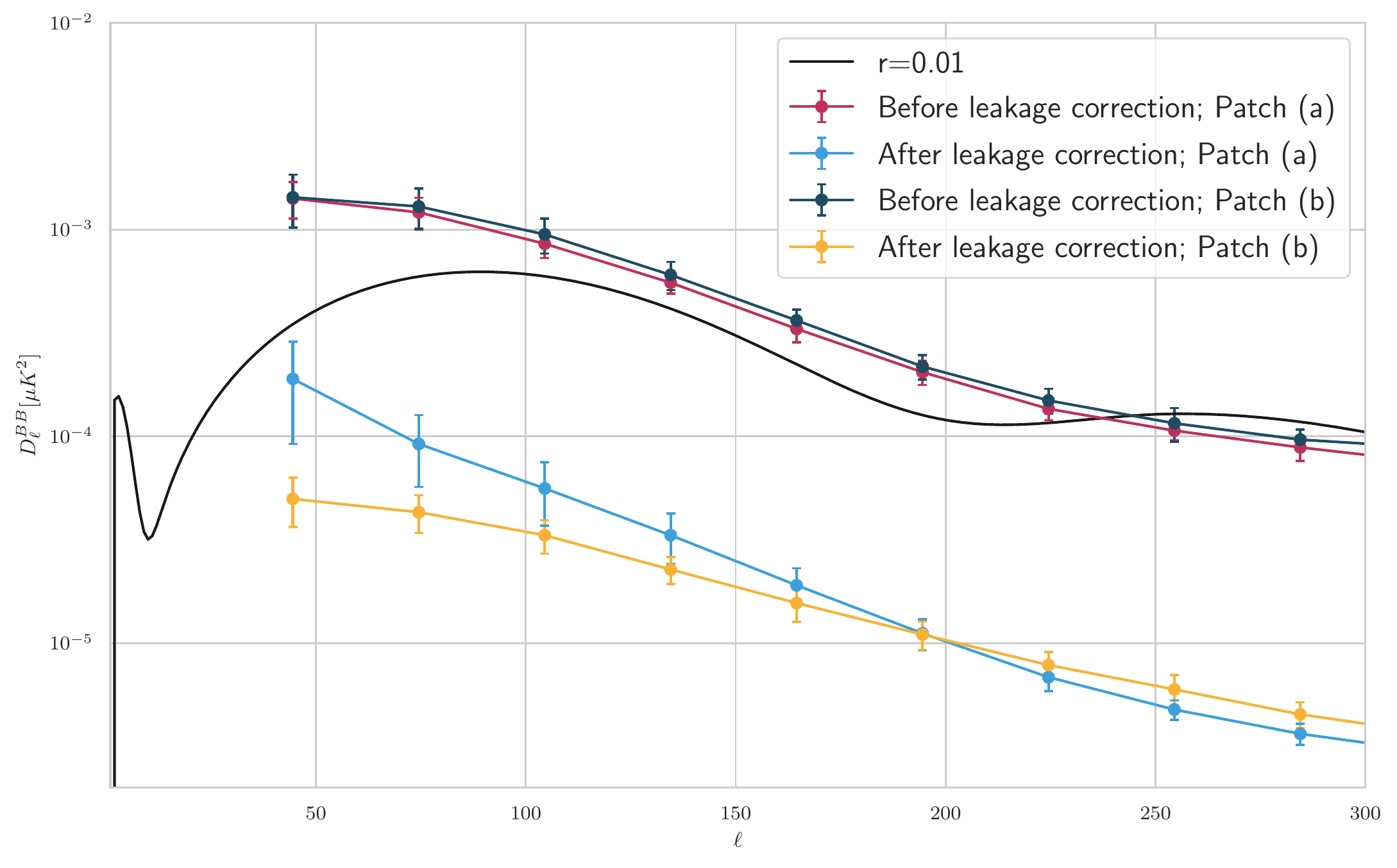}}
    \caption{\textit{(a)} and \textit{(b)}: The leaked $B$ mode signal before and after the correction on the sky patches. The left plots show the leakage due to the limited sky coverage and the right plots show the corresponding residual leakage after the correction. \textit{(c)}: The leaked and residual $B$ mode angular power spectra. The errorbars represent the standard deviation calculated from $30$ different CMB realizations. The solid black line shows the unlensed $B$ mode with $r=0.01$.}
    \label{fig:EBLeakResult}
\end{figure}

The detailed description of the blind $EB$ correction method refers to paper \cite{PhysRevD.100.023538}, we briefly summarize the steps for pure $B$ map construction in the following:
\begin{enumerate}
    \item Transform the incomplete observation $QU$ to $a_{\ell m}^{E},a_{\ell m}^{B}$ directly with \texttt{map2alm} function of healpy. Then do the inverse transformation from ($a_{\ell m}^{E}$, 0) to $QU$ map, denoted as $Q^{(1)}_EU^{(1)}_E$. Use $a_{\ell m}^{B}$ to get the pseudo-scalar $B$ map, denoted as $B^{(1)}$. For now the $B$ map $B^{(1)}$ is contaminated by $EB$ leakage.
    \item Similarly, obtain $B$ map from $Q^{(1)}_EU^{(1)}_E$ by using the same mask, denoted as $B^{(2)}$.
    \item Now $B^{(2)}$ is the leakage template of $B^{(1)}$ in the available region.  
        The final leakage-corrected $B$ map can be represented as 
        \begin{align}
            B_{\rm clean}(p) = B^{(1)}(p) - aB^{(2)}(p), \label{eq:ebleak}
        \end{align}
        where the factor $a$ is obtained by a linear fitting of $B^{(1)}$ and $B^{(2)}$, $p$ denotes the pixel location.
\end{enumerate}
It has been proved that the final output $B_{\rm clean}$ is the best blind estimate of pure B map, which means that $B_{\rm clean}$ contains the smallest error due to the $EB$ leakage, and any improvement requires additional information about the unobserved sky region\cite{2019JCAP...04..046L}. 
Figure \ref{fig:EBLeakResult} shows the results of before and after $EB$ leakage correction by using this method for two different sky patches concerned. It’s clear that, in the power spectrum, the $EB$ leakage due to partial sky is larger than $r=0.01$, which will bias r measurement heavily, especially on the degree scale. After the correction, the residual leakage is an order of magnitude smaller than $r=0.01$. The bias on $r$ eliminated significantly. Notice that, the residual leakage on the trapezoidal patch is slightly larger than the residual on the circular patch. It's because the circular patch is the most ideal case for this correction method. More complicated shapes will result in higher leakage and residuals, and reduce the performance.

\subsection{The ILC methods}

In the ILC framework, it assumes that the observation map is a linear combination of CMB and the foreground emission which scales with frequency, and 
the map can be written in a compact form:
\begin{align}
    \mathbf{y}(p)=s_{\rm CMB}(p)+\mathbf{s}_{FG}(p)+\mathbf{n}(p), \label{eq:obs}
\end{align}
where $\mathbf{y}(p)$ denotes the observation vector with $N_\nu$ elements for each pixel ($p$), $s_{\rm CMB}$ is the signal of CMB, $\mathbf{s}_{FG}$ denotes the foreground signal, and $\mathbf{n}$ denotes for instrumental noise. If the map is in the unit of $\mu K_{\rm CMB}$, from the derivation, it can be seen that CMB will be independent of frequency.

Following ILC \cite{2003ApJS..148...97B, 2004ApJ...612..633E}, we can get CMB through the combination of the maps of different frequency bands in pixel domain, and
the estimator $\hat{s}_{\rm ILC}$ is straightforward:
\begin{align}
    \hat{s}_{\rm ILC} = \frac{\mathbf{1}^T\hat{\mathbf{C}}^{-1}}{\mathbf{1}^T\hat{\mathbf{C}}^{-1}\mathbf{1}}\mathbf{y}, \label{eq:mainILC}
\end{align}
where $\hat{\mathbf{C}}$ is the map-to-map covariance matrix, and $\mathbf{1}^T$ is a column vector of all ones. This method can be extended to different domains, and the key of ILC methods is to estimate the covariance matrix $\hat{\mathbf{C}}$ in the chosen domain. 

There is a big difference between ILC on maps of polarization and temperature. As we know that both $Q$ and $U$ are spinors, minimizing $\left\langle Q^2\right\rangle$ and 
$\left\langle U^2\right\rangle$ separately will break the physical connection of cleaned $Q,~U$. Thus it may introduce 
bias when doing $EB$ leakage correction. The most straightforward way to avoid such problem is to work on pure $E$ and $B$ maps, directly. 
It's trivial for full-sky observations, however, for partial sky patch data processing, the problem of $EB$ leakage needs to be considered carefully. 
The following subsections describe different ILC methods for CMB polarization map reconstruction. One bases on the complex map $Q+iU$ and the others base on leakage-corrected $B$ map ($B_{\rm clean}$ in Eq. \eqref{eq:ebleak}) introduced in last section.

\subsubsection{Pixel domain Polarization ILC}\label{sec:pilc} 

PILC\cite{Fern_ndez_Cobos_2016} is an ILC method works on $QU$ map. Respecting the philosophy of the ILC, the CMB estimation can be written as:
$$\hat{Q}_{\rm CMB}(p)\pm i\hat{U}_{\rm CMB}(p) = \sum_{j=1}^{N_{\nu}}\left[\omega_j^{(R)}\pm i\omega_j^{(I)}\right]\left[Q_{j}(p)\pm iU_{j}(p)\right],$$
where $Q_j(p)$ and $U_j(p)$ denote the value of Stokes parameter at pixel $p$ at frequency band $\nu_j$. Compared with the standard ILC\cite{2003ApJS..148...97B},
the coefficients of the linear combination become complex numbers. To ensure that the estimated CMB signal is unbiased, the complex coefficients must follow 
the constraints:
$$\sum_{j=1}^{N_\nu}\left[\omega_{j}^{(R)} + i\omega_{j}^{(I)}\right] = 1.$$

In the case of standard ILC, the coefficients are estimated by minimizing the variance of the estimated map. 
For PILC, the cost function chosen to be minimized is a covariant quantity $\left\langle|\hat{P}_{\rm CMB}|^2\right\rangle $, where $\hat{P}_{\rm CMB}=\hat{Q}_{\rm CMB}+ i\hat{U}_{\rm CMB}$.
By using Lagrange multipliers, the weights can be calculated as\cite{2004ApJ...612..633E}: 
\begin{align}
    \omega_{j}^{(R)} &= \frac{\lambda_R}{2}\sum_{k=1}^{N_{\nu}}C_{jk}^{-1}+\frac{\lambda_I}{2}\sum_{k=N_{\nu}+1}^{2N_{\nu}}C_{jk}^{-1}, \label{pilcwr}\\
    \omega_{j}^{(I)} &= \frac{\lambda_R}{2}\sum_{k=1}^{N_{\nu}}C_{N_{\nu}+j,k}^{-1}+\frac{\lambda_I}{2}\sum_{k=N_{\nu}+1}^{2N_{\nu}}C_{N_{\nu}+j,k}^{-1}\label{pilcwi},
\end{align}
where 
$$\mathbf{C}\equiv \left[\begin{matrix}
    \mathbf{C}^{(+)}&-\mathbf{C}^{(-)}\\\mathbf{C}^{(-)}&\mathbf{C}^{(+)}
\end{matrix}\right],$$
with the definition
\begin{align}
    C_{jk}^{(+)}\equiv\langle Q_{j}(p)Q_{k}(p)+U_{j}(p)U_{k}(p)\rangle,\\
    C_{jk}^{(-)}\equiv\langle Q_{j}(p)U_{k}(p)-U_{j}(p)Q_{k}(p)\rangle,
\end{align}
and the angle brackets $\langle\cdot\rangle$ here denote the average over all pixels $p$ in the map.

The Lagrange multipliers $\lambda_R, \lambda_I$ in Eq. \eqref{pilcwr} and \eqref{pilcwi} are 
\begin{align}
    \frac{\lambda_R}{2} &= \frac{S_+}{S_+^2+S_-^2}, &\frac{\lambda_I}{2} &= \frac{-S_-}{S_+^2+S_-^2}.
\end{align}
where
\begin{align}
    S_+&\equiv\sum_{j,k=1}^{N_\nu}C_{jk}^{-1}, & S_-&\equiv\sum_{j=1}^{N_\nu}\sum_{k=N_\nu+1}^{2N_\nu}C_{jk}^{-1}.
\end{align}

PILC gives a foreground-cleaned map in $QU$ base, and it is a natural extension of ILC in temperature maps analysis, and the simplest and direct way is to treat Q and U as scalars and operate them like ILC on T maps.

In the literature, there has been some tries to carry out ILC on the $Q$ and $U$ maps to do the foreground cleaning\cite{Fern_ndez_Cobos_2016,2016MNRAS.463.2310R}. However, the result after ILC can not truly reflect the primordial polarization signal, as there is still further process is needed to obtain $E$ and $B$ mode from the resulting $QU$ maps, and those process may introduce extra contamination, for example $EB$ leakage introduced from $QU$ decomposition in an incomplete sky patch. Thus, a better way is to transform $QU$ maps to $EB$ maps, then treating them as an independent scalar like $T$, and it is introduced in the next subsection.




\subsubsection{Pixel domain ILC on $B$ map}
The term ``pixel domain" means that the final CMB signal is estimated as\cite{2003ApJS..148...97B}:
\begin{align}
    \hat{X}_{\rm CMB}(p) = \sum_{j}^{N_{\nu}} \omega_j X_j(p).
\end{align}
Here and after, $X$ represents any scalar or pseudo-scalar map. The weights are exactly the fraction term of Eq.\eqref{eq:mainILC}. And it's obvious to see that the sum of the weights equals $1$. The covariance matrix $\mathbf{C}$ can be expressed as,
\begin{align}
    C_{ij} = \left\langle X_i(p)X_j(p)\right\rangle - \left\langle X_i(p)\right\rangle\left\langle X_j(p)\right\rangle, 
\end{align}
and also the angle brackets denote taking the average over pixel $p$. $\mathbf{C}$ is exactly the covariance matrix between 
different observation maps. That's a direct result by minimizing the variance of the resulting map.
\subsubsection{Harmonic domain ILC on $B$ map}

ILC method on harmonic domain is quite similar to the pixel domain. The only difference is that ILC deals with objects that are the vectors of spherical harmonic coefficient $a_{\ell m}$.  
The linear combination becomes 
\begin{align}
    \hat{a}_{\ell m}^{X} = \sum_{j}^{N_{\nu}} \omega^j_{\ell} a^{Xj}_{\ell m}.
\end{align}

The weights $\omega^{j}_{\ell}$ is calculated by minimizing $\left\langle |\hat{a}_{\ell m}|^2 \right\rangle$ for every single $\ell$ with the constraint
$\sum_j w^j_{\ell}=1$. The weights can also be represented as the fraction term in Eq.\eqref{eq:mainILC}. In harmonic space, the covariance matrix becomes
\begin{align}
    C_{\ell}^{ij} = \frac{1}{2\ell + 1} \sum_{m=-\ell}^{\ell} a_{\ell m}^{X*i}a_{\ell m}^{Xj},
\end{align}
which is a matrix-valued cross-power spectrum\cite{2003PhRvD..68l3523T}.

\subsubsection{Needlet domain ILC on $B$ map}

The term `needlet' means that the object of processing is the wavelets on the two-dimensional sphere. The most important property of the needlets is that 
the transformation can keep the information of spatial domain and harmonic domain simultaneously. The needlets provide good localization in both pixel domain and harmonic domain, which makes this method very useful in pixel domain (foreground emission and noise level distributions are inhomogeneous) as well as harmonic domain (contaminants do not have the same angular power spectra)\cite{2009A&A...493..835D, 2013MNRAS.435...18B}.

The objectives in needlet framework are a set of functions which can be written as:
\begin{align}
    \Psi_{jk}(\hat{n})=\sqrt{\lambda_{jk}}\sum_{\ell=0}^{\ell_{max}}\sum_{m=-\ell}^{\ell}h_{\ell}^jY_{\ell m}^*(\hat{n})Y_{\ell m}(\hat{\xi}_{jk})
\end{align}
where $h_{\ell}^{j}$ is the filters, and they keep the harmonic information. $\hat{\xi}_{jk}$ is a set of grid points on sphere and $\lambda_{jk}$ is the weight for every grid point. In practice, if the maps are pixelized in HEALPix\cite{2005ApJ...622..759G} pixelization scheme, we can choose the pixel center as the grid point $\hat{\xi}_{jk}$.

Using the needlet base, the analysis and synthesis operations for a scalar map $X$ are:
\begin{align}
    Analysis: & & \beta_{jk} =& \int X(\hat{n})\Psi_{jk}(\hat{n}){\rm d}\Omega_{\hat n},\\
    Synthesis: & & X^{j}(\hat{n}) &=  \sum_{k=1}^{N_{\rm pix}}\beta_{jk}\Psi_{jk}(\hat{n}).
\end{align}

The ILC estimate of needlet coefficients of reconstructed CMB map can be written as linear combination of needlet coefficients calculated from observed maps: 
\begin{align}
    \hat{\beta}_{jk} = \sum_{c=1}^{N_{\nu}}w_{jk}^{c}\beta_{jk}^{c}
\end{align}

Here, $c$ is the channels index, and $N_{\nu}$ means number of channels. From the analysis and synthesis of needlet transformation, trivial calculation shows that the relationship between $X^j$ and $X$ in harmonic space is $X^{j}_{\ell m} = (h_{\ell}^j)^2 X_{\ell m}$.
Thus there are two constraints here to preserve CMB response:
\begin{align}
    \sum_{c}^{N_{\nu}}\omega_{jk}^{c}&=1, & \sum_{j}(h^{j}_{\ell})^2 &= 1
\end{align}

Unlike the ILC working on pixel and harmonic domain, there is no intuitive estimator to estimate the variance of resulting map. In practice, the variance of one specific pixel $k$ at scale $j$ is estimated over a domain $\mathcal{D}_k$ centered at $k$ and including some neighbouring pixels\cite{2013MNRAS.435...18B}. Thus, the empirical covariance $C_{jk}^{cc'}$ is calculated from taking the average of the coefficients over $\mathcal{D}_k$, written as:
\begin{align}
    \hat{C}_{jk}^{cc'} = \frac{1}{N_{\mathcal{D}}}\sum_{k'}\omega_{j}(k,k')\beta_{jk'}^{c}\beta_{jk'}^{c'}.
\end{align}

%% file: simulation.tex
We investigate our foreground removal methods using simulated multi-frequency data sets at map level observed by a future ground based CMB experiment. In the following, we will describe the sky model and the configuration of the experiment to generate the datasets, then show how to do data processing.

\subsection{Sky model}

We adopt the sky model which is as close to the true sky as possible. It contains CMB signal, diffuse foreground emission composed of synchrotron, thermal dust, free-free and spinning dust, we also include the point sources to study their effects. The polarized foreground maps are simulated by using the Planck Sky Model\cite{2013A&A...553A..96D} (PSM). The details for the sky model is as follows:

\begin{enumerate}
    \item \textbf{CMB:} the input CMB maps are Gaussian realizations from a particular power spectrum obtained from the Boltzmann code CAMB, using  Planck 2018 best-fit cosmological parameters with $r = 0$. The CMB dipole was not considered in the simulations.
    
    The effect of CMB lensing was also considered in our simulation. The weak lensing distorts a part of $E$-mode signal to $B$-mode. CMB lensing is not a linear effect and therefore it is non-Gaussian. The precise way to simulate lensing effects is to rearrange the pixel on the map level. For this part, the package \texttt{lenspyx}\footnote{\url{https://github.com/carronj/lenspyx}} is used to simulate it. the algorithm is presented in the reference\cite{2005PhRvD..71h3008L}, which distorts the primordial signal given a realization of lensing potential map from $C_{\ell}^{\phi\phi}$.
    
    \item \textbf{Synchrotron:} Galactic synchrotron radiation arises from charged relativistic cosmic-rays accelerated by the Galactic magnetic field. For low frequency ($<$ 80 GHz), synchrotron is the most important polarized foreground. The effects of Faraday rotation were ignored in our simulations. In the simplest model and in the CMB frequency range, the synchrotron frequency dependence is also a power-law (in Rayleigh-Jeans unit):
    \begin{align}
        T_{sync}(\nu)\propto \nu^{\beta_s}.
    \end{align}
    The typical value for spectral index $\beta_s$ is equal to $-3$. For polarization, the WMAP 23 GHz channel polarization maps $Q_{23}$, $U_{23}$ are used as templates with the same synchrotron spectral index.
    \item \textbf{Thermal dust:} For higher frequency ($>$ 80 GHz), the thermal emission from heated dust grains dominates the foreground signal. In PSM, the emission of thermal dust is modeled as the sum of two modified blackbodies,
    \begin{align}
        I_{\nu}=\sum_{i=1}^{2}A_i\nu^{\beta_i}B_{\nu}(T_i),
    \end{align}
    where $B_{\nu}(T_i)$ is the Planck function at temperature $T_i$. For polarization, the Stokes $Q$ and $U$ parameters are 
    \begin{align}
        Q_{\nu}(\hat{\mathbf{n}})&=f(\hat{\mathbf{n}}) I_{\nu}\cos(2\gamma_d(\hat{\mathbf{n}})), & 
        U_{\nu}(\hat{\mathbf{n}})&=f(\hat{\mathbf{n}}) I_{\nu}\sin(2\gamma_d(\hat{\mathbf{n}})).
    \end{align}
    The template maps $f(\hat{\mathbf{n}})$ and $\gamma_d(\hat{\mathbf{n}})$ are calculated from the observational data.
    
    \item \textbf{free-free:} Free-free emission arises from electron–ion scattering in interstellar plasma. Generally, it is fainter than the synchrotron or thermal dust emission. Free-free is also modeled as power-law and the spectral index of free-free $\beta_{ff}$ is roughly $-2.12$. Free-free emission is intrinsically unpolarized. It contributes only to temperature measurements.
    
    \item \textbf{Spinning dust:} The small rotating dust grain can also produce microwave emission. It can also be modeled as a power law. Spinning dust emission is weakly polarized. In PSM, the emission is assumed to be unpolarized.
    
    \item \textbf{Point sources:} Radio sources are mainly modeled based on the surveys of radio sources of frequencies between $0.85$ GHz and $4.85$ GHz. The catalogues are detailed in the reference\cite{2013A&A...553A..96D}. The point sources are also modeled as a power law to approximate the spectra as $S\propto\nu^{-\alpha}$. The spectral indices $\alpha$ for each point sources are obtained from the large-scale surveys. The polarization of each source attributing to a polarisation degree is randomly drawn from the observed distributions for flat- and steep-spectrum sources at 20 GHz\cite{2004A&A...415..549R}, and a polarization angle is randomly drawn from a uniform distribution.
\end{enumerate}

\subsection{Instrumental configuration}

\begin{figure}
    \begin{center}
        \subfloat[The northern sky patch\label{subfig:skypatch}]{\includegraphics[width=0.24\textwidth]{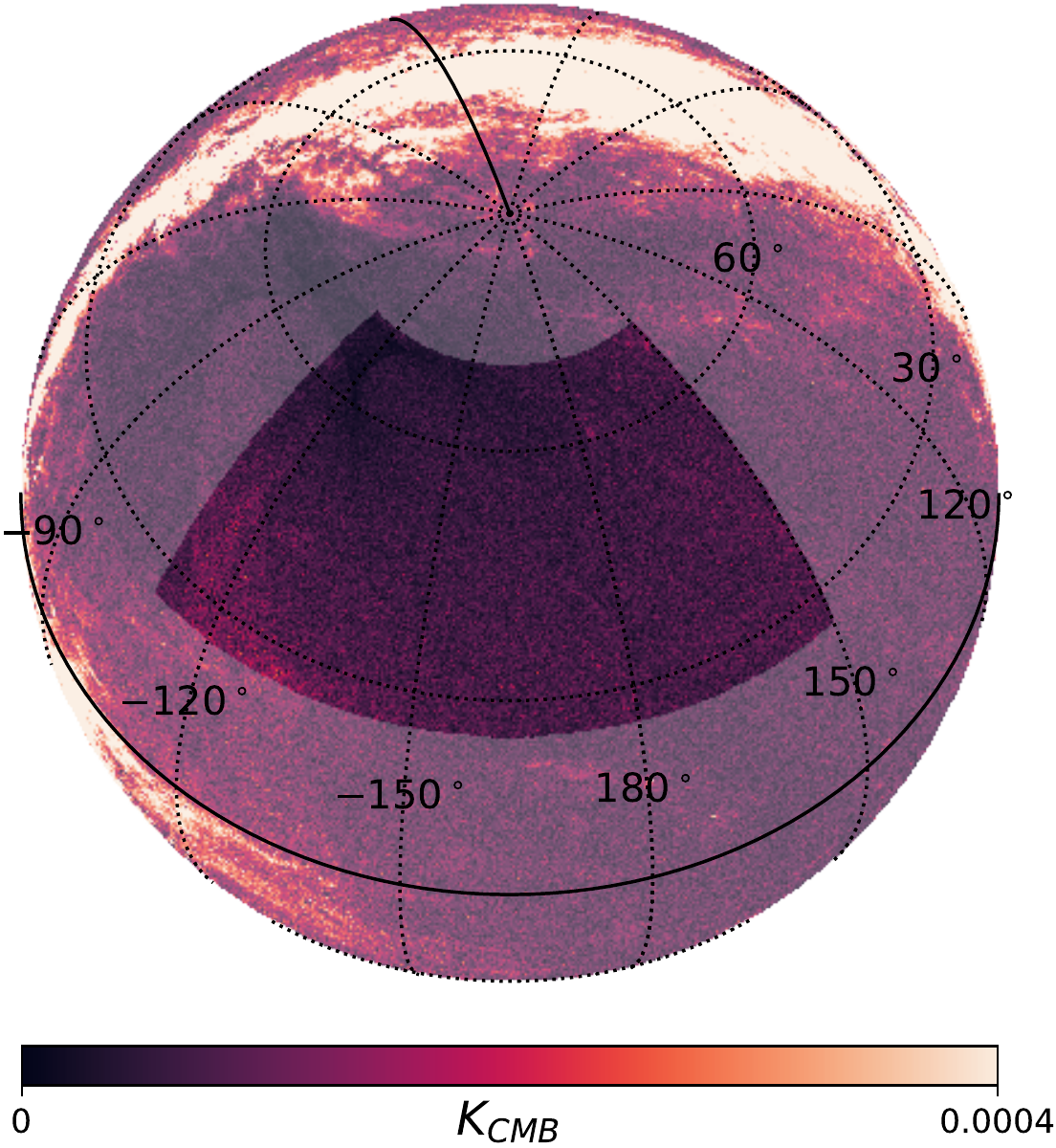}} \hskip 0.01\textwidth
        \subfloat[The northern masks\label{subfig:mask}]{\includegraphics[width=0.24\textwidth]{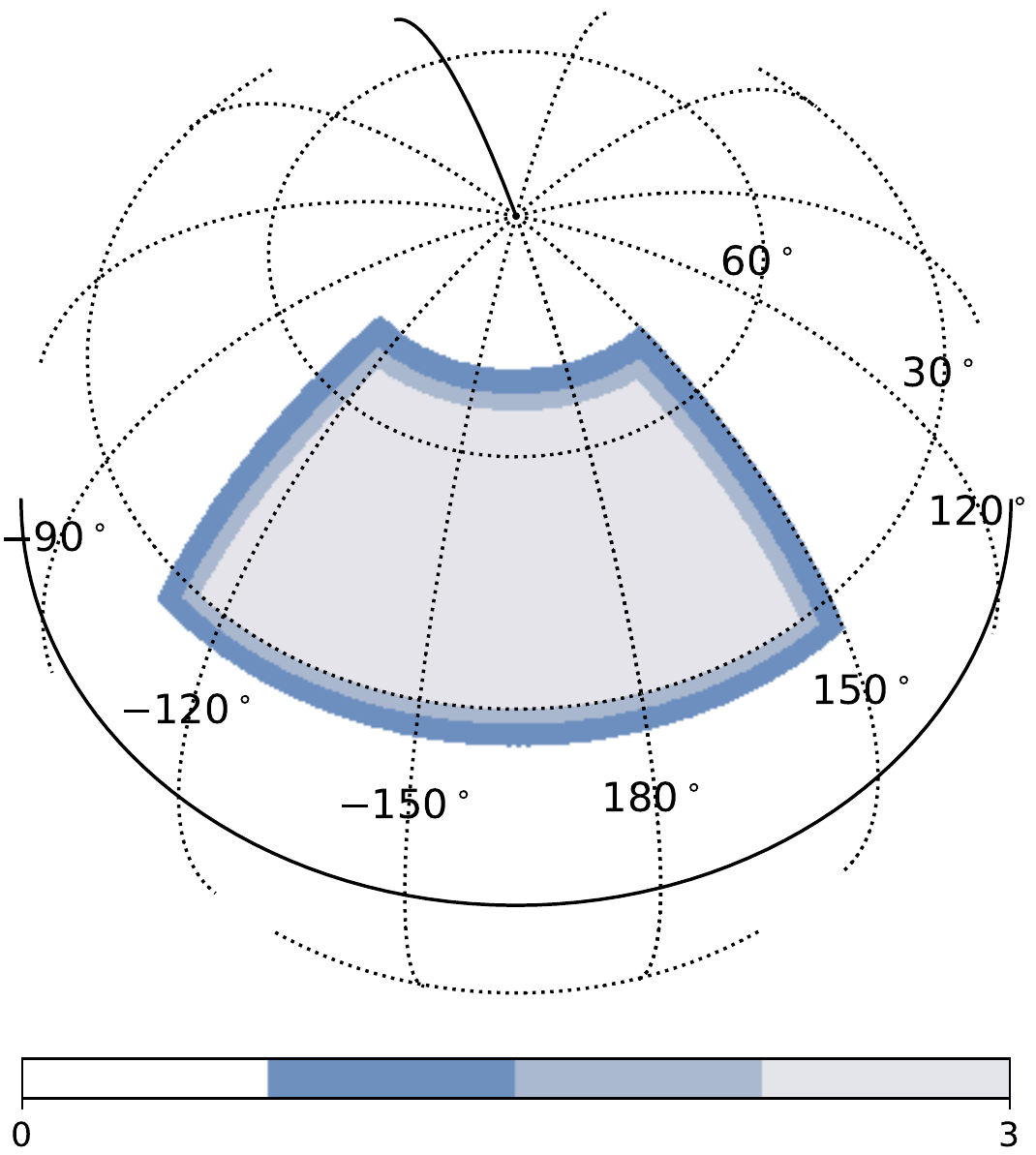}}\hskip 0.01\textwidth
        \subfloat[The southern sky patch\label{subfig:skypatchCMBS4}]{\includegraphics[width=0.24\textwidth]{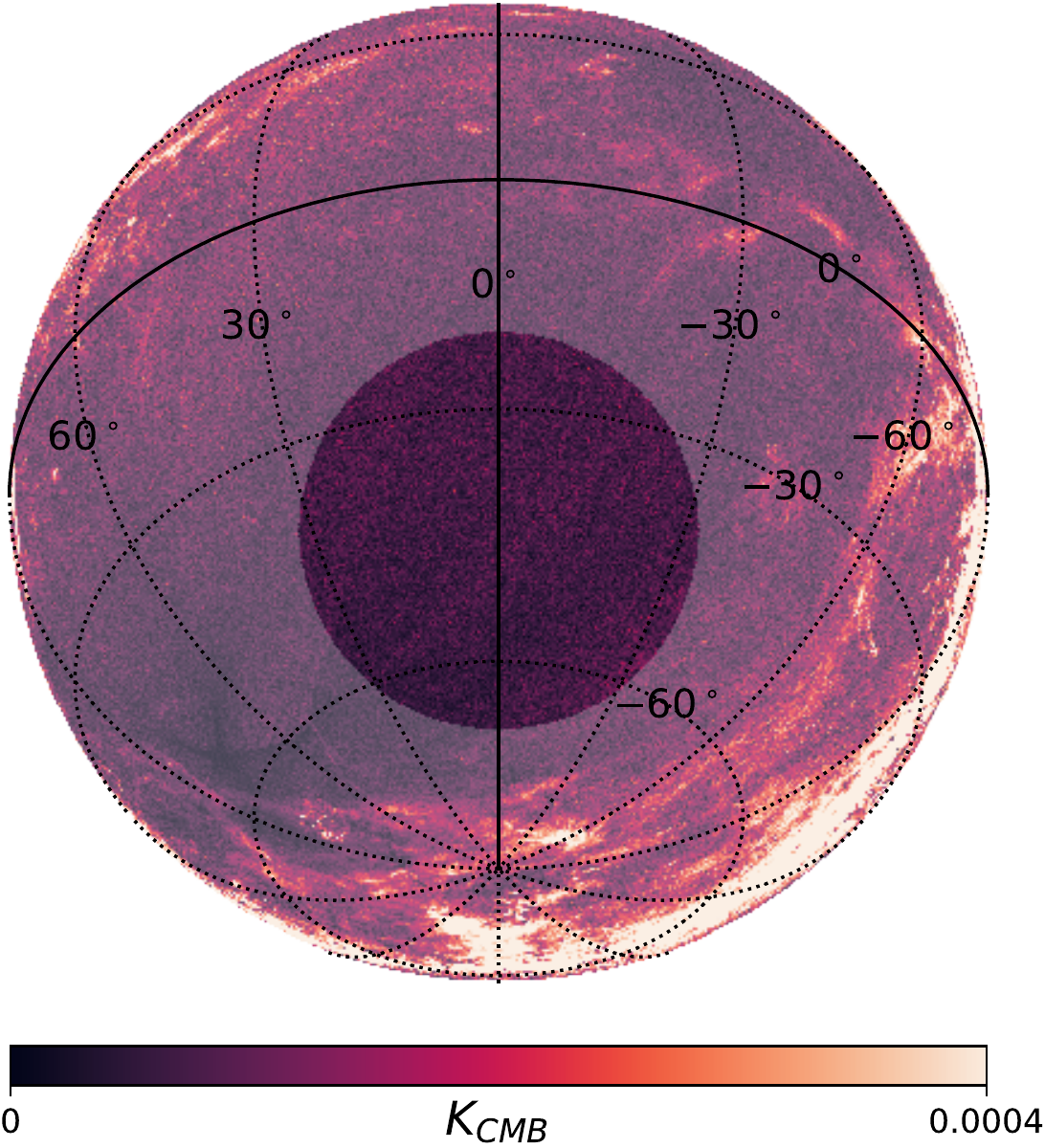}} \hskip 0.01\textwidth
        \subfloat[The southern masks\label{subfig:maskCMBS4}]{\includegraphics[width=0.24\textwidth]{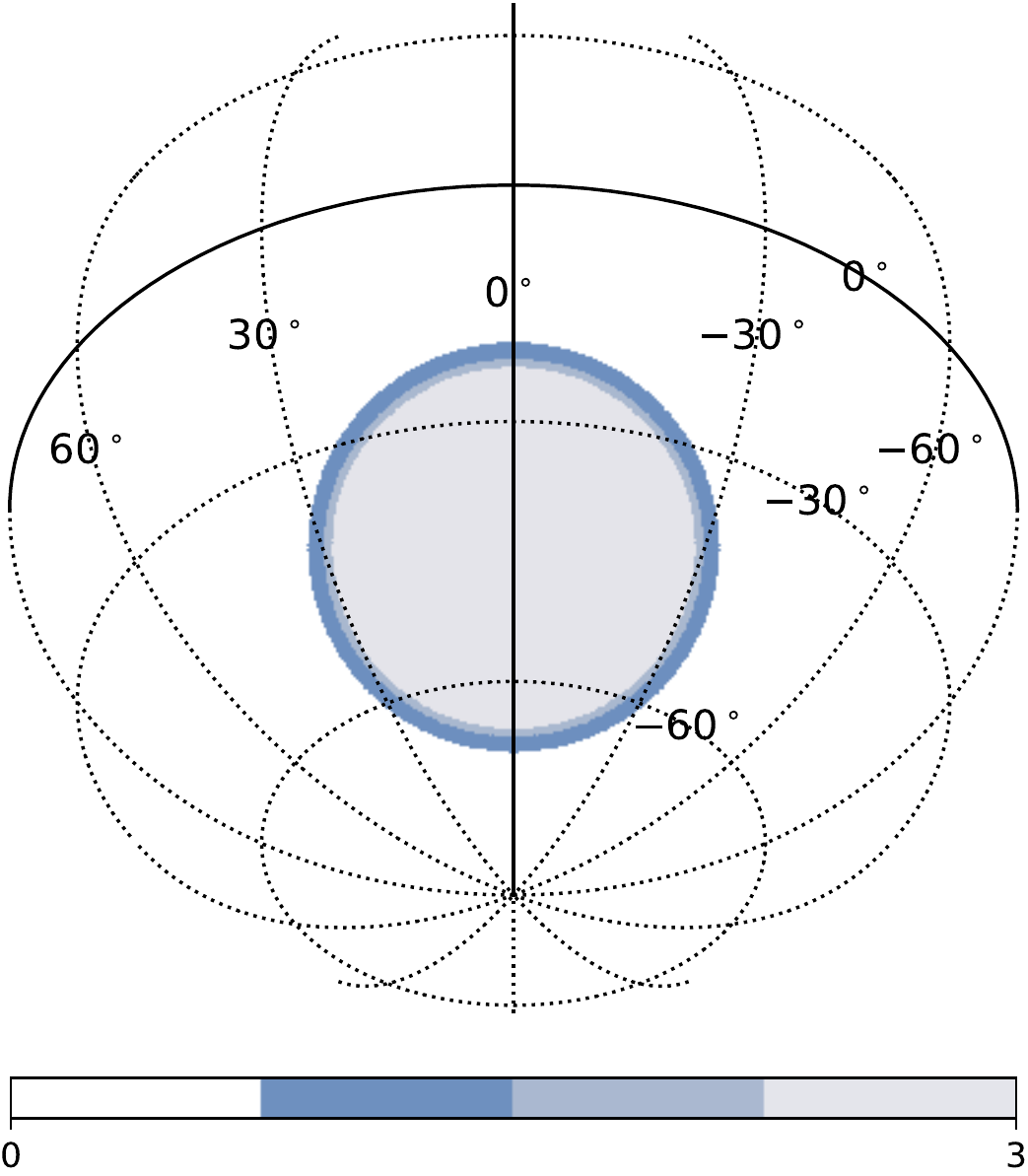}}
    \end{center}
    \caption{Orthogonal projection of the sky patches and masks in Equatorial coordinate. The left two plots show the selected northern sky patch, centered at ($RA = 197^\circ$, $Dec = 55^\circ$), and the masks, respectively. The patch extends from $RA=150^\circ$ to $RA=250^\circ$, and from $Dec=25^\circ$ to $Dec=70^\circ$. From outside to inside of the masks, they are the mask for smoothing all the input maps to the same resolution, for $EB$ leakage correction, and for the ILC methods. The right two plots show the southern sky patch and the masks. The sky patch is circular-shaped centered at $RA=0^\circ, Dec=-45^\circ$, which was used in the CMB-S4 tensor-scalar ratio forecasting. Both the background of $(a)$ and $(c)$ are the polarization intensity map $P=\sqrt{Q^2+U^2}$ observed by \textit{Planck} at 353 GHz.}\label{fig:skypatch_mask}
\end{figure}

Two configurations are adopted in this work, and both of them are assumed as future ground-based CMB experiments. For primordial gravitational waves detection, it should carry out a deep survey on a clean sky patch and accumulate enough scanning, and along this guidance we select two clean patches, which are distributed in northern hemisphere and southern hemisphere respectively. For the northern sky patch, we choose the area centered at ($RA\ 13.13h$, $Dec\ 55^\circ$), and it is about  $7\%$ sky coverage where it appears to be the patch with lowest galactic foreground according to the Planck 353 GHz polarization intensity map. For the southern sky patch, a circular patch was choosed with roughly $3\%$ sky coverage which is adopted in CMB-S4 primordial $B$ mode forecasting. The sky patches with Planck 353 GHz polarization emission intensity map is shown in Figure \ref{fig:skypatch_mask}\subref{subfig:skypatch} and \ref{fig:skypatch_mask}\subref{subfig:skypatchCMBS4}.

Due to the limitation from atmospheric emission and absorption effect, there are only a few frequency bands available for ground-based experiments. The frequency bands and the related beam resolution are listed in Table \ref{tab:setup}. The noise distribution is assumed to be homogeneous among the patch. The map depth for each frequency band is also listed in the same table.
The final observed multi-frequency maps were generated by adding the noise map realized from map depth and the simulated sky maps at each frequency. 
All the maps are pixelized in HealPix format at $\texttt{NSIDE}=512$. We also use the same map depth to generate 500 noise-only simulations which can be used to debias the noise at angular power spectral level. 

\begin{table}[bthp]
\caption{Instrumental specifications. For the northern patch, the map-depth and beam-size are taken from the reference\cite{2017PhRvD..95d3504A} with $f_{sky}\simeq7\%$. The configurations of CMB-S4 are taken from the reference\cite{CMBS4Forecast}.} 
    \centering
    \begin{threeparttable}
        \begin{tabular}{c|cc|cc|cc}
            \hline\hline
            Frequency&\multicolumn{2}{c|}{Map-depth T}&\multicolumn{2}{c|}{Map-depth P}&\multicolumn{2}{c}{Beam size}\\
            (GHz)&\multicolumn{2}{c|}{($\mu$K$\cdot$arcmin)}&\multicolumn{2}{c|}{($\mu$K$\cdot$arcmin)}&\multicolumn{2}{c}{(arcmin)}\\
            &Nothern&CMB-S4&Nothern&CMB-S4 &Nothern&CMB-S4\\\hline
            $20$\tnote{**}&  --   &$ 9.90$&  --   &$14.00$&  --   &$11.0$   \\
            $30$\tnote{*} &$7.49$ &$ 6.15$&$10.60$&$ 8.70$&$ 67.0$&$77.0$\\
            $40$          &$7.49$ &$ 5.80$&$10.60$&$ 8.20$&$ 63.0$&$58.0$\\
            $85$\tnote{*} &$1.21$ &$ 1.13$&$ 1.72$&$ 1.60$&$ 40.0$&$27.0$\\
            $95$          &$0.96$ &$ 0.92$&$ 1.35$&$ 1.30$&$ 30.0$&$24.0$\\
            $145$\tnote{*}&$0.90$ &$ 1.41$&$ 1.27$&$ 2.00$&$ 19.0$&$16.0$\\
            $155$         &$0.87$ &$ 1.41$&$ 1.24$&$ 2.00$&$ 17.0$&$15.0$\\
            $220$         &$1.34$ &$ 3.68$&$ 1.90$&$ 5.20$&$ 11.0$&$11.0$\\
            $270$         &$1.16$ &$ 5.02$&$ 1.64$&$ 7.10$&$  9.0$&$ 8.5$\\
            \hline\hline
        \end{tabular}
        \begin{tablenotes}
            \footnotesize
            \item[$*$] These frequency channels are only used in $8$ band case on northern sky.
            \item[$**$] The $20$ GHz channel only exist in CMB-S4 configuration. 
            \end{tablenotes}
        \end{threeparttable}
    \label{tab:setup}
\end{table}

\subsection{Data Processing}\label{sec:dataproc}

After we get the multi-frequency maps, we construct the 3 data sets: \textbf{(1)} 5 frequency bands without point sources, we will call it \textbf{Baseline}; \textbf{(2)} 5 frequency bands with point sources; \textbf{(3)} 8 frequency bands with point sources. All the data sets will go through the procedure list in the following:

\begin{enumerate}
    \item \textbf{PILC:} We will first make sure the maps for all frequencies should share the same beam size, so we deconvolve their own beam size and convolve a beam size that is larger than beam sizes for all the frequency bands (here we choose the final beam size as $70$ arcmin for northern sky patch and $80$ arcmin for CMB-S4 sky patch), we also multiply an apodized mask (most outside) shown in Figure \ref{fig:skypatch_mask}\subref{subfig:mask} before we smooth the maps to avoid numerical precision problems at the edge area. Then we go through the PILC procedure to construct the clean $QU$ maps and the foreground residual maps described in section \ref{sec:pilc}. Finally, we use NaMaster\cite{2019MNRAS.484.4127A} to estimate the angular power spectra from clean $QU$ maps as well as residual maps.
    \item \textbf{ILC on $B$ maps:} We do the same thing as in PILC method to smooth the multi-frequency maps into the same beam size, then we construct the $EB$ leakage-free $B$ maps using the template cleaning described in section \ref{sec:bilc}, followed by three different methods (pixel, harmonic, needlet) on the pseudo-scalar $B$ maps, then we get the clean $B$ maps and foreground residual $B$ maps with ILC methods in three domains. Finally, we estimate the angular power spectra from clean and residual $B$ maps.
    \item \textbf{Angular power spectra estimation:} The pseudo-power spectra are generated by using the python package of NaMaster (\texttt{pymaster\footnote{\url{https://github.com/LSSTDESC/NaMaster}}}). A $6$ degree C2 type apodization was applied on the innermost mask of Figure \ref{fig:skypatch_mask}\subref{subfig:mask}. The power spectra are binned every $40$ multipole moments from $\ell=40$ to $280$, according to the scale of sky patch.
    \item \textbf{Noise debias:} We use the 500 noise only simulations to go through the ILC processing while keeping the weights unchanged to reconstruct 500 reconstructed noise only maps, then estimate the mean value over these noise simulations to get the noise angular power spectra $N_l$, finally subtract it from the clean spectra calculated in items 1 and 2 to get the unbiased clean angular power spectra.
    \item \textbf{constraint on $r$:} To show how good our methods are, we also use the calculated unbiased clean spectra to constraint the tensor to scalar ratio $r$. The full posterior for the individual band powers is non-Gaussian. However, for high enough multipoles (usually $\ell\ge30$) the central limit theorem justifies a Gaussian approximation\cite{2008PhRvD..77j3013H}. The lowest multipole $\ell$ in our binning scheme is $30$ thus the Gaussian approximation is valid:
\begin{align}\label{eq:likelihood}
    -2\ln\mathcal{L} = (\hat{C}_{\ell_b}^{BB}-C_{\ell_b}^{BB})^T{\rm Cov}(\hat{C}_{\ell_b}^{BB},\hat{C}_{\ell_b}^{BB})^{-1}(\hat{C}_{\ell_b}^{BB}-C_{\ell_b}^{BB}) +  const.
\end{align}
where the observed power spectrum is denoted as $\hat{C}_{\ell_b}^{BB}$. $C_{\ell_b}^{BB}$ denotes the theoretical power spectrum which was calculated by \texttt{CAMB} code. The likelihood only concerns the $BB$ spectra at $\ell\ge30$, where the primordial gravitational wave and lensing effect matters, so we fixed all the parameters to the best fit value from Planck-2018\cite{2018arXiv180706209P}, while only free $r$ and $A_L$. The covariance term is estimated from the signal+noise simulations.
We sample this likelihood for these two parameters using the \texttt{CosmoMC}\cite{2002PhRvD..66j3511L}\footnote{\url{https://cosmologist.info/cosmomc/}} package. The posteriors are summarized by \texttt{GetDist}\cite{Lewis:2019xzd}\footnote{\url{https://pypi.org/project/GetDist/}} package.
\end{enumerate}

%% file: results.tex
%


In this section, we present both analytical and numerical results obtained from ILC analysis in different cases. 
First, We focus on the northern sky patch ILC analysis, the results of which are summarized in the Section \ref{subsec:baseline}  and \ref{subsec:pointsources}.  Then we provide the results of the southern sky patch of the CMB-S4  in Section \ref{subsec:CMBS4}.
We construct the cleaned CMB maps estimated by ILC component separation, as well as residual maps of different frequency bands, which is the difference map between the estimated foreground maps and the simulated input foreground maps. We compute the angular power spectra of each of these maps and compare them to see its efficiency.  We discuss the influence of point sources on ILC analysis methods. Finally, we discuss the foreground residual from ILC methods and their influence on uncertainty on $r$ in Section \ref{sec:parameter}. 

\subsection{Results for baseline case} \label{subsec:baseline}

\begin{figure}
    \begin{center}
        \subfloat{\includegraphics[width=0.2\textwidth]{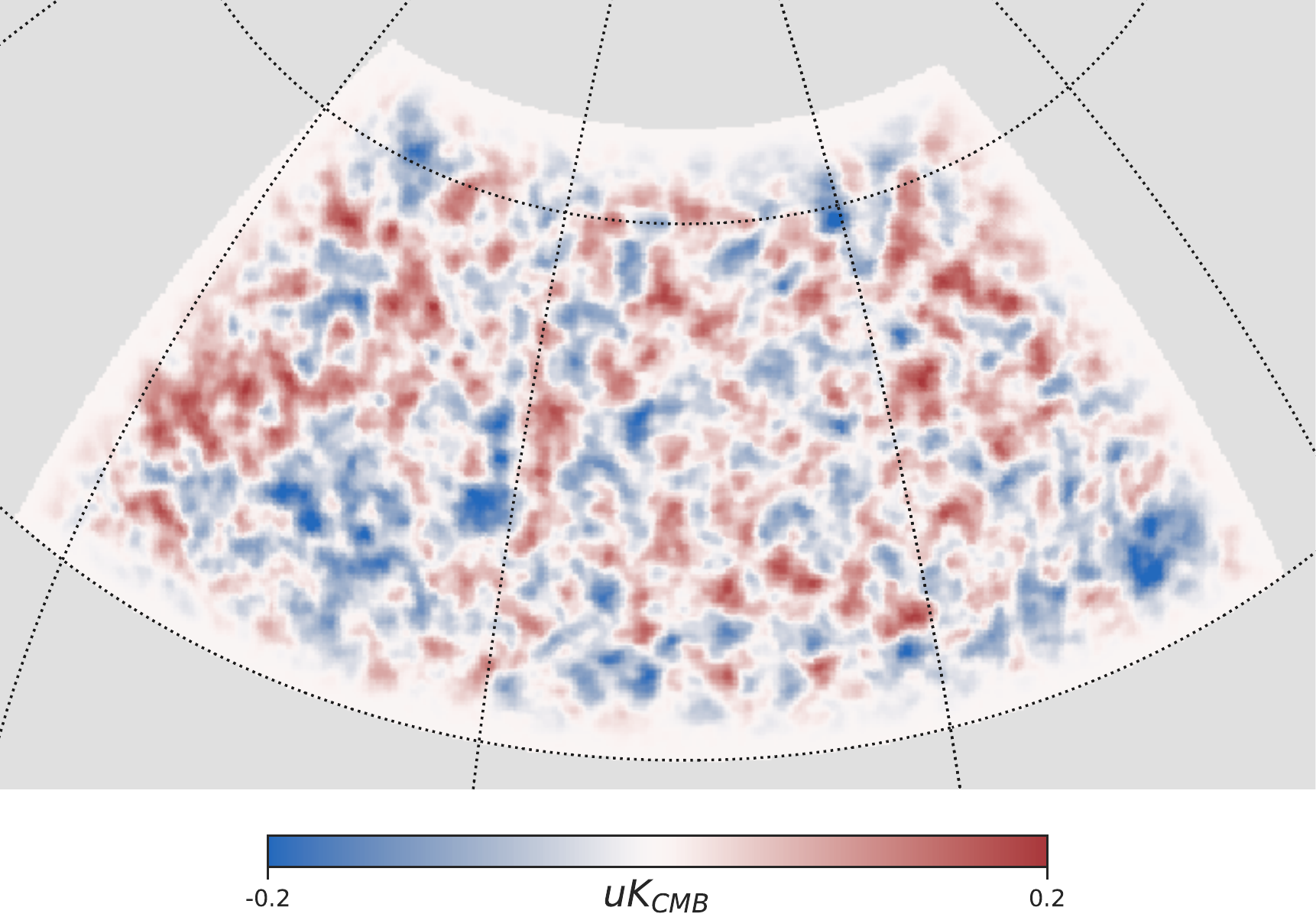}}\hskip 0.5em
        \subfloat{\includegraphics[width=0.2\textwidth]{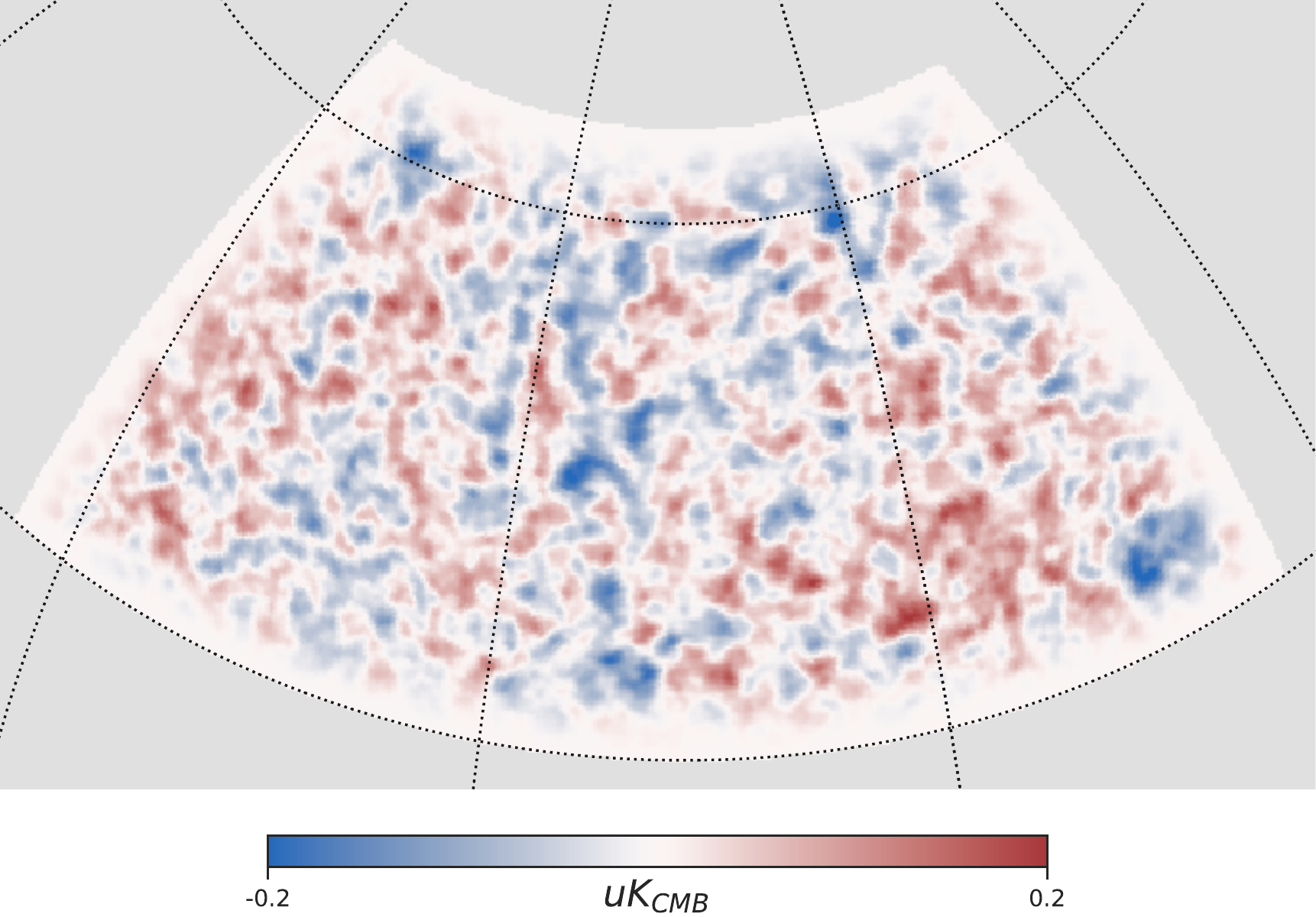}}\hskip 0.5em
        \subfloat{\includegraphics[width=0.2\textwidth]{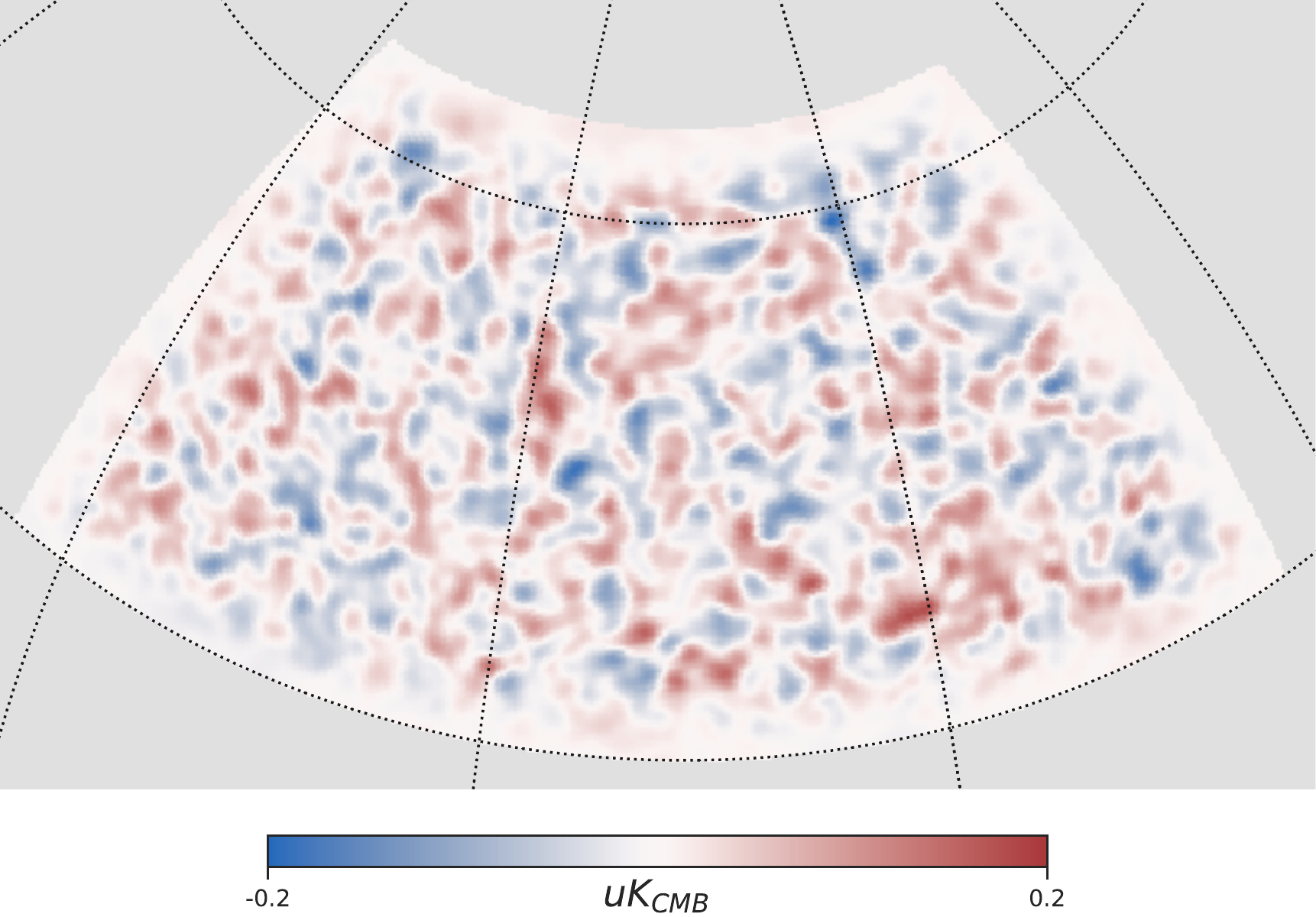}}\hskip 0.5em
        \subfloat{\includegraphics[width=0.2\textwidth]{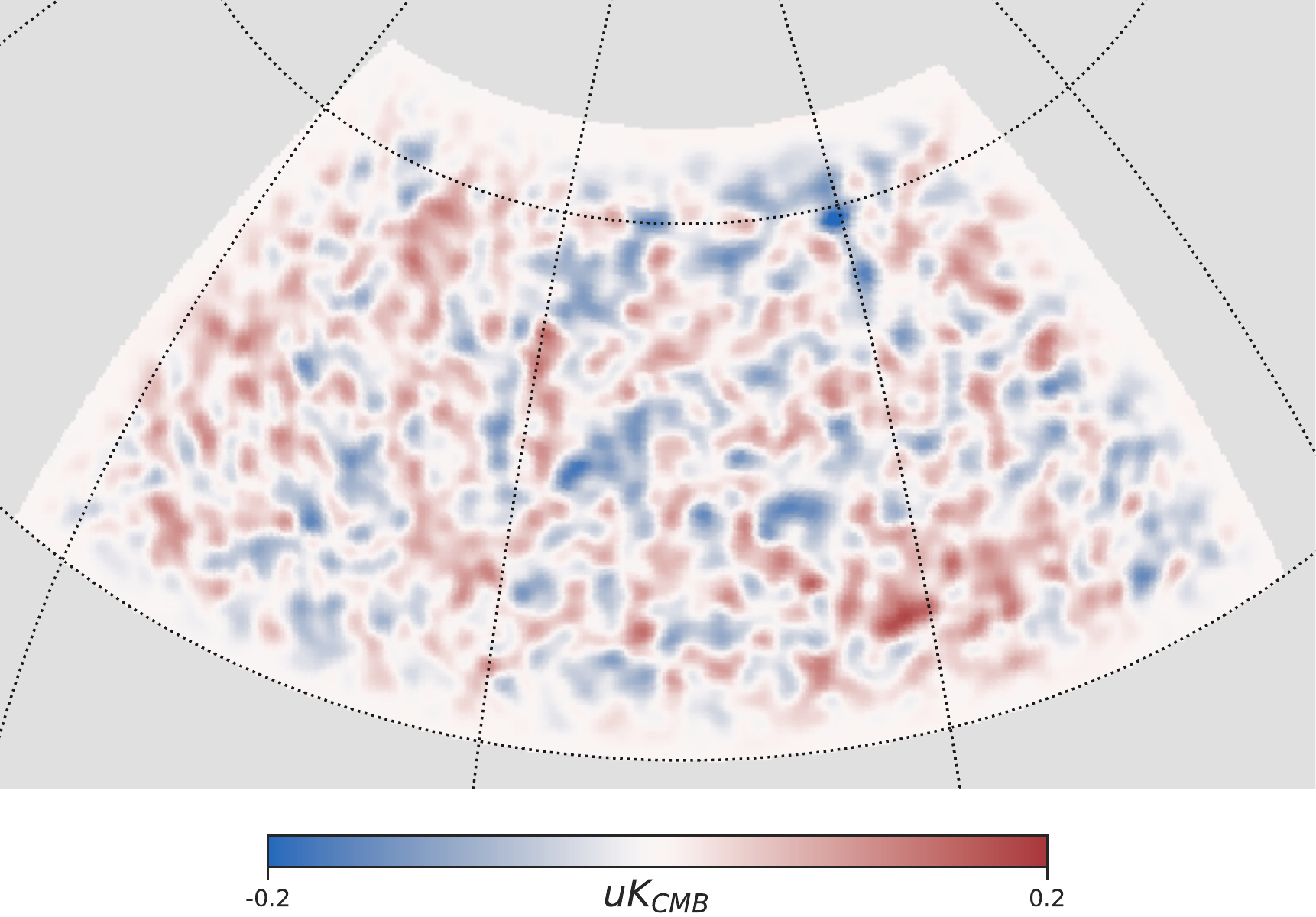}}\\
        \subfloat{\includegraphics[width=0.2\textwidth]{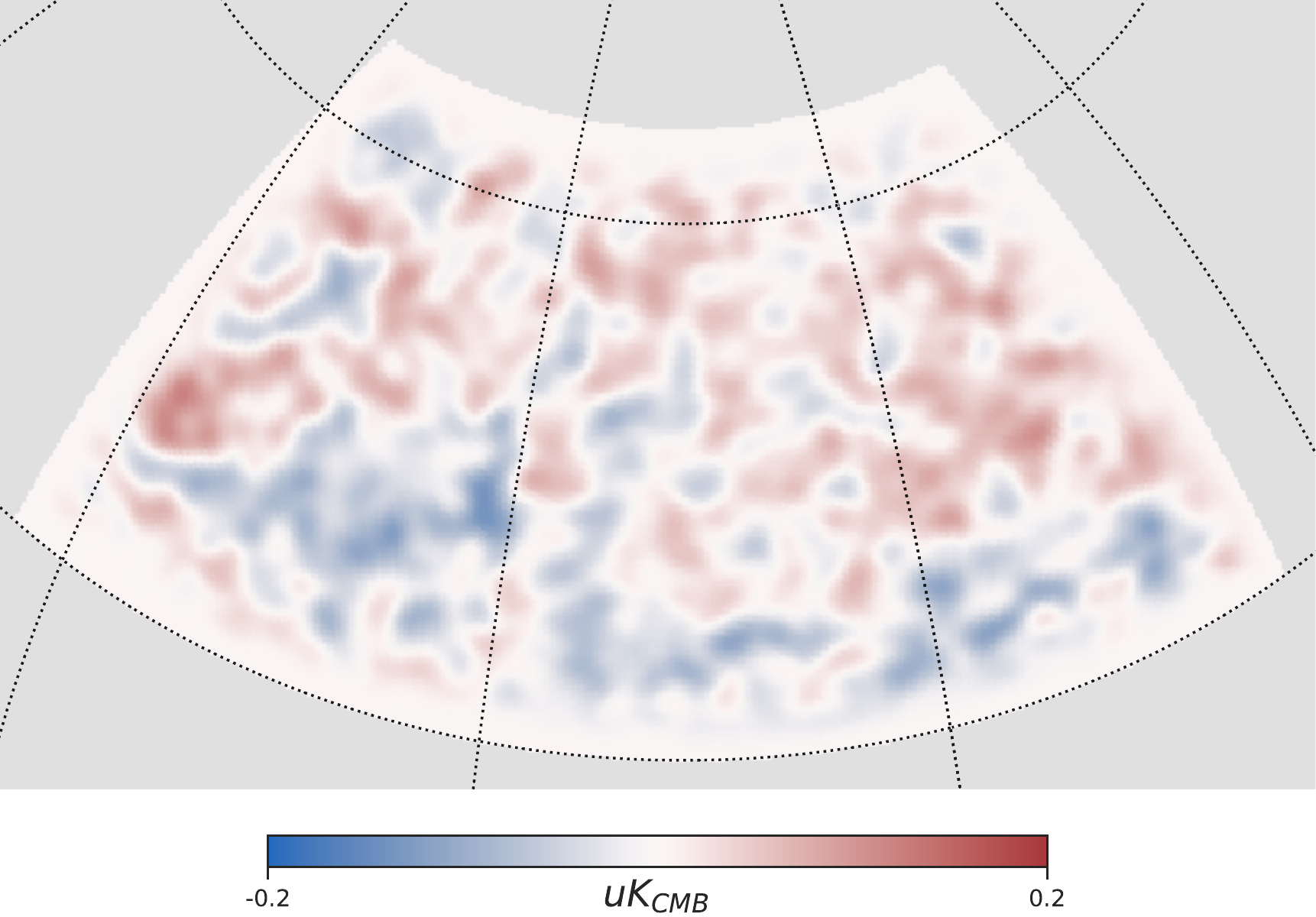}}\hskip 0.5em
        \subfloat{\includegraphics[width=0.2\textwidth]{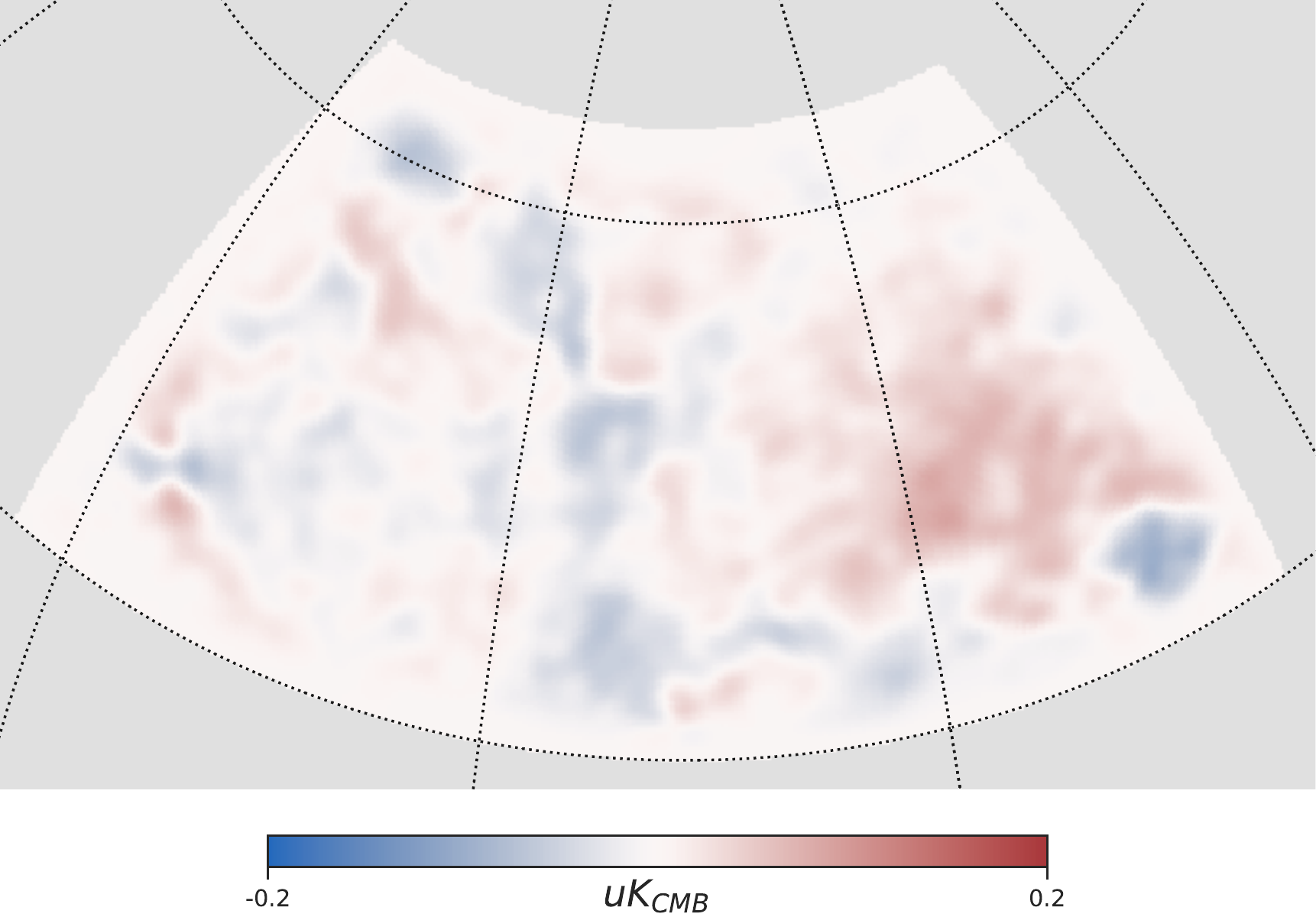}}\hskip 0.5em
        \subfloat{\includegraphics[width=0.2\textwidth]{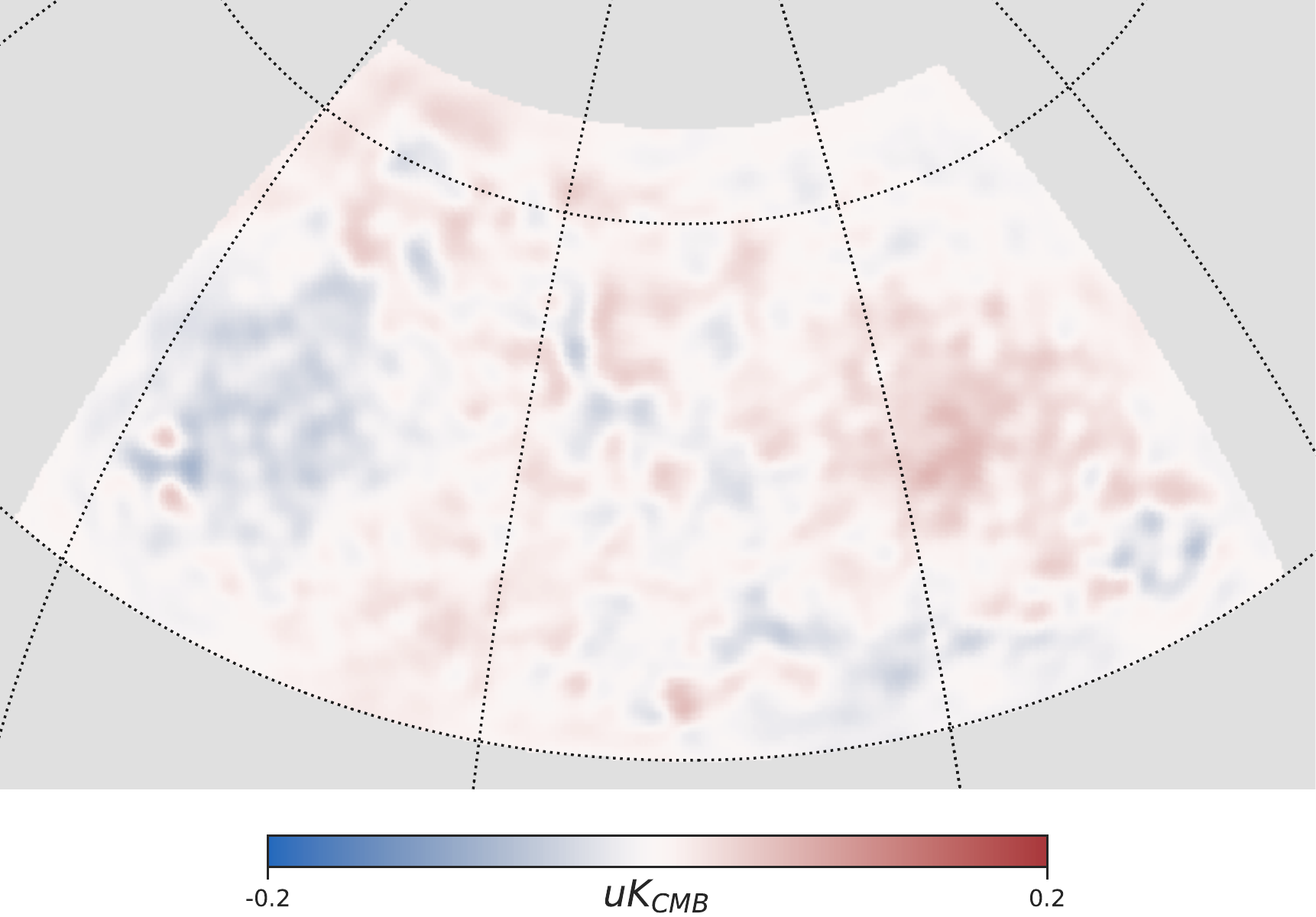}}\hskip 0.5em
        \subfloat{\includegraphics[width=0.2\textwidth]{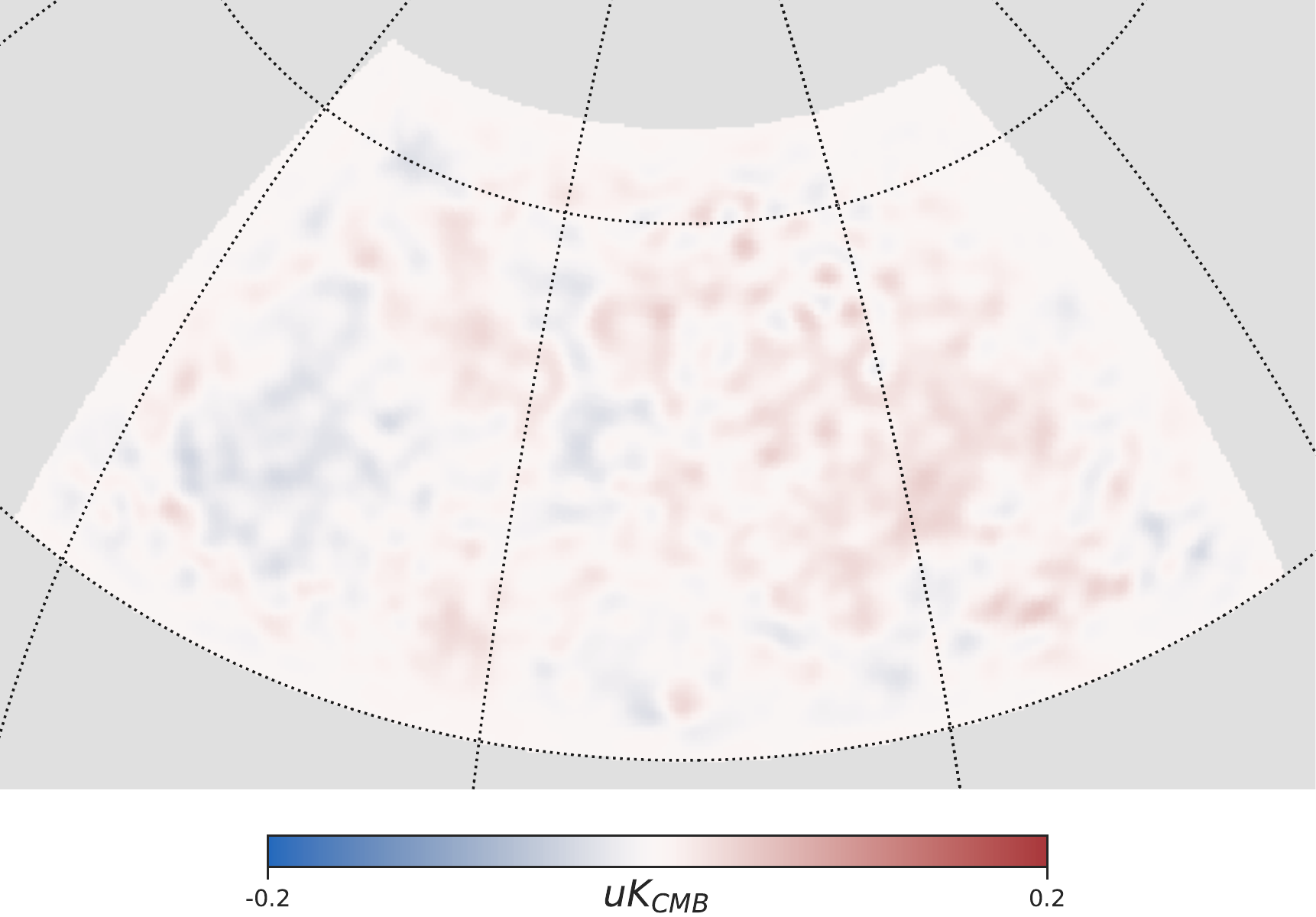}}
    \end{center}
    \caption{$B$ maps of the results of ILC (first row) and foreground residual (second row). Each column represents a kind of ILC method. From left to right, They correspond PILC, pixelILC, harmonicILC and NILC  respectively. The edges of all the maps are apodized by 6 degree. }\label{fig:5band_wops}
\end{figure}

The ILC results for our baseline case, in which the inputted foreground model do not contain point sources, are presented in Figure \ref{fig:5band_wops}, the first row provide cleaned $B$ maps after ILC analysis with 5 channels, and in the second row, it shows the residual foreground $B$ maps. For the PILC method, the output $QU$ maps were transformed to $B$ map by the same method mentioned in Section \ref{sec:bilc}. We plot all maps after smoothing with a Gaussian beam of $70$ arcmin. The cleaned maps in the upper panels are, in fact, a sum of CMB, residual foreground and residual noise. Because the sky patch chosen here avoids the Galactic plane, it is hard to distinguish the residual foreground contamination directly. The bottom panel shows the residual foreground maps for each ILC method from the same linear combination but inputting foreground only. From the colors of the 4 foreground residual maps, it can be seen that PILC has a larger foreground residual, which is very different from pixelILC, harmonicILC, and NILC. 

\begin{figure}[bthp]
    \begin{center}
        \includegraphics[width=0.8\textwidth]{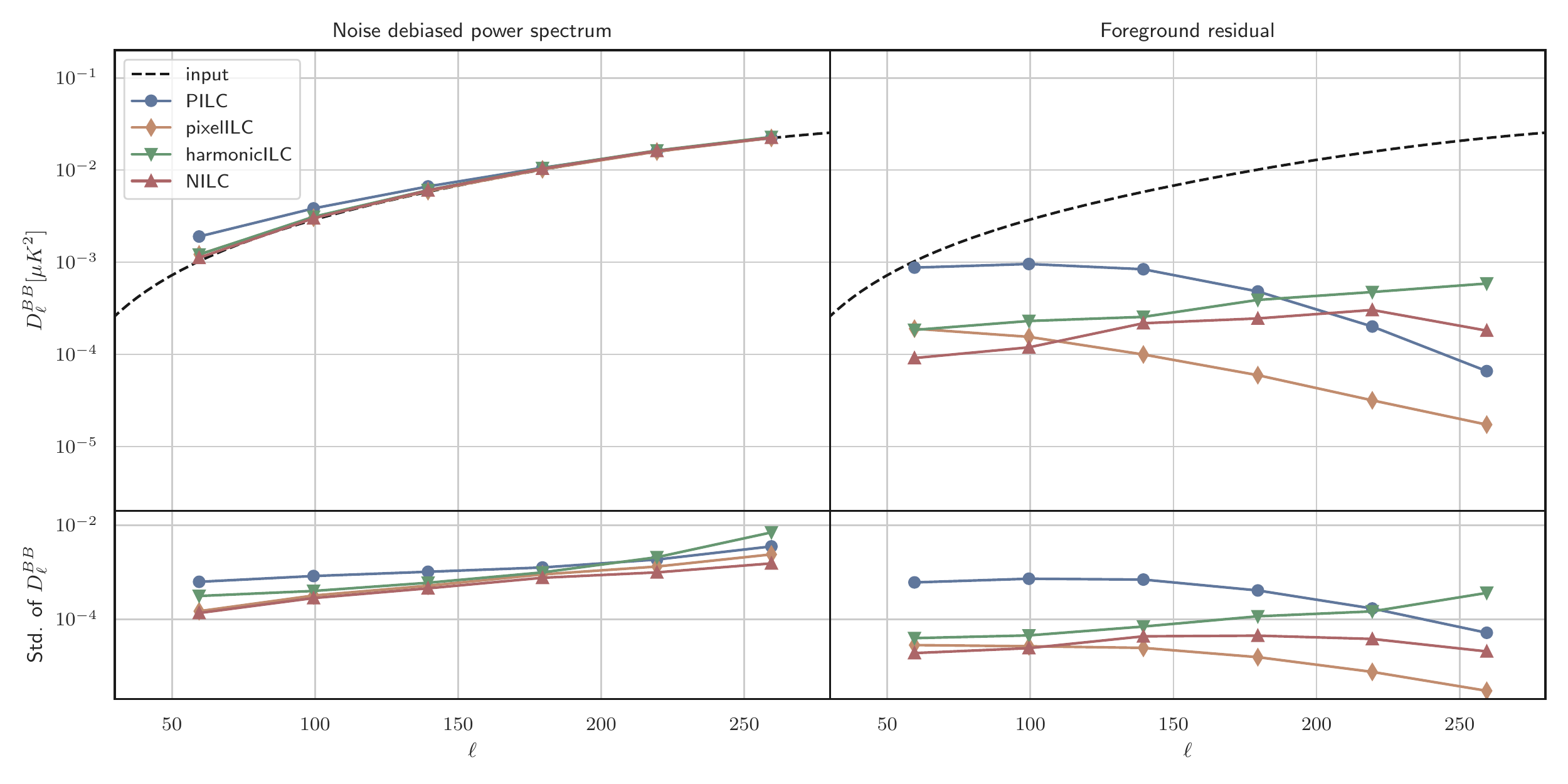}
    \end{center}
    \caption{The noise de-biased $B$ mode power spectra of resulting and residual maps shown in Figure \ref{fig:5band_wops}. The bottom panel shows the standard deviation of the de-biased power spectra, which was estimated from 200 different realizations considering the randomness of both CMB and noise.}\label{fig:powers_noPs}
\end{figure}

A straightforward way to see the effectiveness of different ILC methods is to check the power spectrum of residual foreground, which were calculated directly from the residual maps shown in the bottom panel of Figure \ref{fig:5band_wops}. The noise de-biased pseudo-power spectra are shown in Figure \ref{fig:powers_noPs}. The power spectra calculation procedure is described in item 3 of section \ref{sec:dataproc}. Among them, the results of PILC are analyzed on the $QU$ maps, and we use NaMaster to correct $EB$ leakage and the effect caused by mask, the results of Other ILC methods are based on the input pure B map, so we only do the corrections of mask effect.


Figure \ref{fig:powers_noPs} shows the estimated $BB$ power spectra from different ILC methods. The right panel of Figure \ref{fig:powers_noPs} shows the comparison of residual foreground power spectra. We can see that ILC methods working in the pixel domain, for example, the cases of PILC and pixelILC, have larger foreground residuals on large scale than those on small scale. That's reasonable because the signal on different scales share the same weights in pixel domain ILC. Generally speaking, minimizing the variance of the resulting map can be approximated as minimizing $\sum_{\ell}(2\ell+1)C_{\ell}$, where $C_{\ell}$ is the power spectrum of the resulting map, and it leads to a situation that in the ILC analysis in the pixel domain, the small scale signal dominates the minimization, so the weights were chosen to reduce more small scale foreground automatically. For harmonic ILC and NILC, they did not have such property since they can apply different weights on different scales. Comparing PILC and pixel ILC, it's clear that working on $B$ map is better than working on $QU$ maps. Among all the methods, the power spectrum of residual foreground from NILC is about $1 \sim 2$ orders of magnitude smaller than the signal on all scales, this is due to the fact that NILC can track the signals on both large and small scales and in different spatial positions for different modes while keeping the CMB unaffected. We also find that the estimated power spectrum from NILC is the closest to the input CMB power spectrum from the left panel of Figure \ref{fig:powers_noPs}. All these indicate that NILC on $B$ map is the most promising method for CMB $B$ modes signal reconstruction.

Overall, the power spectrum reconstructed by four ILC methods is basically consistent with the input power spectrum, except for some obvious deviation on large scale. The power of foreground residual is on the order of $10^{-3}$ to $10^{-4}$ of $D_{\ell}^{BB}$, and the bias of the ratio of tensor to scalar $r$ that caused by the residual is in the order of $10^{-3}$.

\subsection{Results of the case involving point sources} \label{subsec:pointsources}

\begin{figure}
    \begin{center}
        \subfloat{\includegraphics[width=0.2\textwidth]{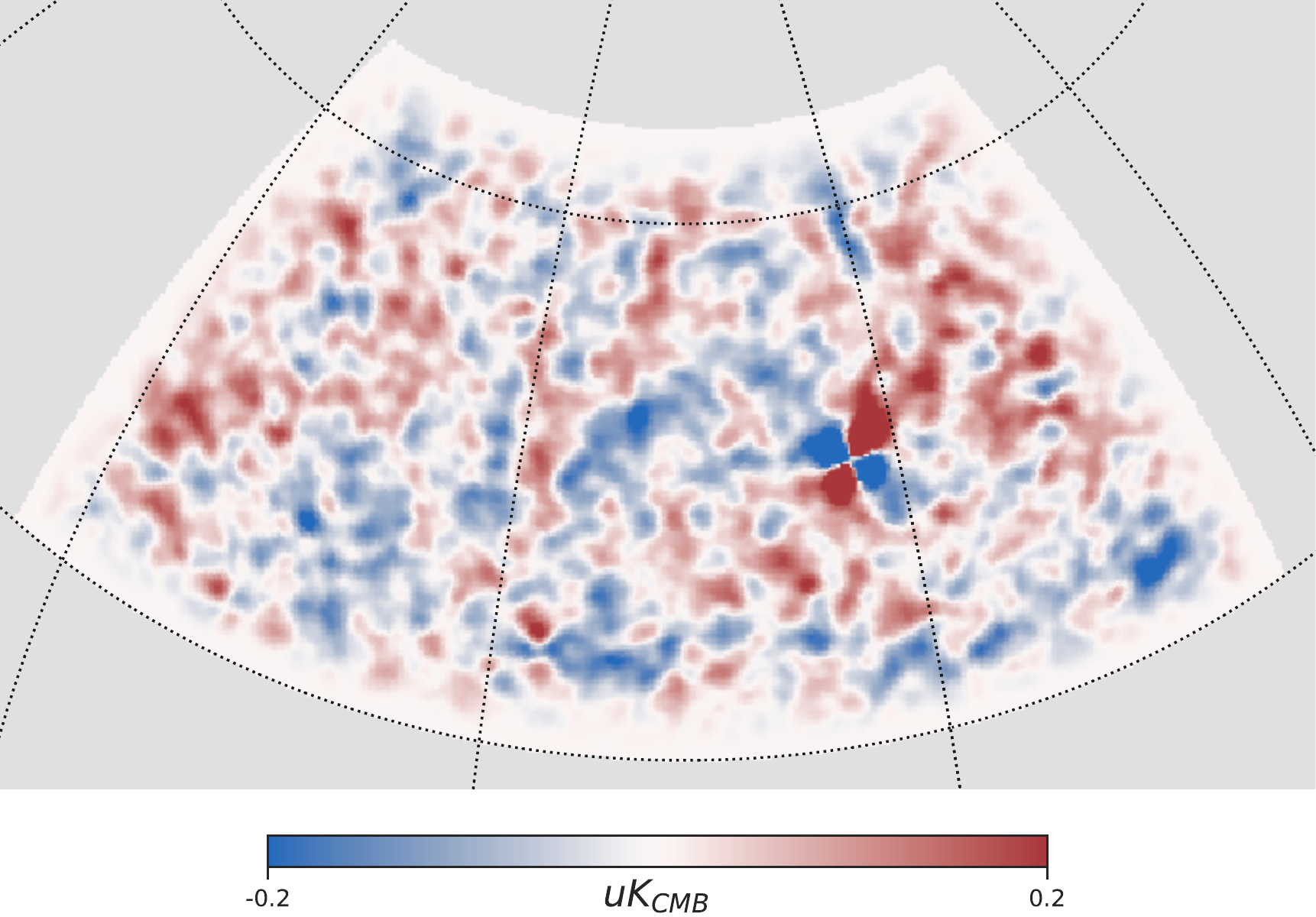}}\hskip 0.5em
        \subfloat{\includegraphics[width=0.2\textwidth]{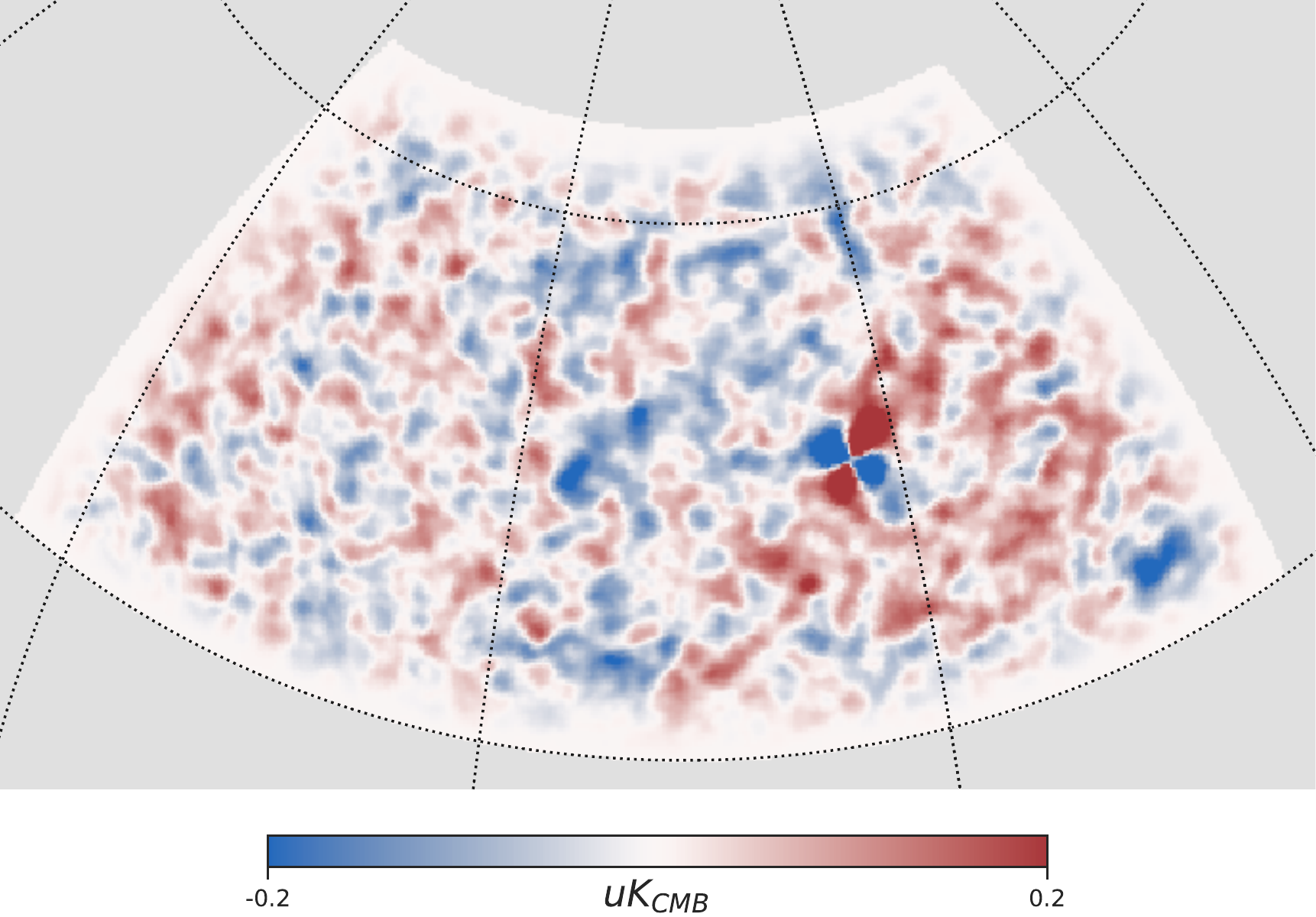}}\hskip 0.5em
        \subfloat{\includegraphics[width=0.2\textwidth]{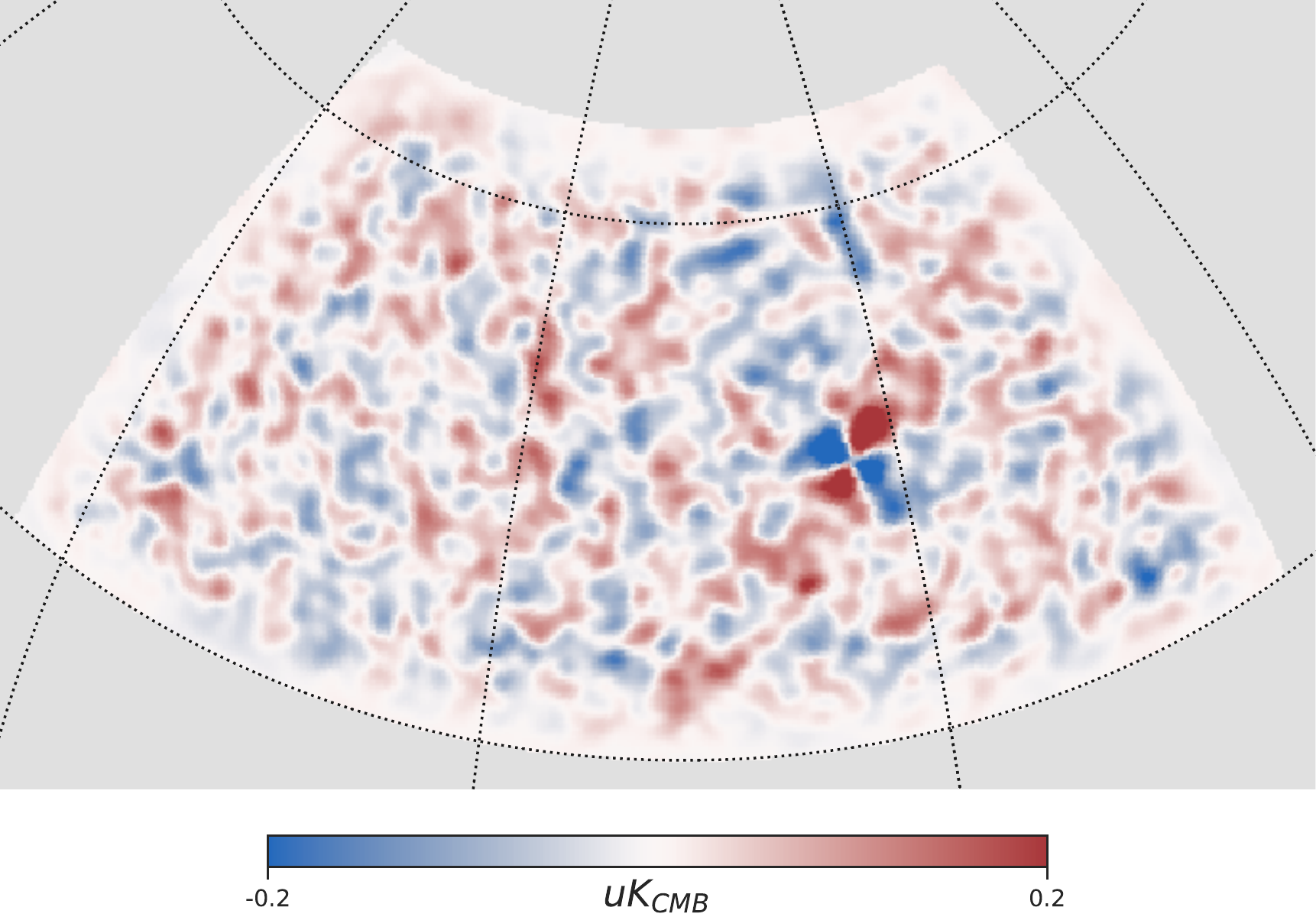}}\hskip 0.5em
        \subfloat{\includegraphics[width=0.2\textwidth]{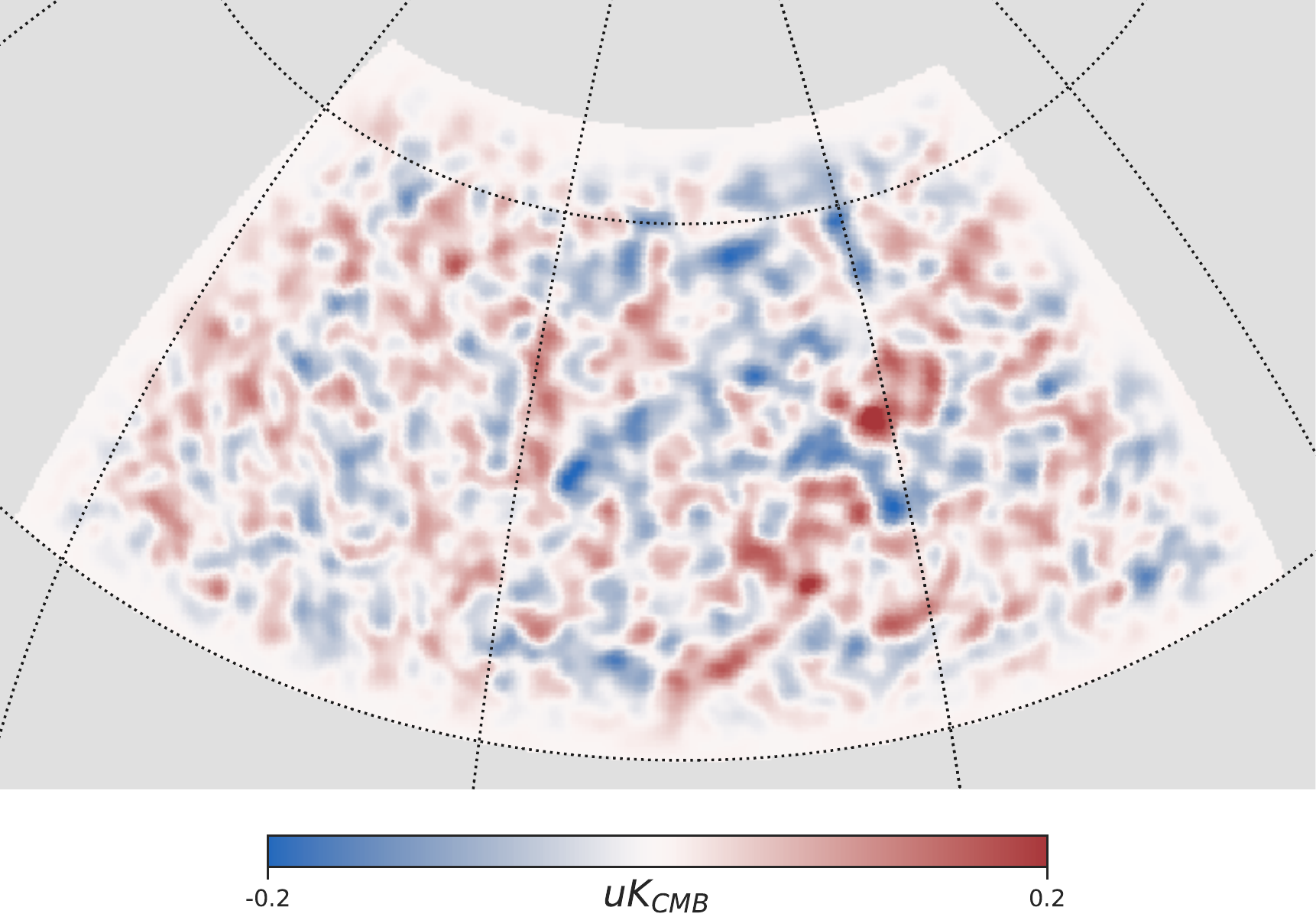}}\\
        \subfloat{\includegraphics[width=0.2\textwidth]{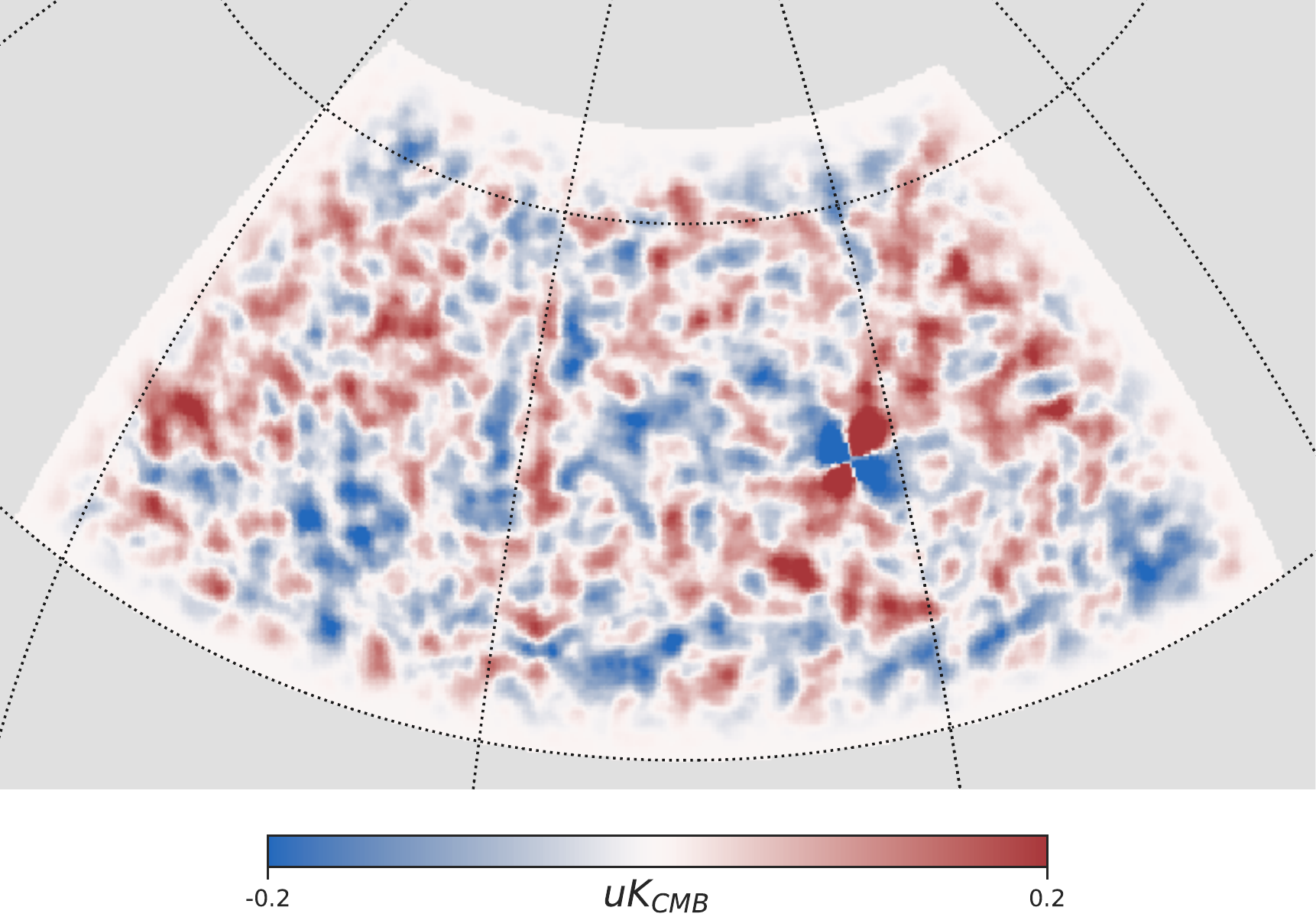}}\hskip 0.5em
        \subfloat{\includegraphics[width=0.2\textwidth]{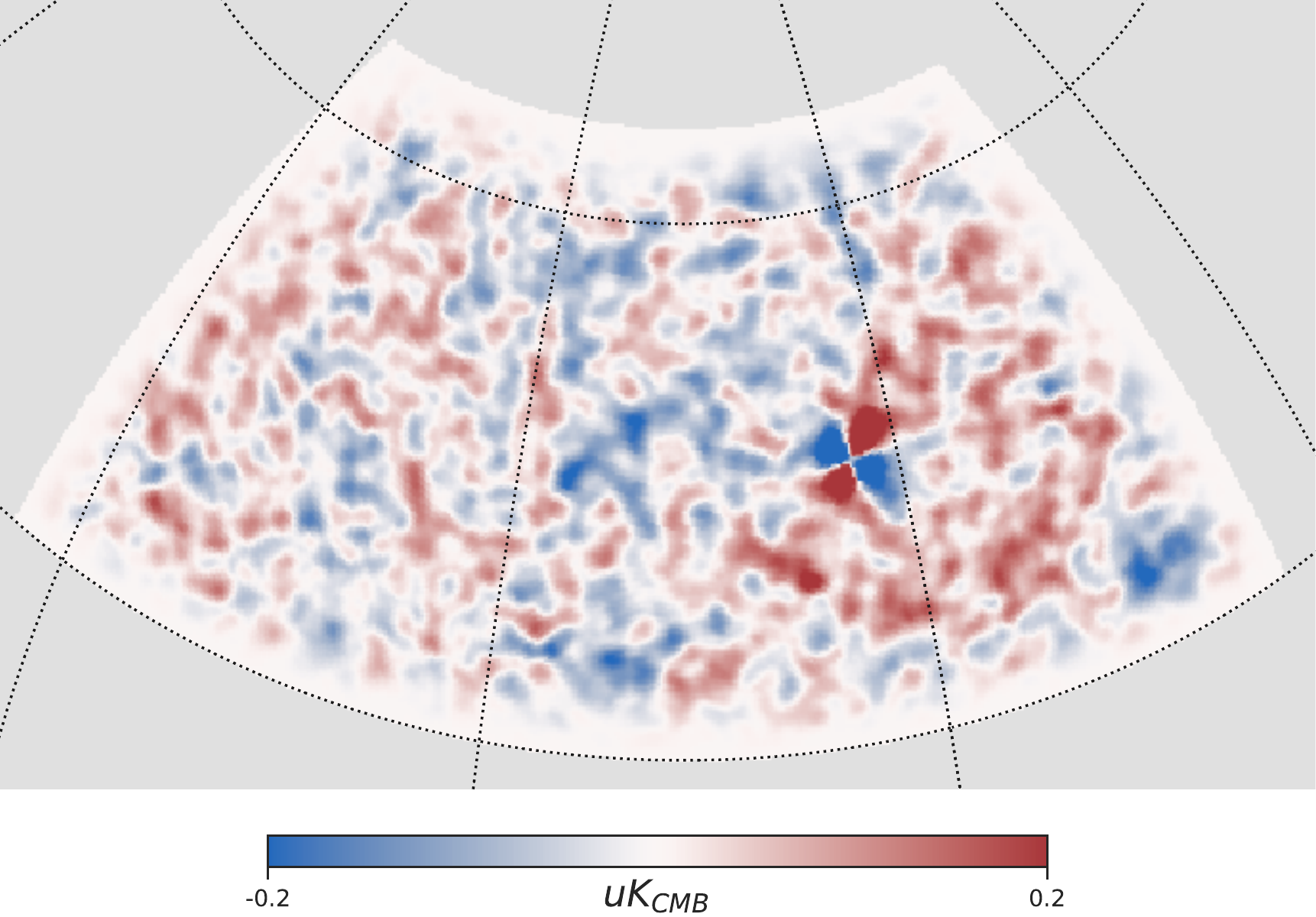}}\hskip 0.5em
        \subfloat{\includegraphics[width=0.2\textwidth]{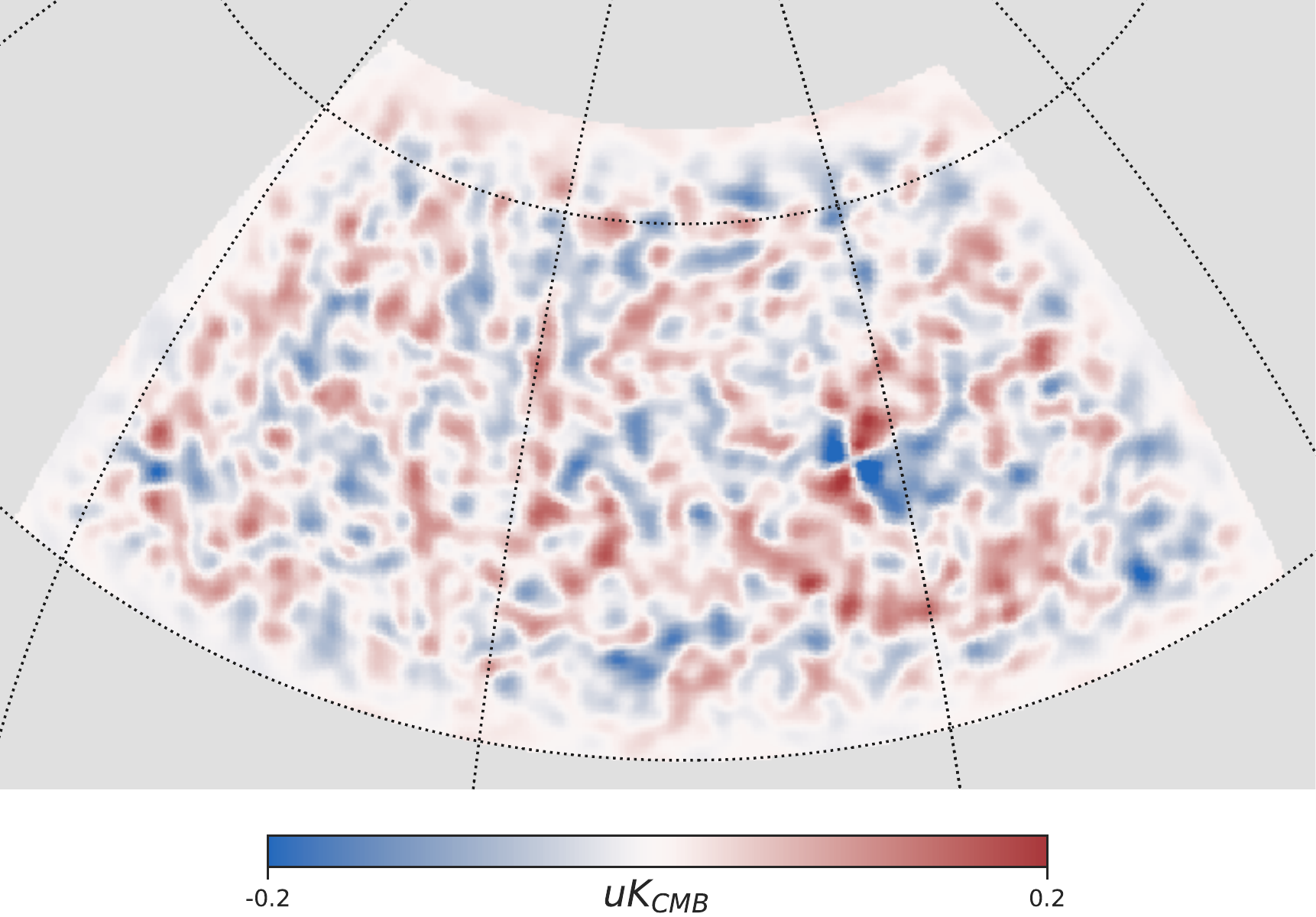}}\hskip 0.5em
        \subfloat{\includegraphics[width=0.2\textwidth]{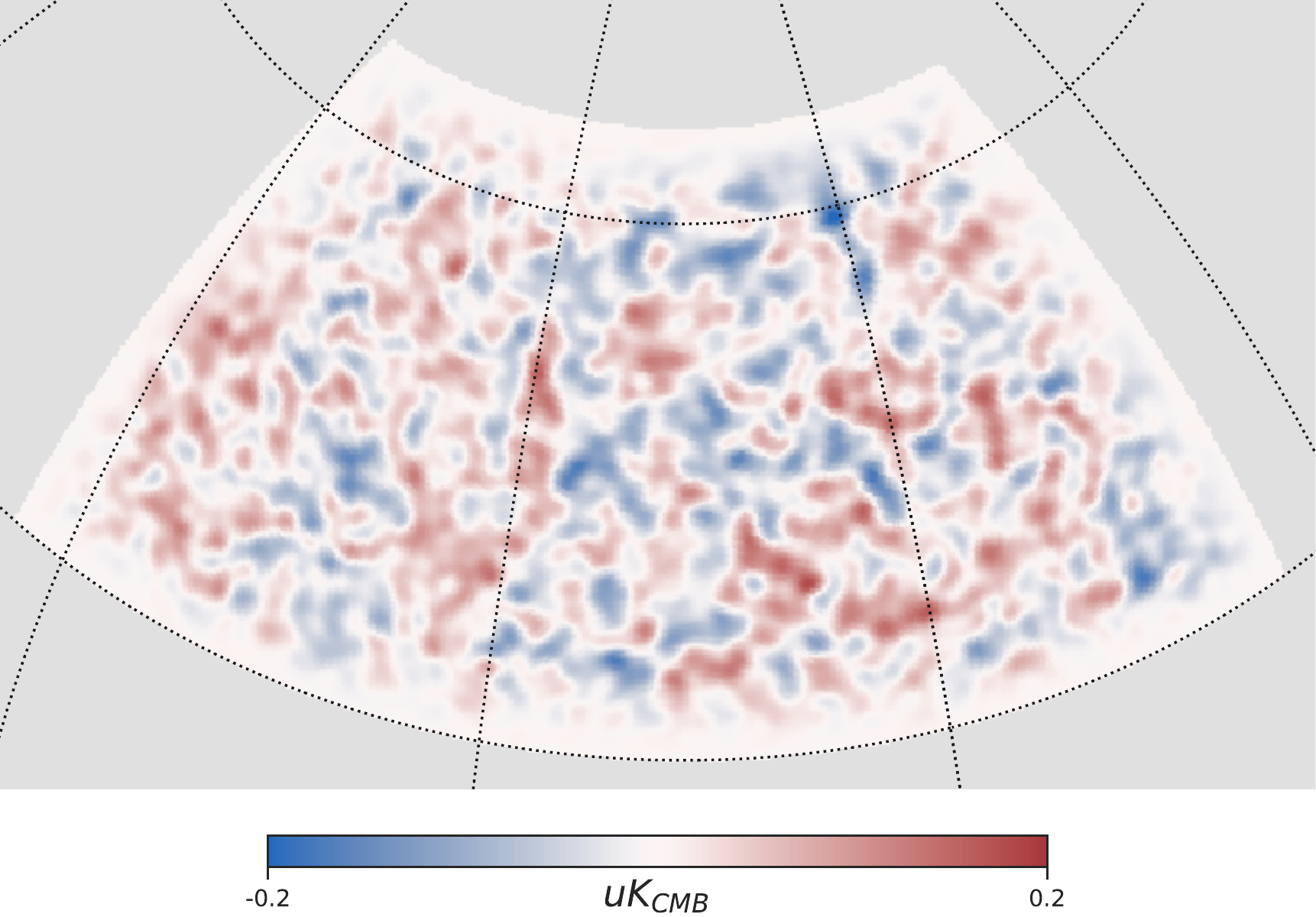}}
    \end{center}
    \caption{The resulting $B$ maps. The point sources are considered in the inputting foreground. The edge of each map was apodized by 6 degree. Each column represents a different ILC method. They correspond PILC, pixelILC, harmonicILC and NILC from left to right, respectively. The first row shows the results of 5 channels and the second row is the results of 8 channels. 
    }\label{fig:resulting_maps_ps}
\end{figure}

\begin{figure}
    \begin{center}
        \subfloat{\includegraphics[width=0.2\textwidth]{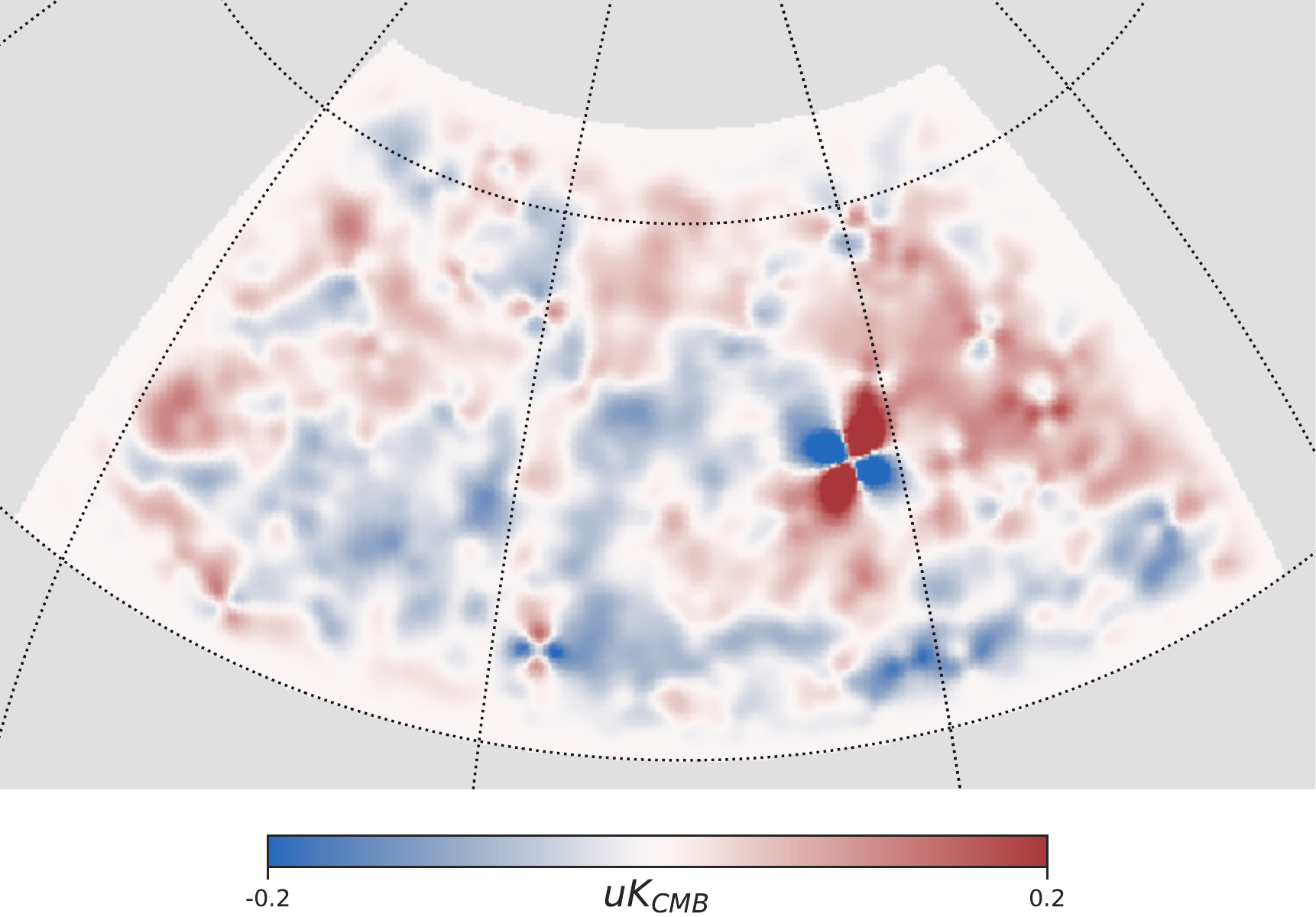}}\hskip 0.5em
        \subfloat{\includegraphics[width=0.2\textwidth]{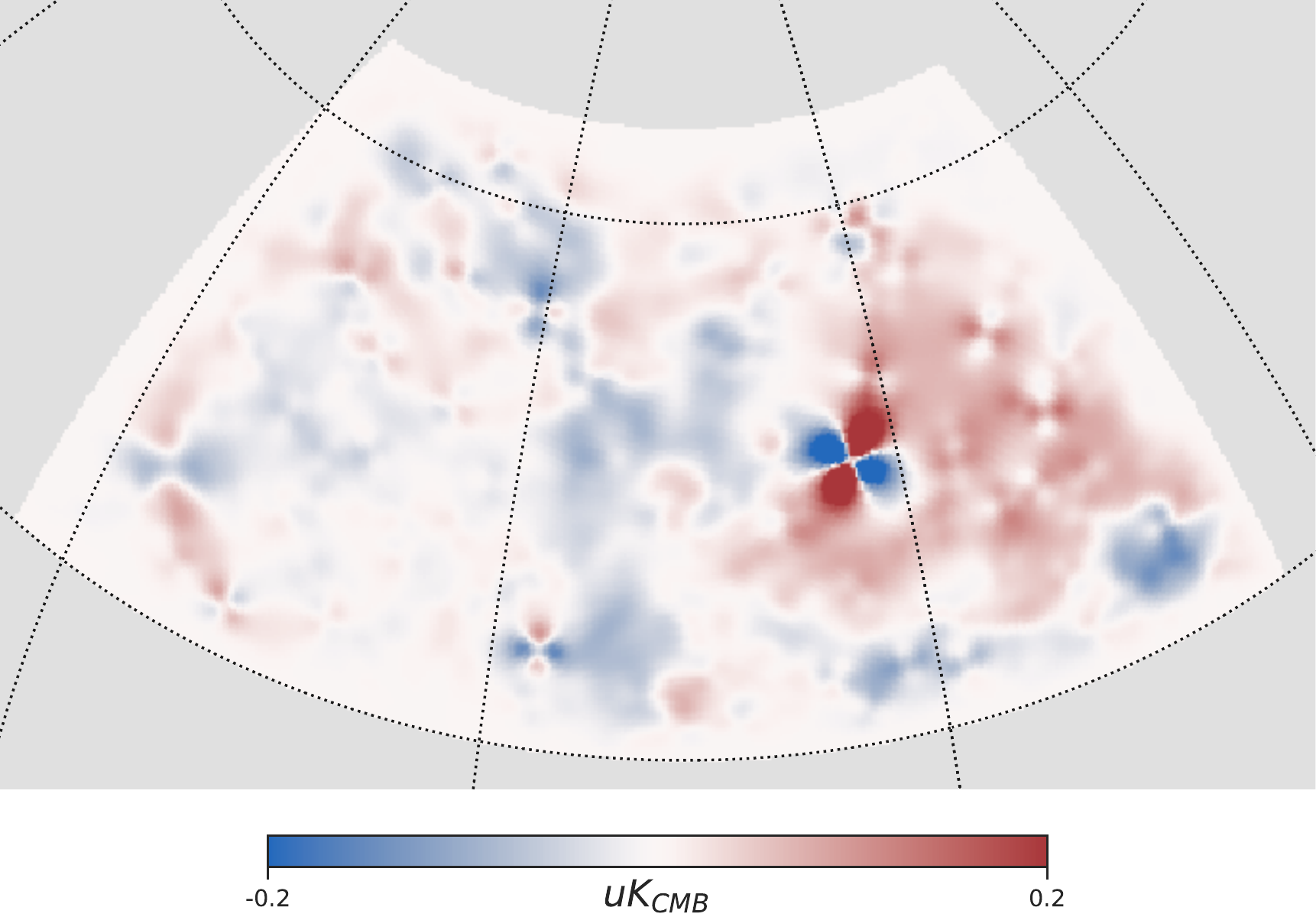}}\hskip 0.5em
        \subfloat{\includegraphics[width=0.2\textwidth]{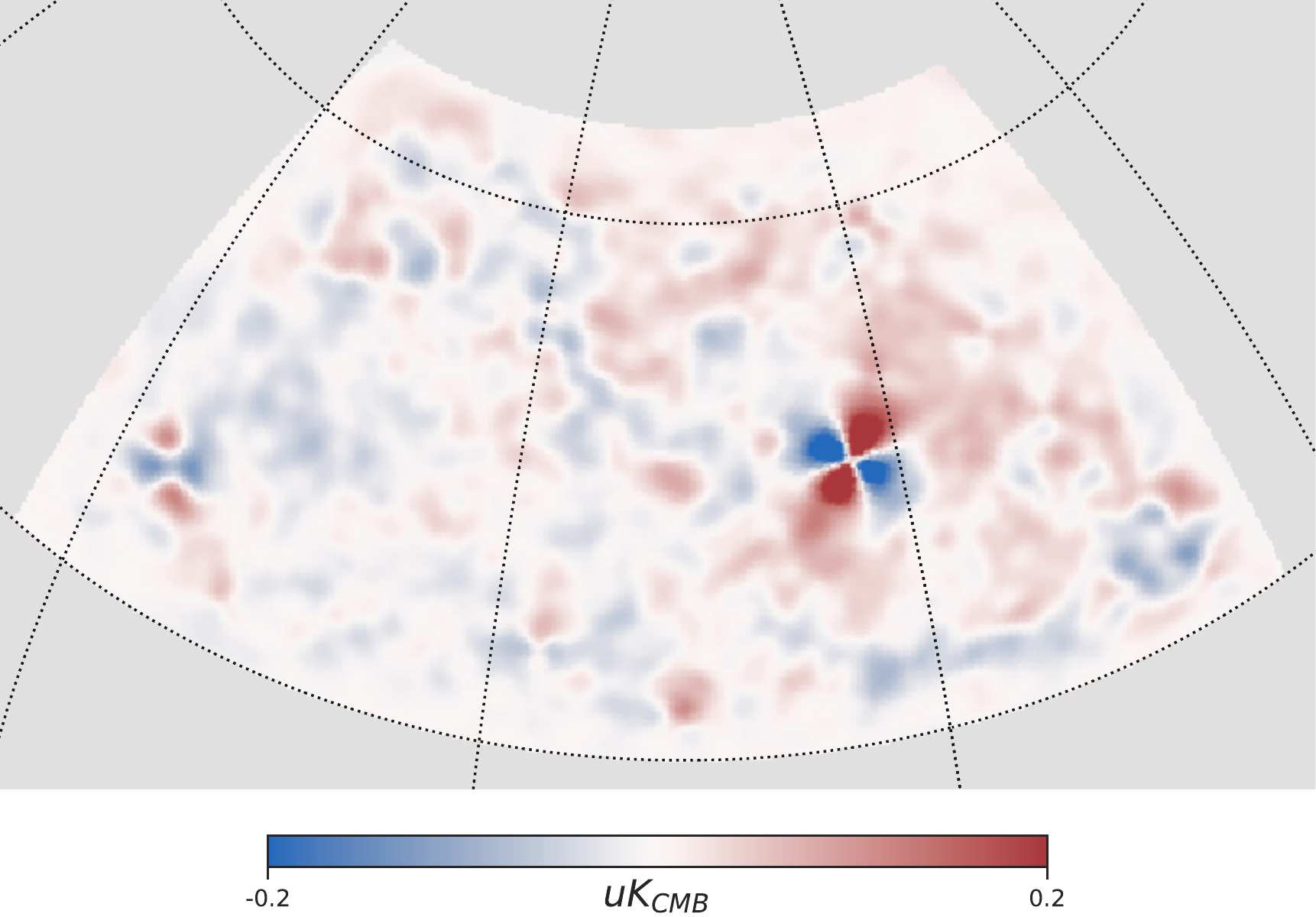}}\hskip 0.5em
        \subfloat{\includegraphics[width=0.2\textwidth]{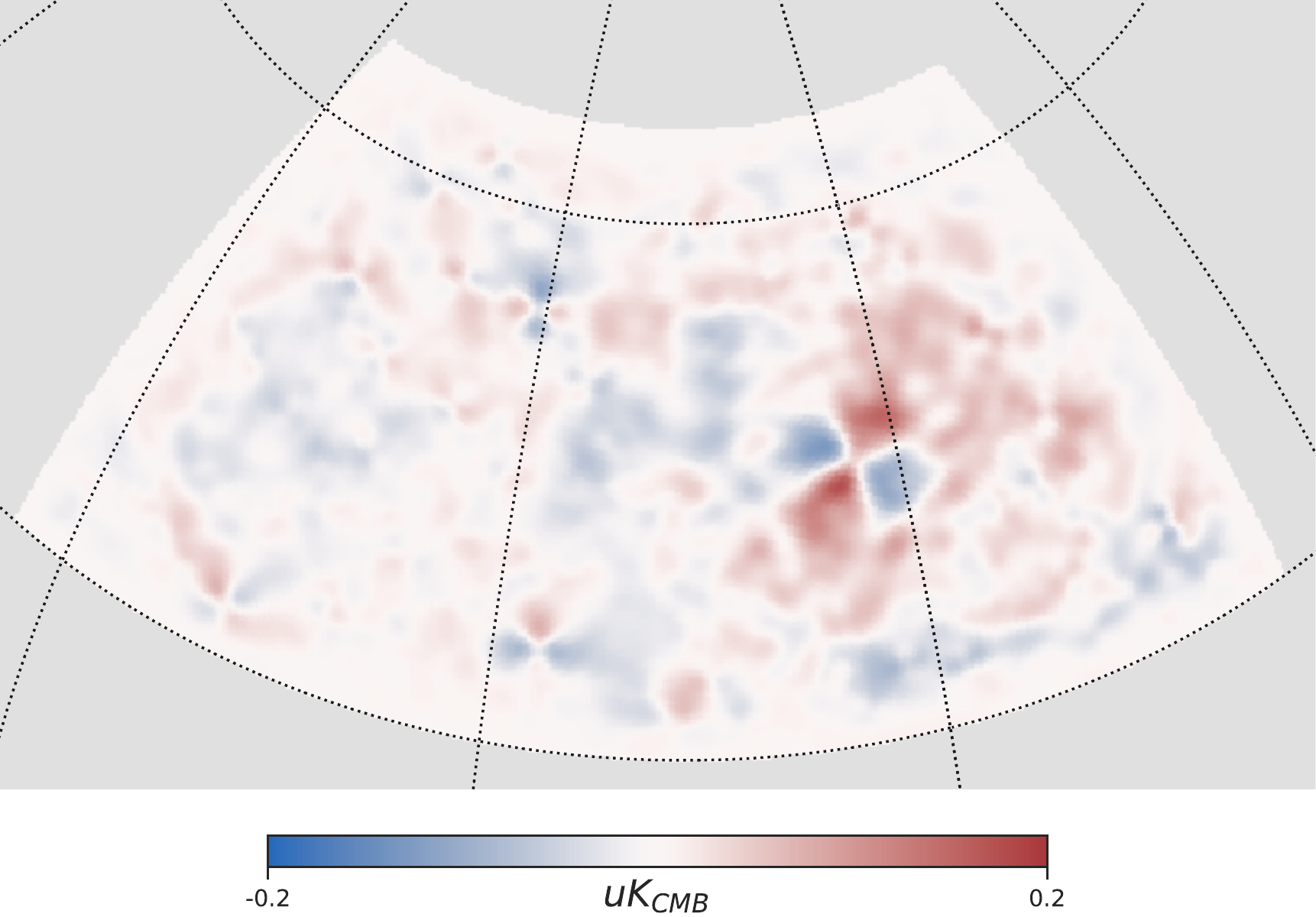}}\\
        \subfloat{\includegraphics[width=0.2\textwidth]{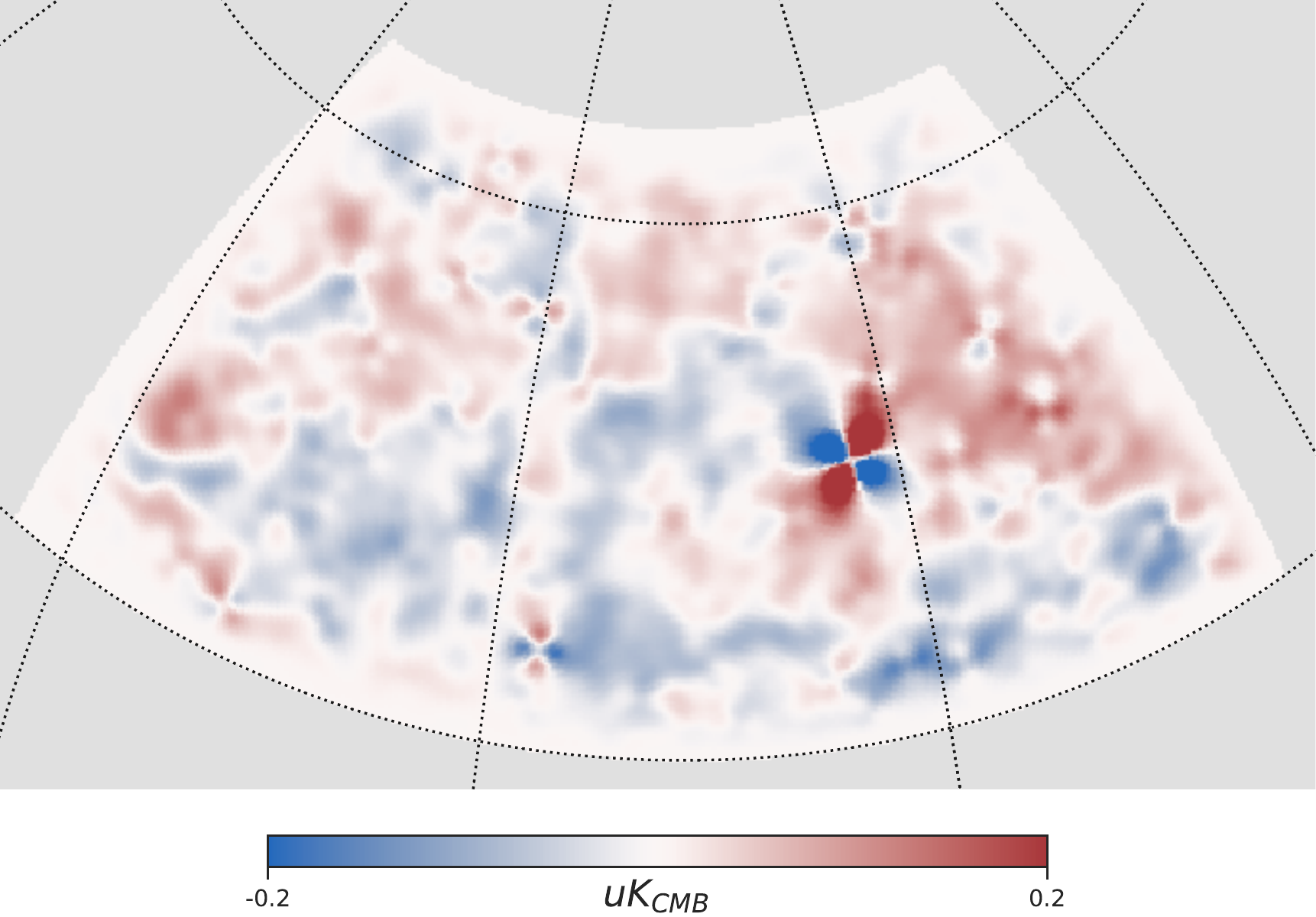}}\hskip 0.5em
        \subfloat{\includegraphics[width=0.2\textwidth]{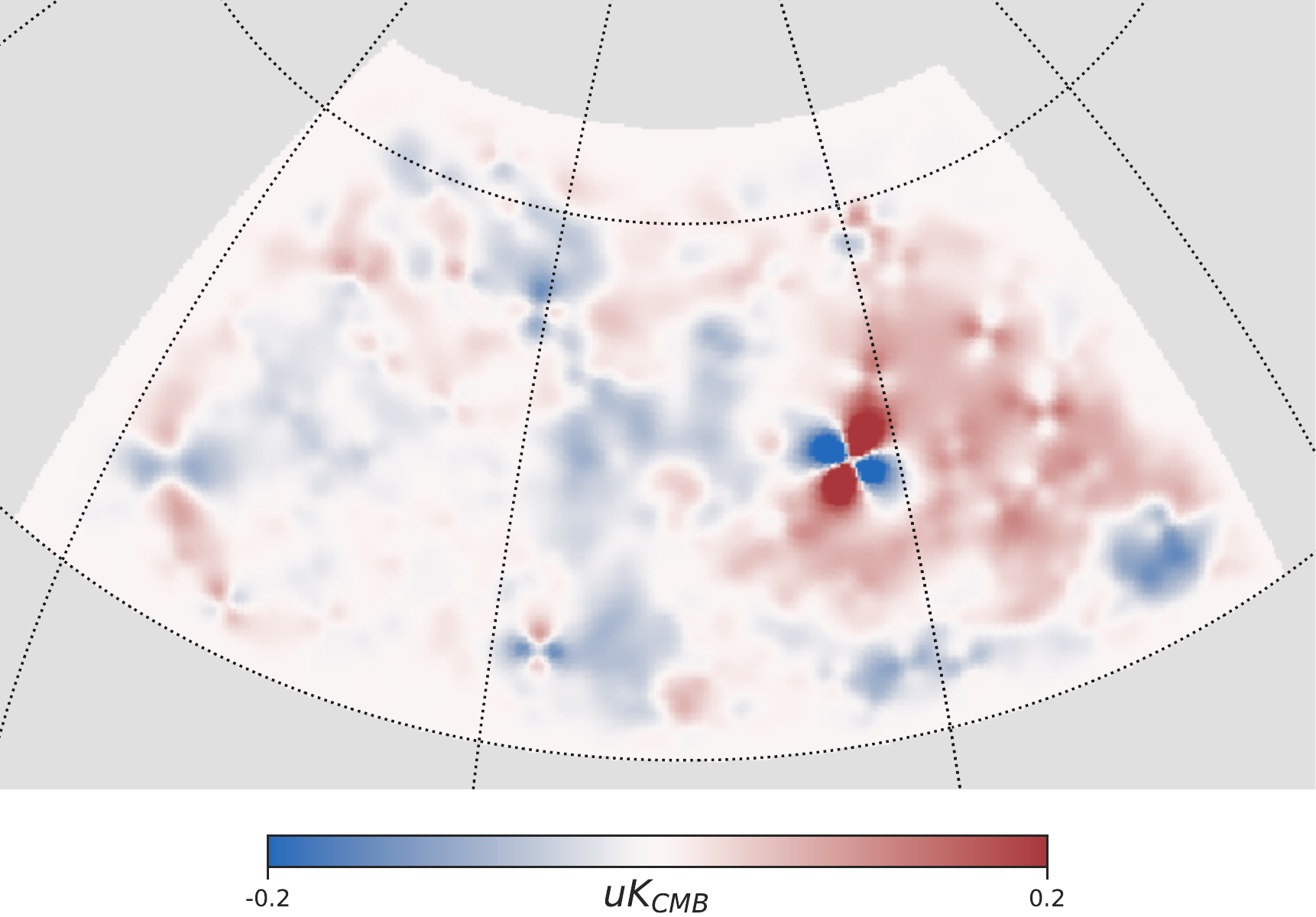}}\hskip 0.5em
        \subfloat{\includegraphics[width=0.2\textwidth]{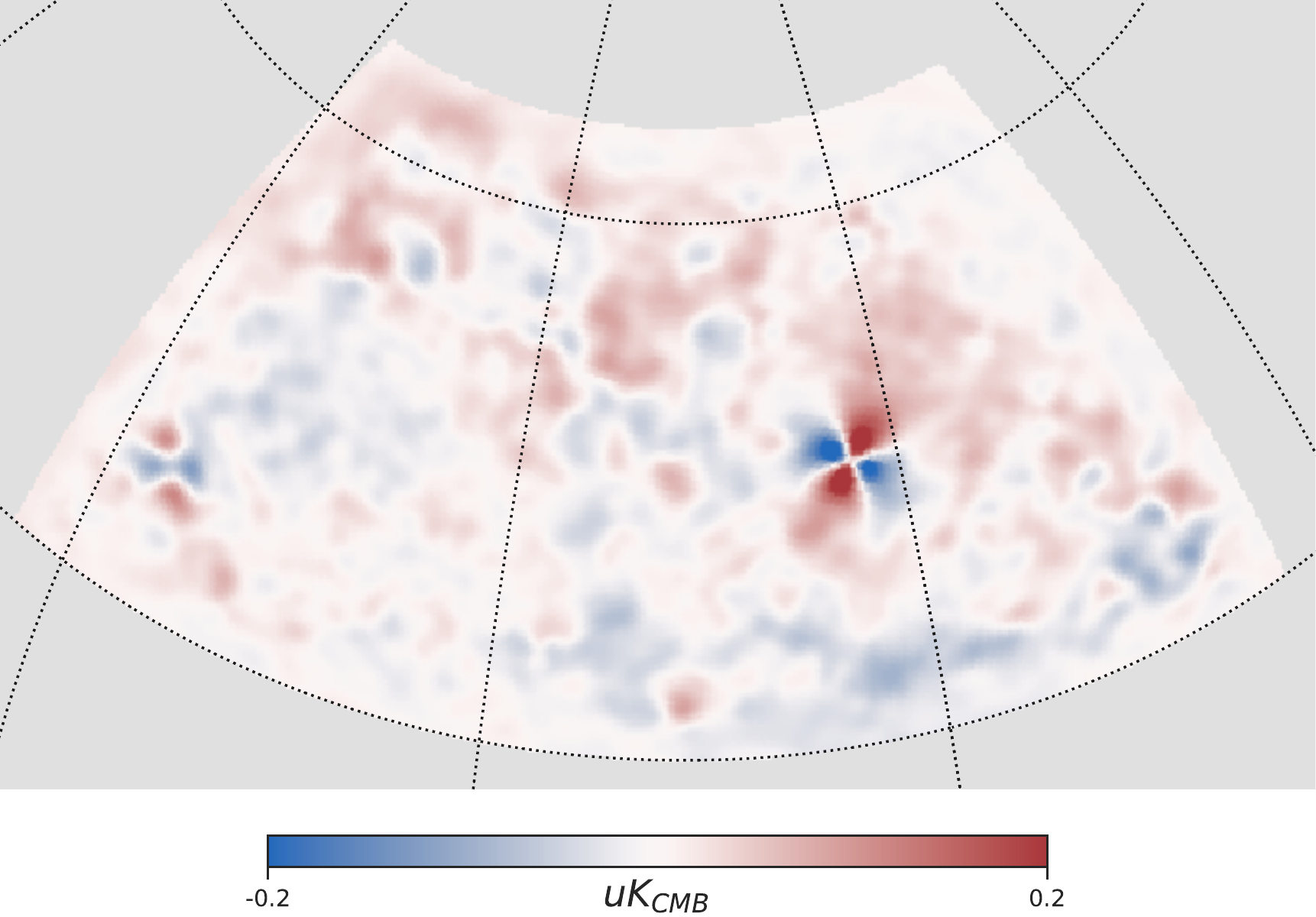}}\hskip 0.5em
        \subfloat{\includegraphics[width=0.2\textwidth]{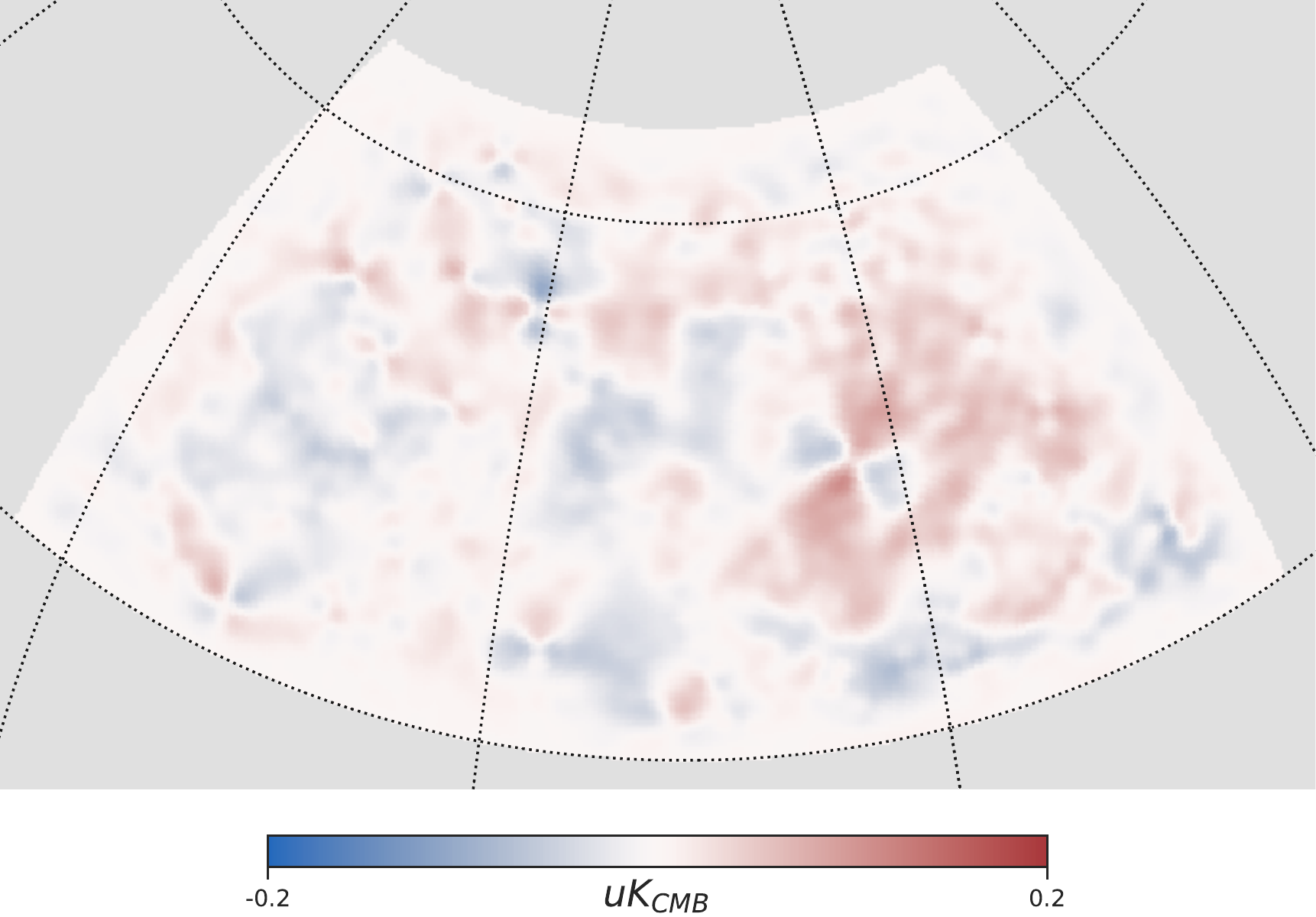}}
    \end{center}
    \caption{The foreground residual $B$ maps. Each map corresponds to the maps shown in Figure \ref{fig:resulting_maps_ps}. The point sources can be easily distinguished from the residual maps.}\label{fig:fg_residual_maps_ps}
\end{figure}

In this subsection, we will discuss the ILC analysis for the case of mock data involving point sources in the foreground model. We show the results in figure \ref{fig:resulting_maps_ps}, and we compare two cases of different input: 5 channels in the top row, and 8 channels in the bottom row. Figure \ref{fig:fg_residual_maps_ps} shows their corresponding foreground residual maps. The residual maps show that the point sources can not be obviously eliminated, there appears a very strong point source (the butterfly-shaped structure on foreground residual $B$ map) on the resulting maps, which shows the ILC methods can not handle the point sources very well. Comparing the analysis results of four ILCs, NILC is the best for the foreground removal. Increasing the frequency is helpful to improve the foreground cleaning, but there is no significant improvement in the elimination of point sources.

\begin{figure}[bthp]
    \begin{center}
        \includegraphics[width=0.8\textwidth]{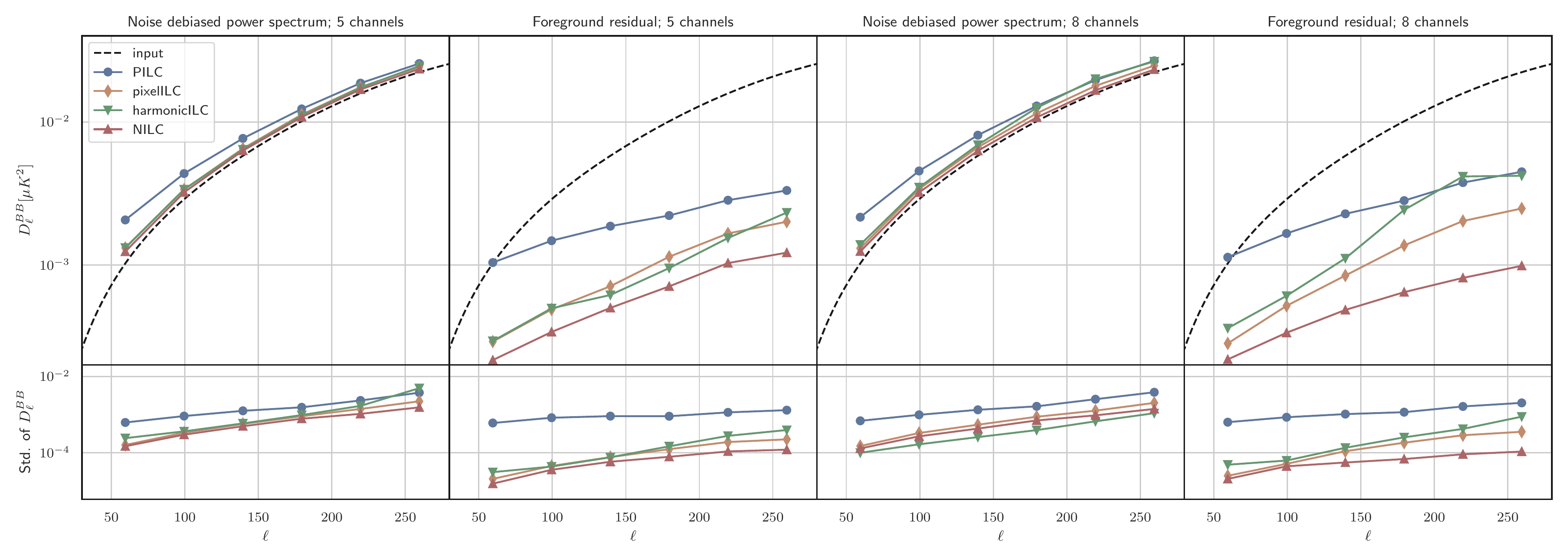}
    \end{center}
    \caption{The noise de-biased $B$ mode power spectra and their standard deviation} of resulting and residual maps shown in Figure \ref{fig:resulting_maps_ps} and Figure \ref{fig:fg_residual_maps_ps}.\label{fig:powers_Ps}
\end{figure}

Since the point sources are a kind of small scale structure, the existence of them on maps will bias the reconstructed CMB power spectra at high $\ell$s. Figure \ref{fig:powers_Ps} shows the de-biased resulting and residual foreground power spectra. It can be easily figured out that they have more small scale foreground residual compared to the power spectra shown in Figure \ref{fig:5band_wops}. The bias introduced by the point sources will bias the final $r$ constraint, which will be shown in Section \ref{sec:parameter}.

\subsection{Results on CMB-S4 patch}\label{subsec:CMBS4}
\begin{figure}[bthp]
    \begin{center}
        \subfloat{\includegraphics[width=0.2\textwidth]{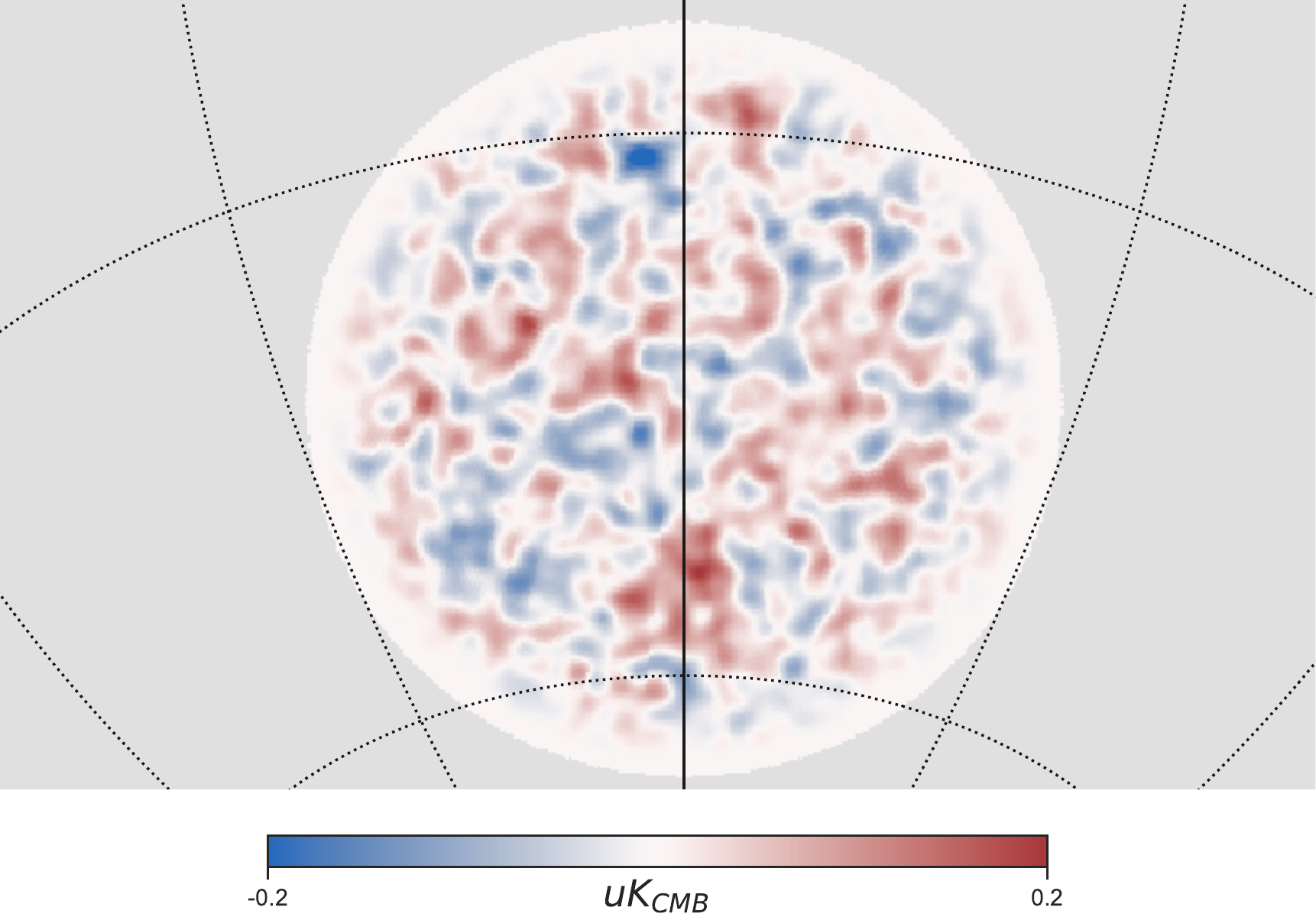}}\hskip 0.5em
        \subfloat{\includegraphics[width=0.2\textwidth]{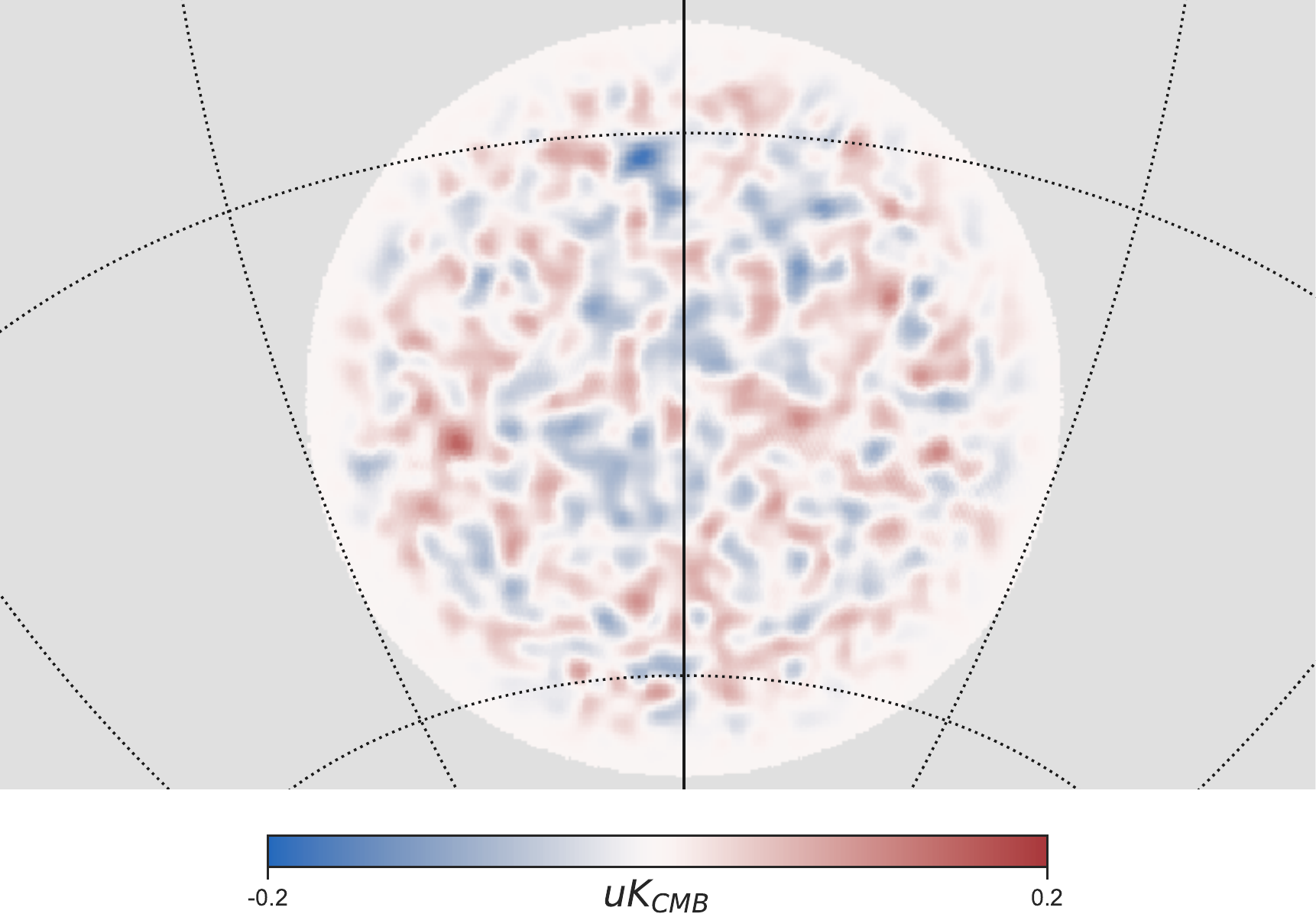}}\hskip 0.5em
        \subfloat{\includegraphics[width=0.2\textwidth]{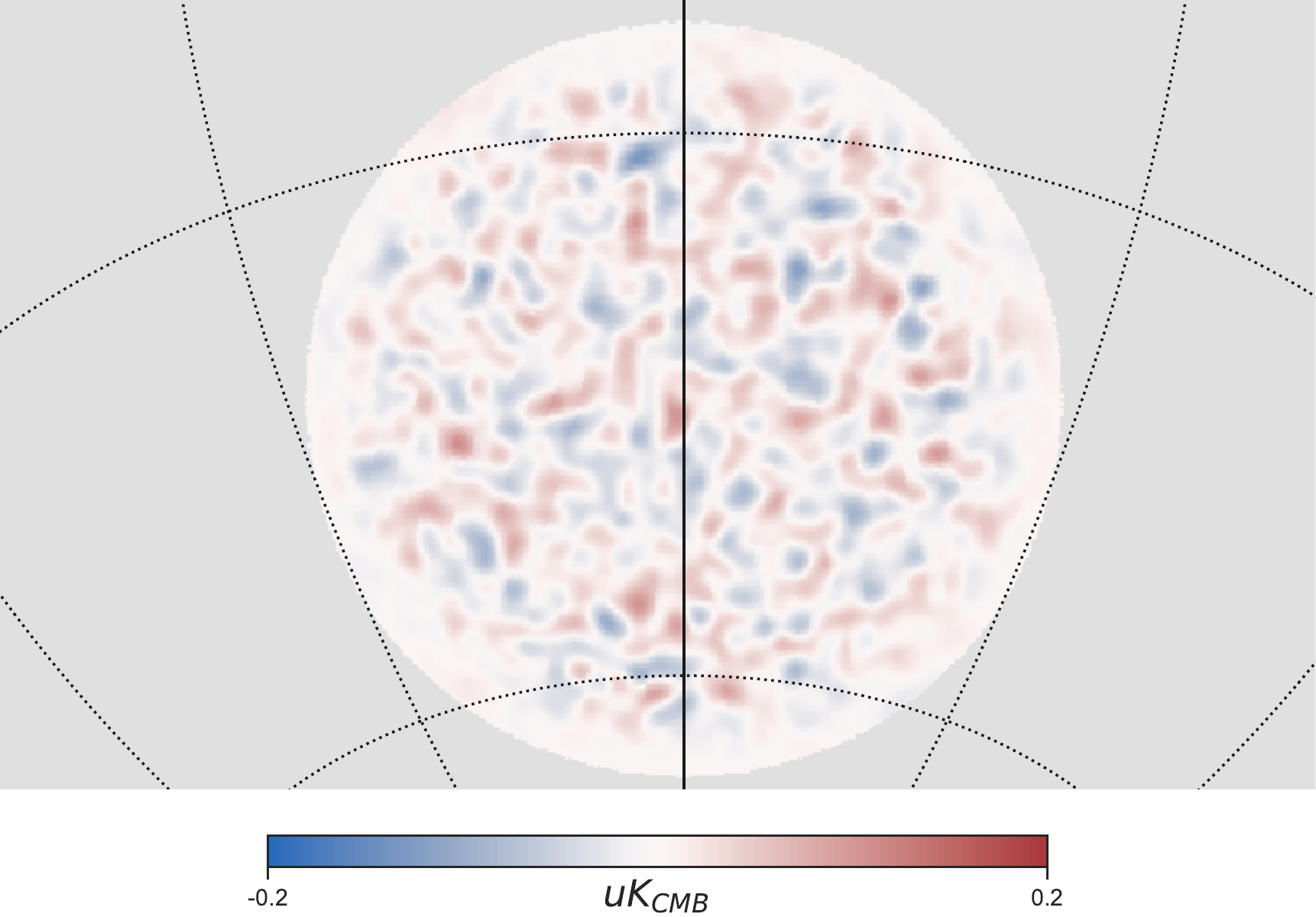}}\hskip 0.5em
        \subfloat{\includegraphics[width=0.2\textwidth]{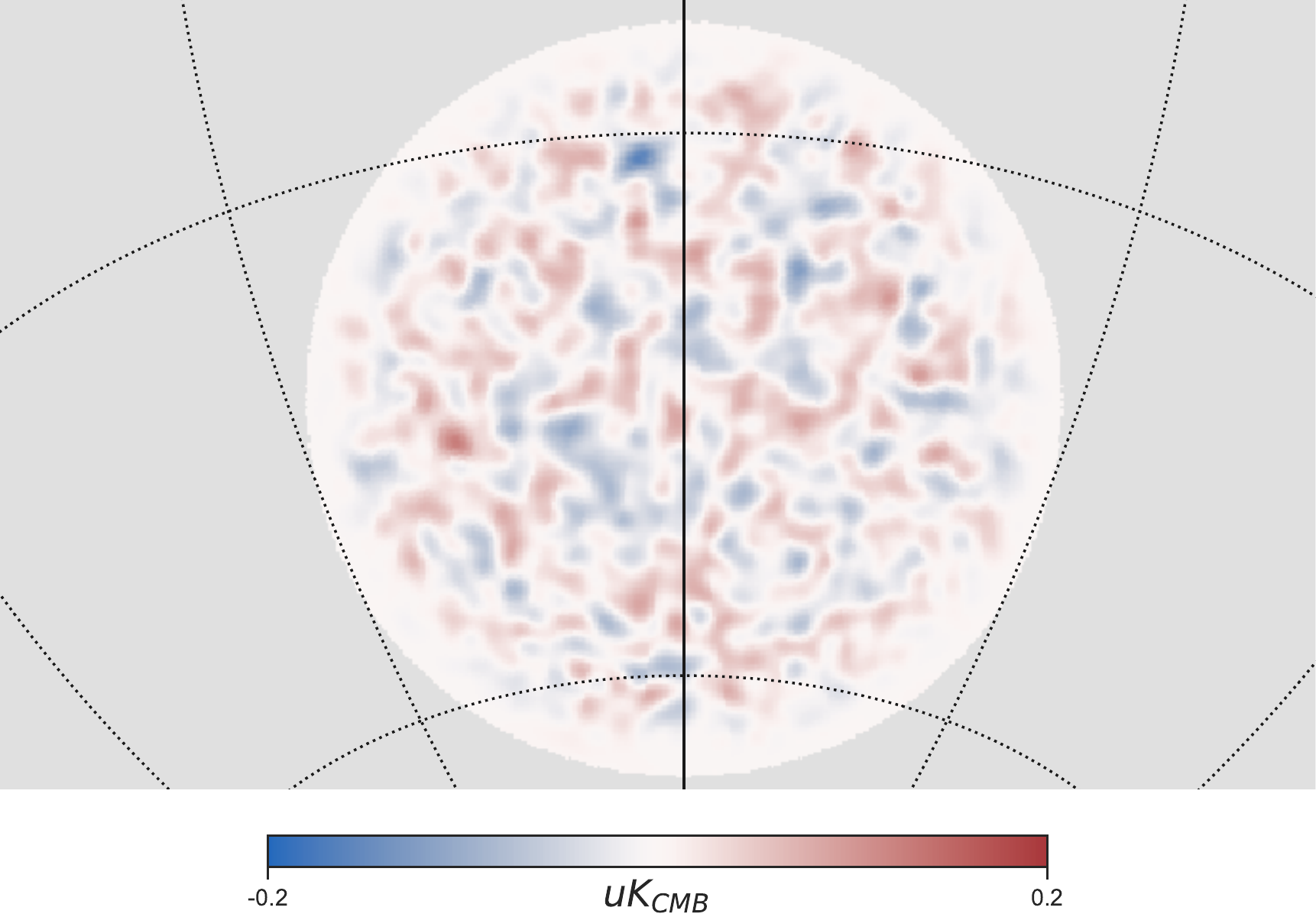}}\\
        \subfloat{\includegraphics[width=0.2\textwidth]{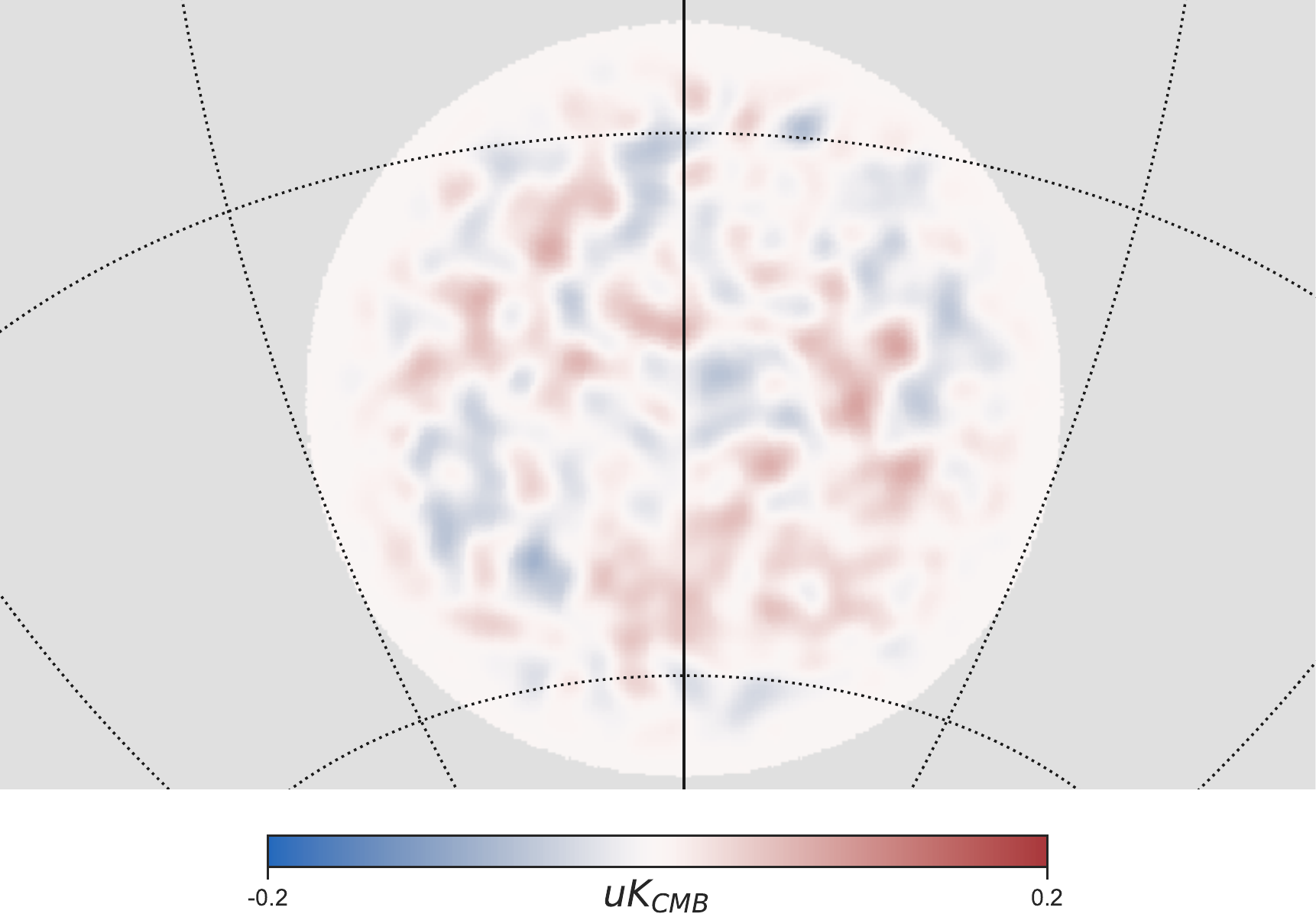}}\hskip 0.5em
        \subfloat{\includegraphics[width=0.2\textwidth]{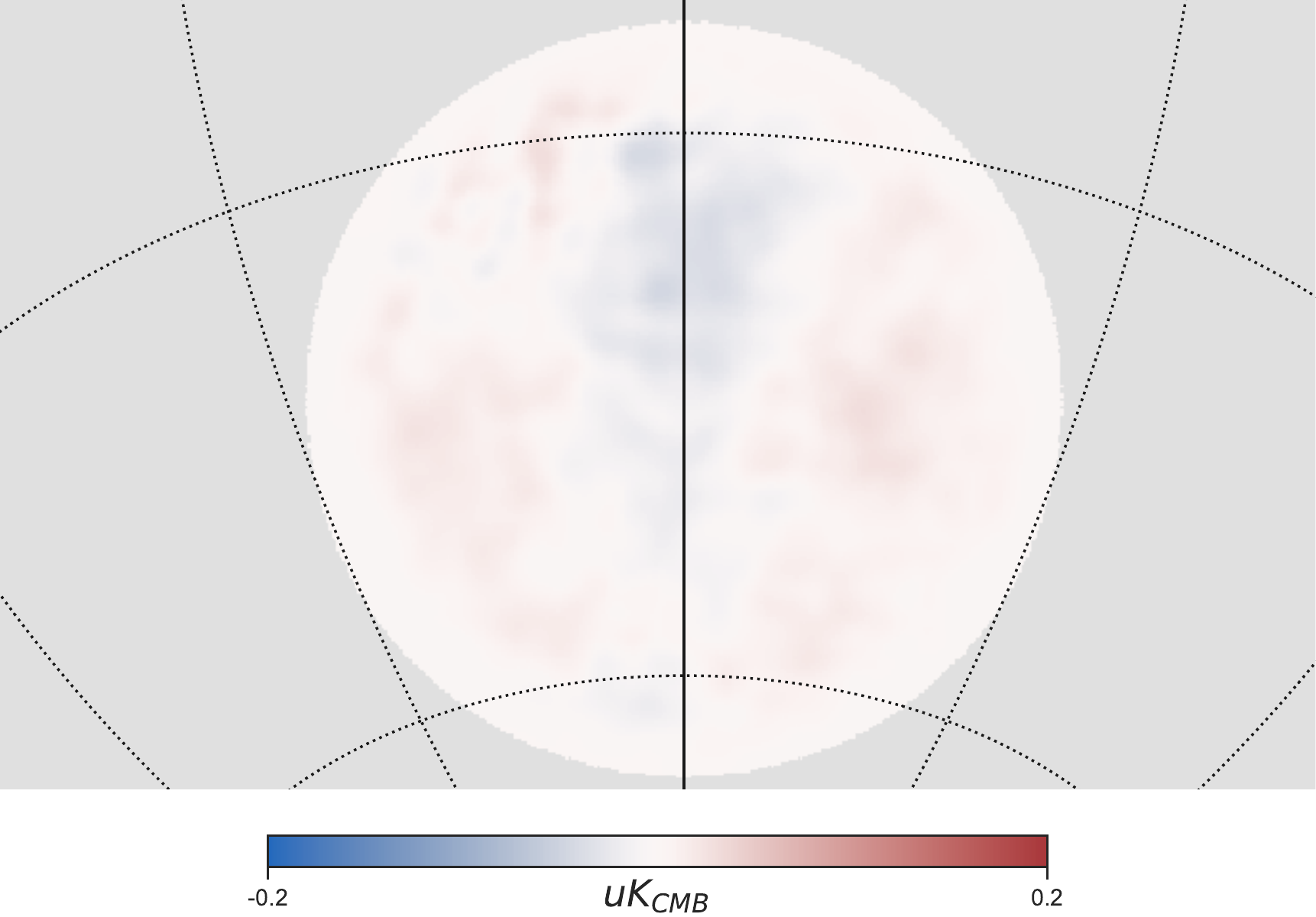}}\hskip 0.5em
        \subfloat{\includegraphics[width=0.2\textwidth]{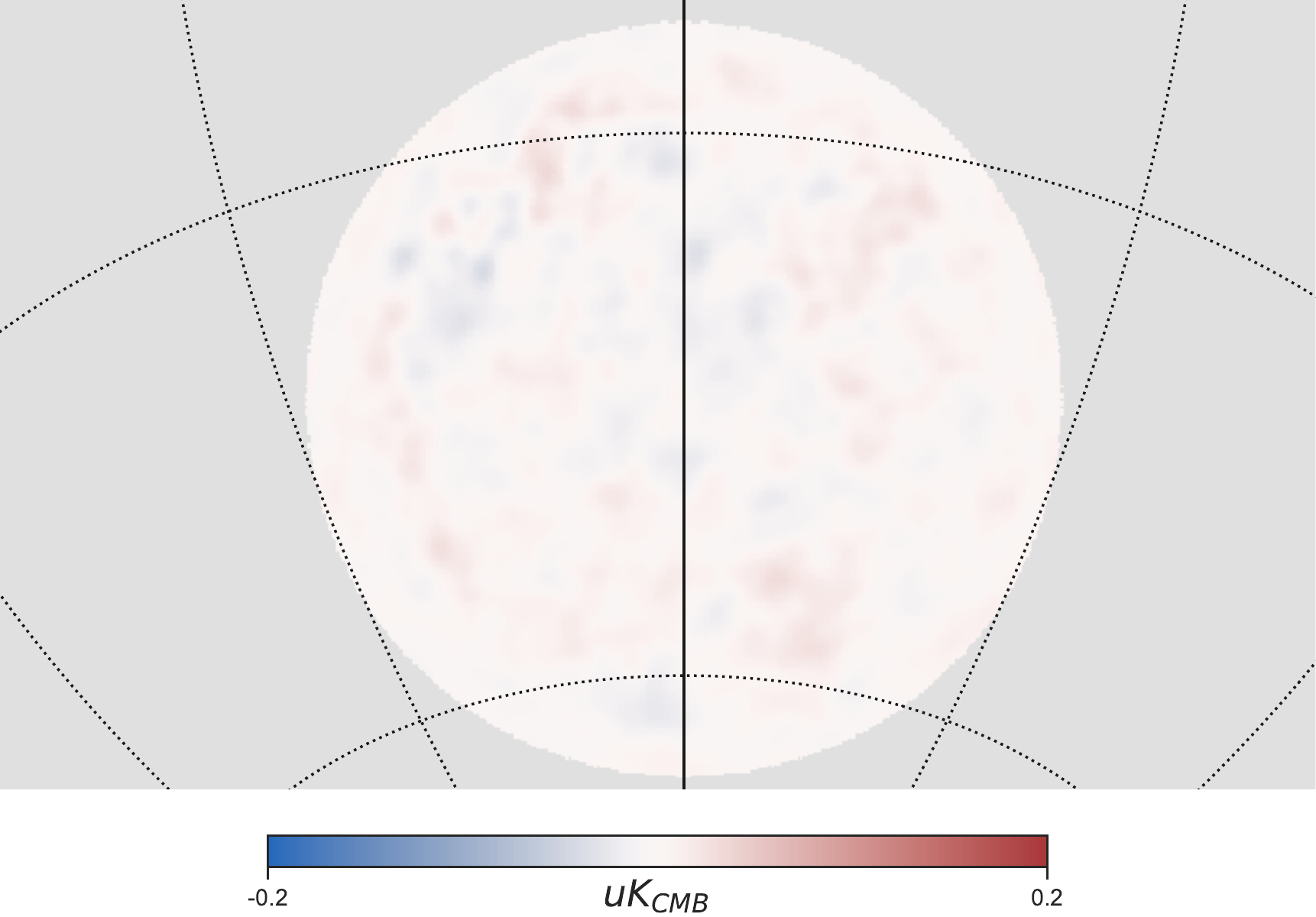}}\hskip 0.5em
        \subfloat{\includegraphics[width=0.2\textwidth]{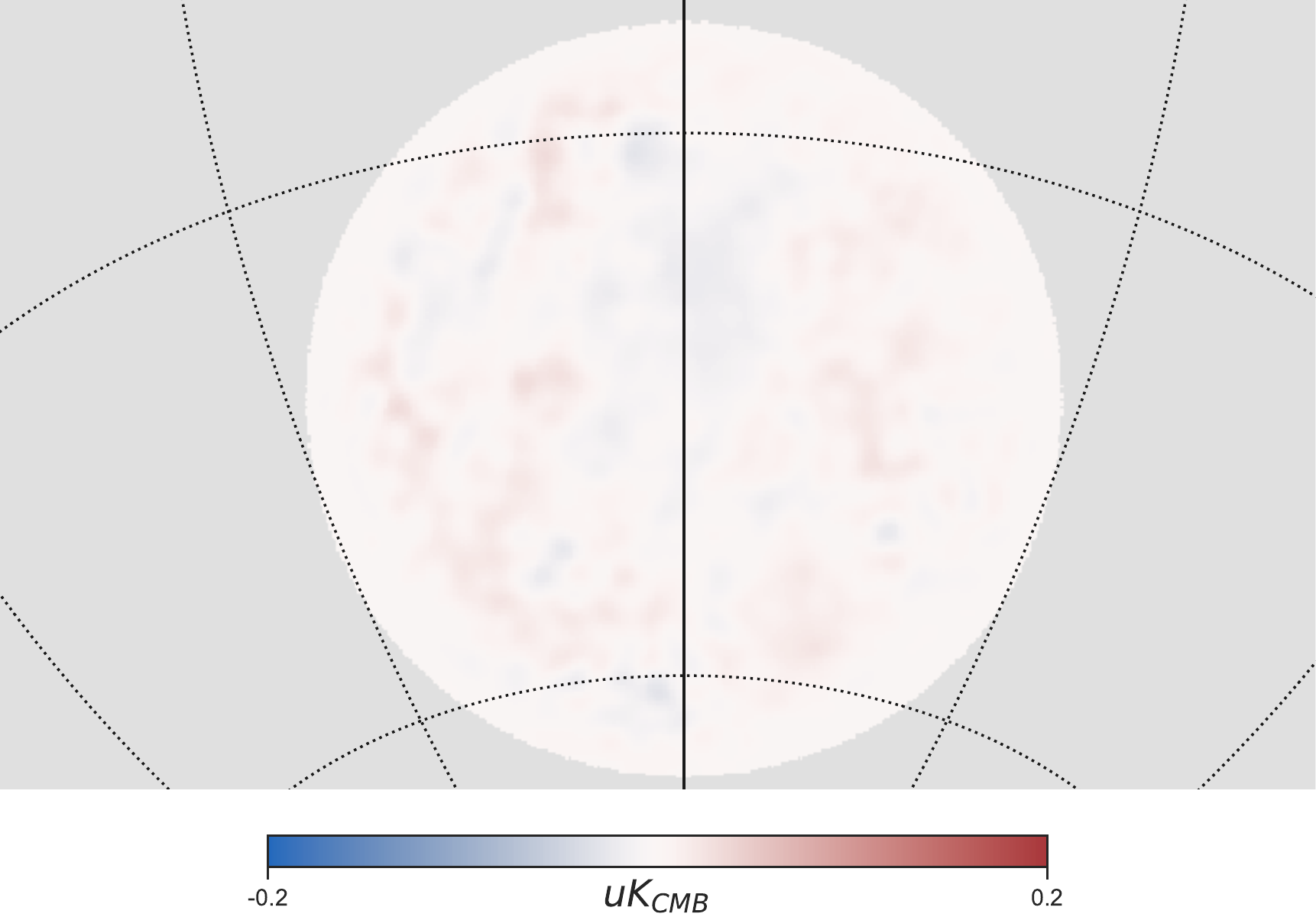}}\\
        \subfloat{\includegraphics[width=0.2\textwidth]{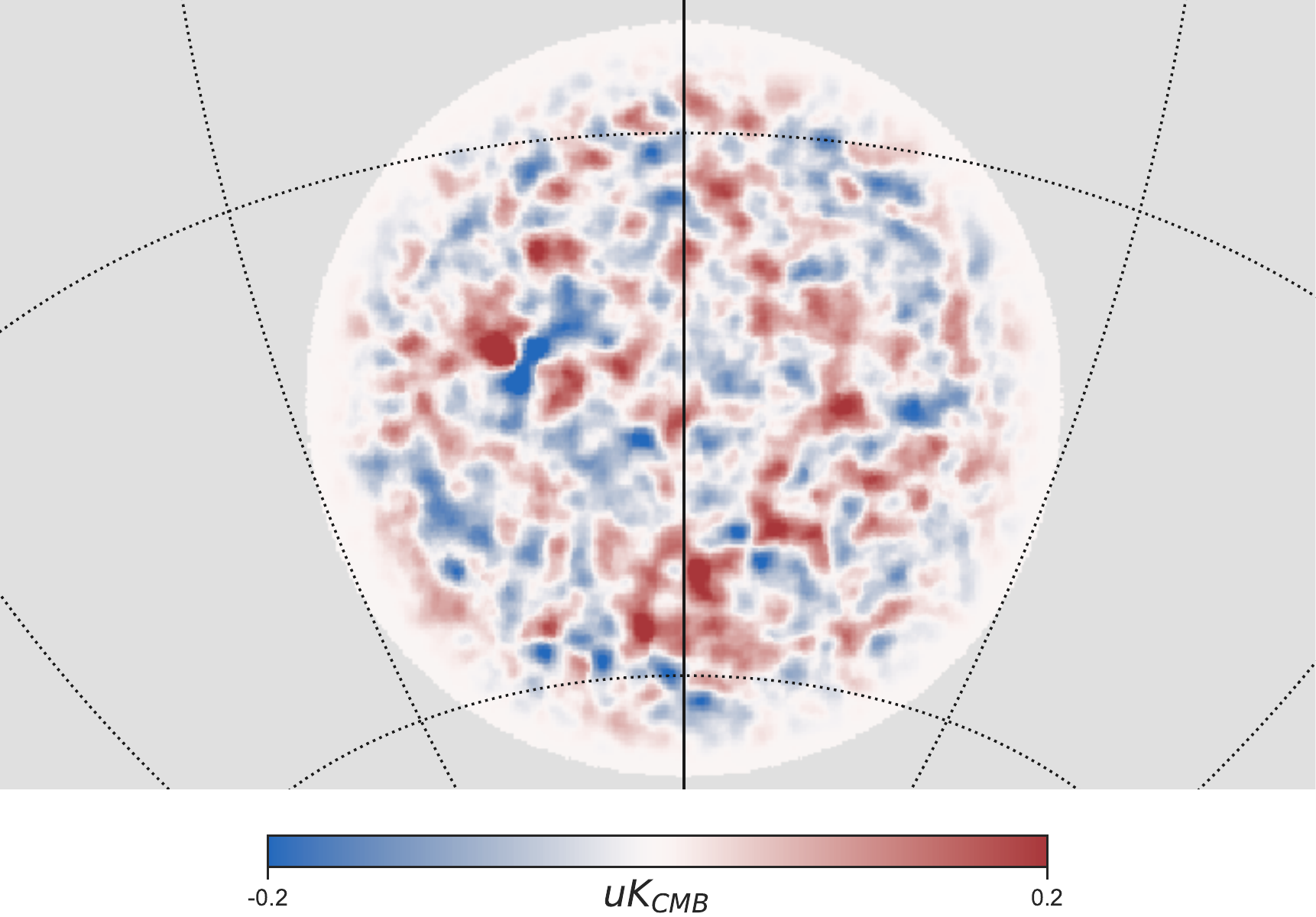}}\hskip 0.5em
        \subfloat{\includegraphics[width=0.2\textwidth]{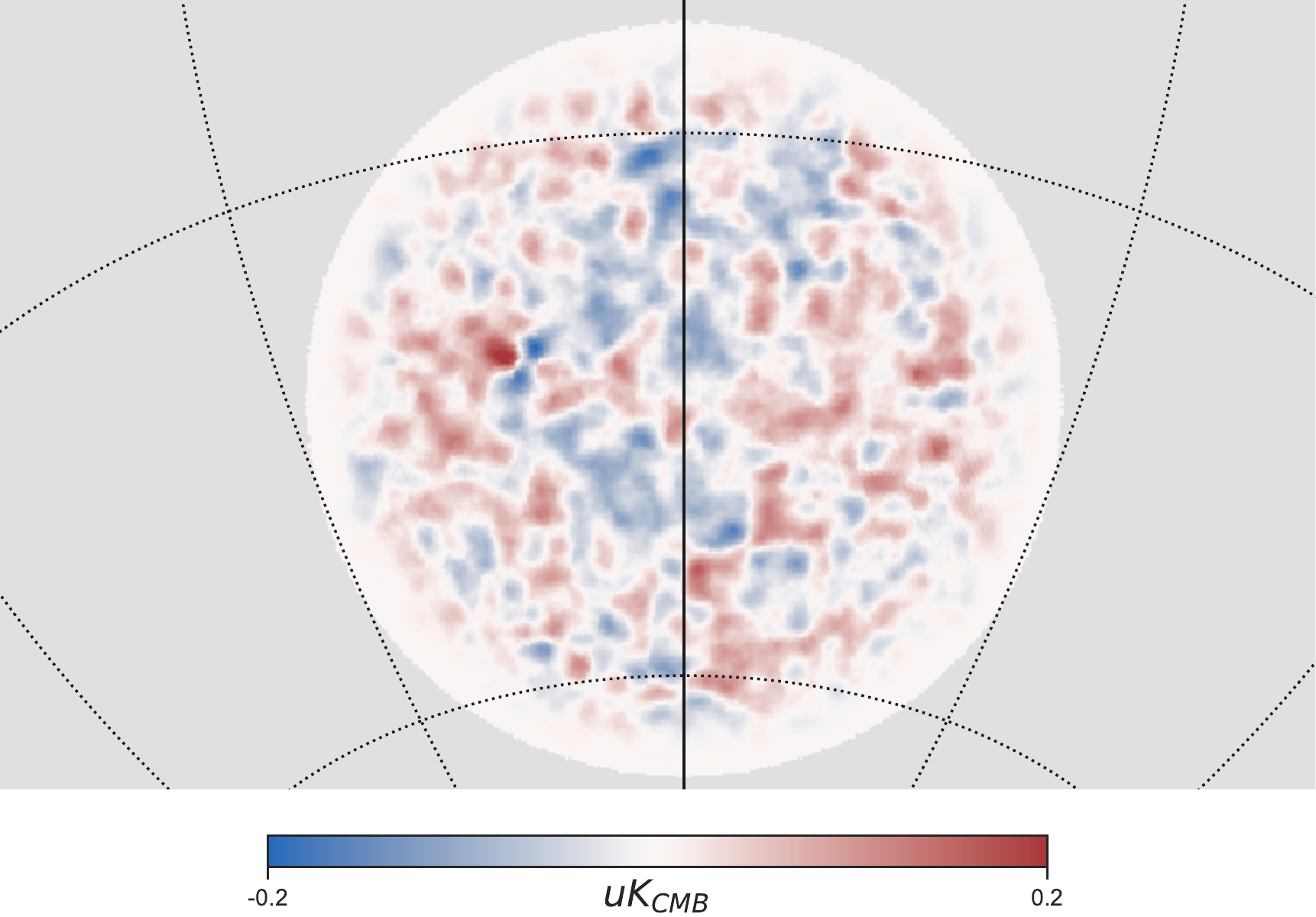}}\hskip 0.5em
        \subfloat{\includegraphics[width=0.2\textwidth]{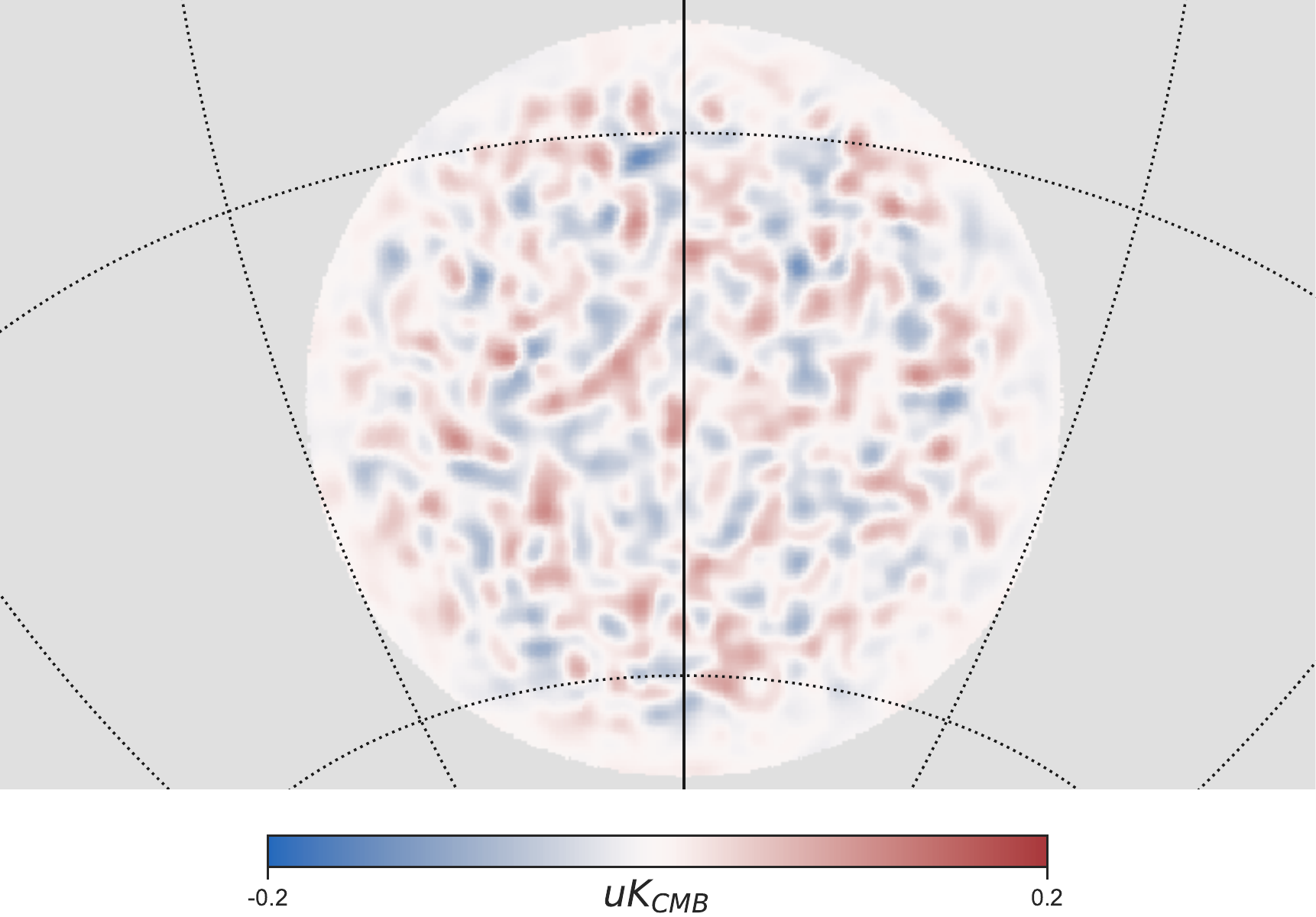}}\hskip 0.5em
        \subfloat{\includegraphics[width=0.2\textwidth]{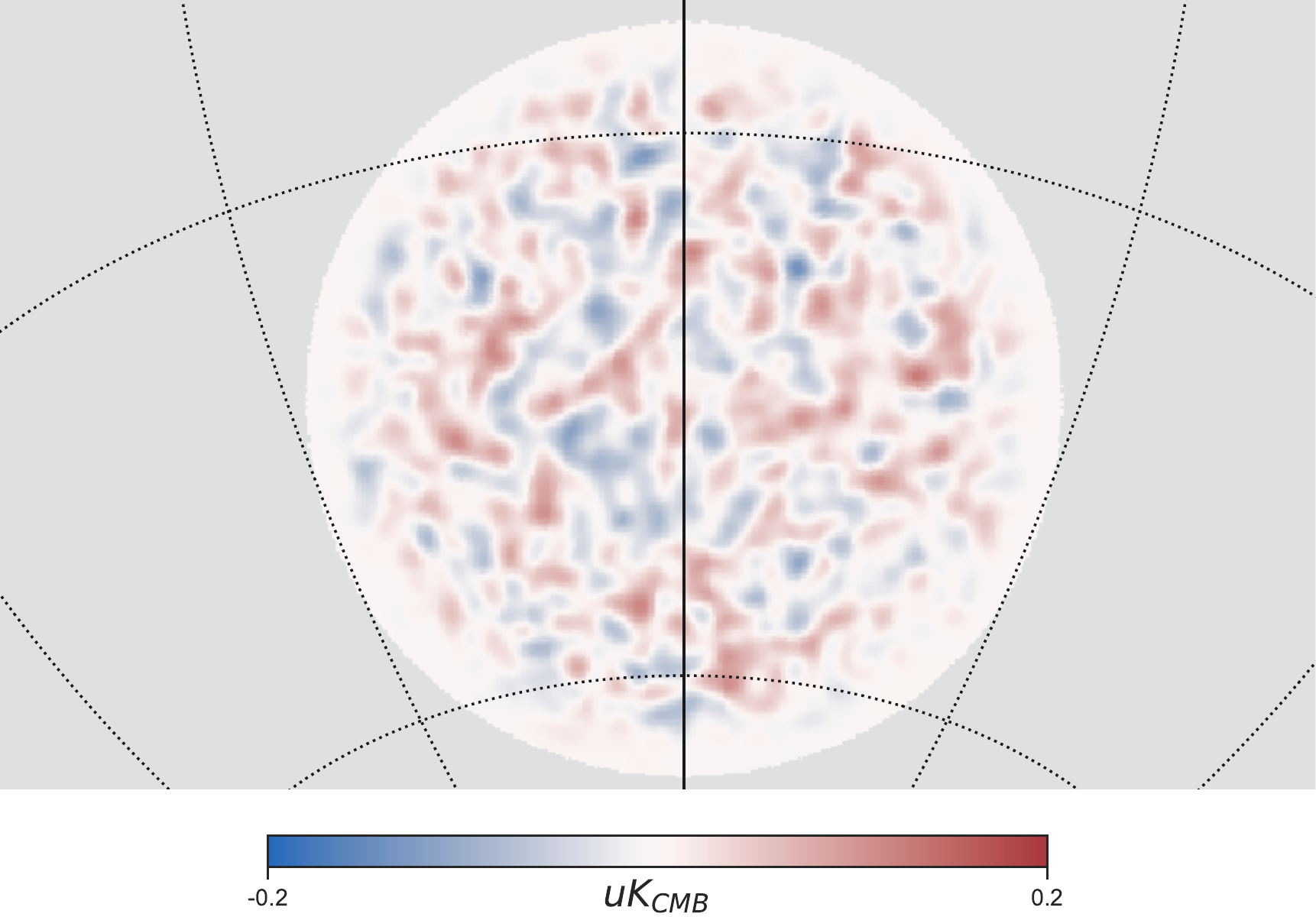}}\\
        \subfloat{\includegraphics[width=0.2\textwidth]{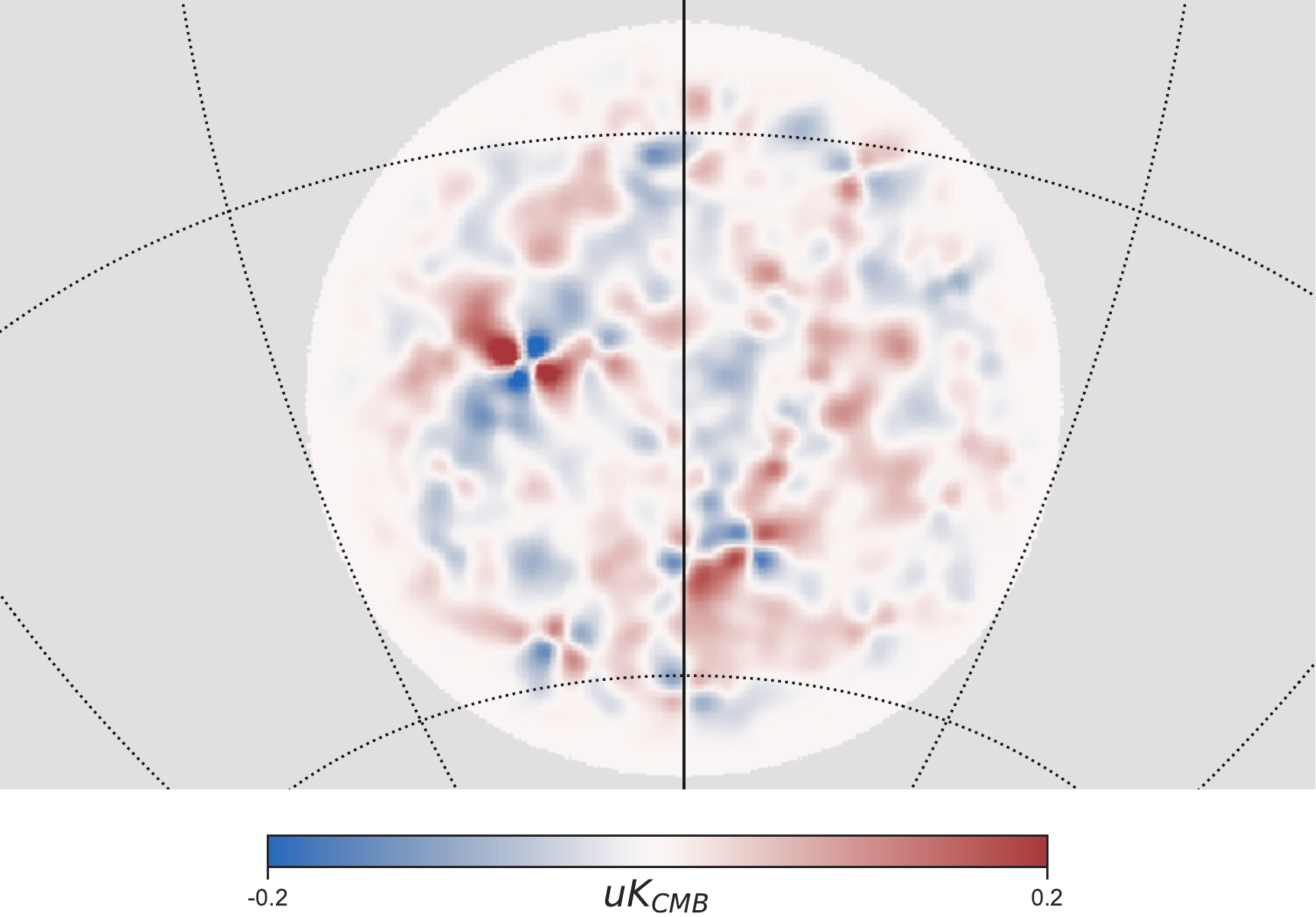}}\hskip 0.5em
        \subfloat{\includegraphics[width=0.2\textwidth]{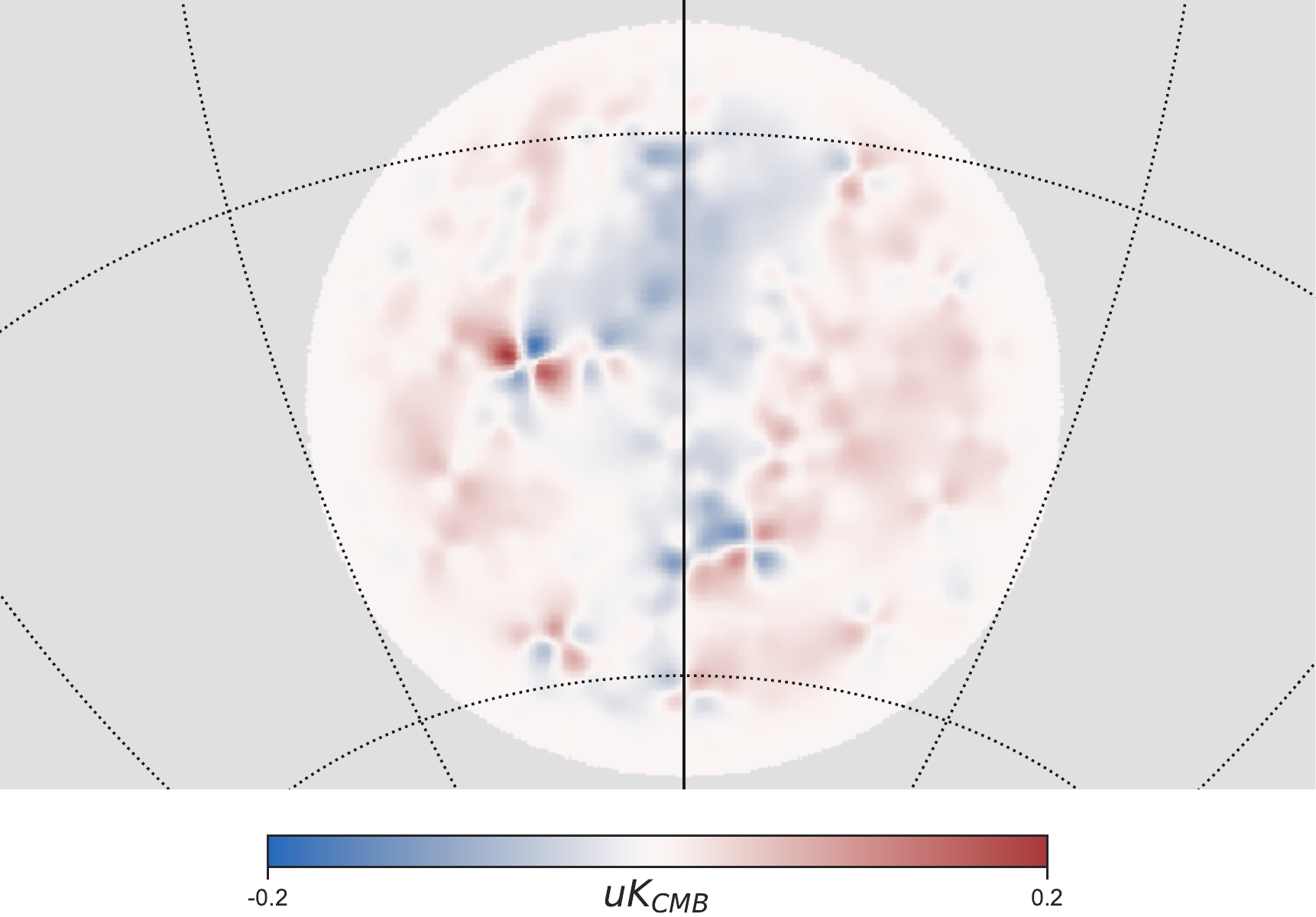}}\hskip 0.5em
        \subfloat{\includegraphics[width=0.2\textwidth]{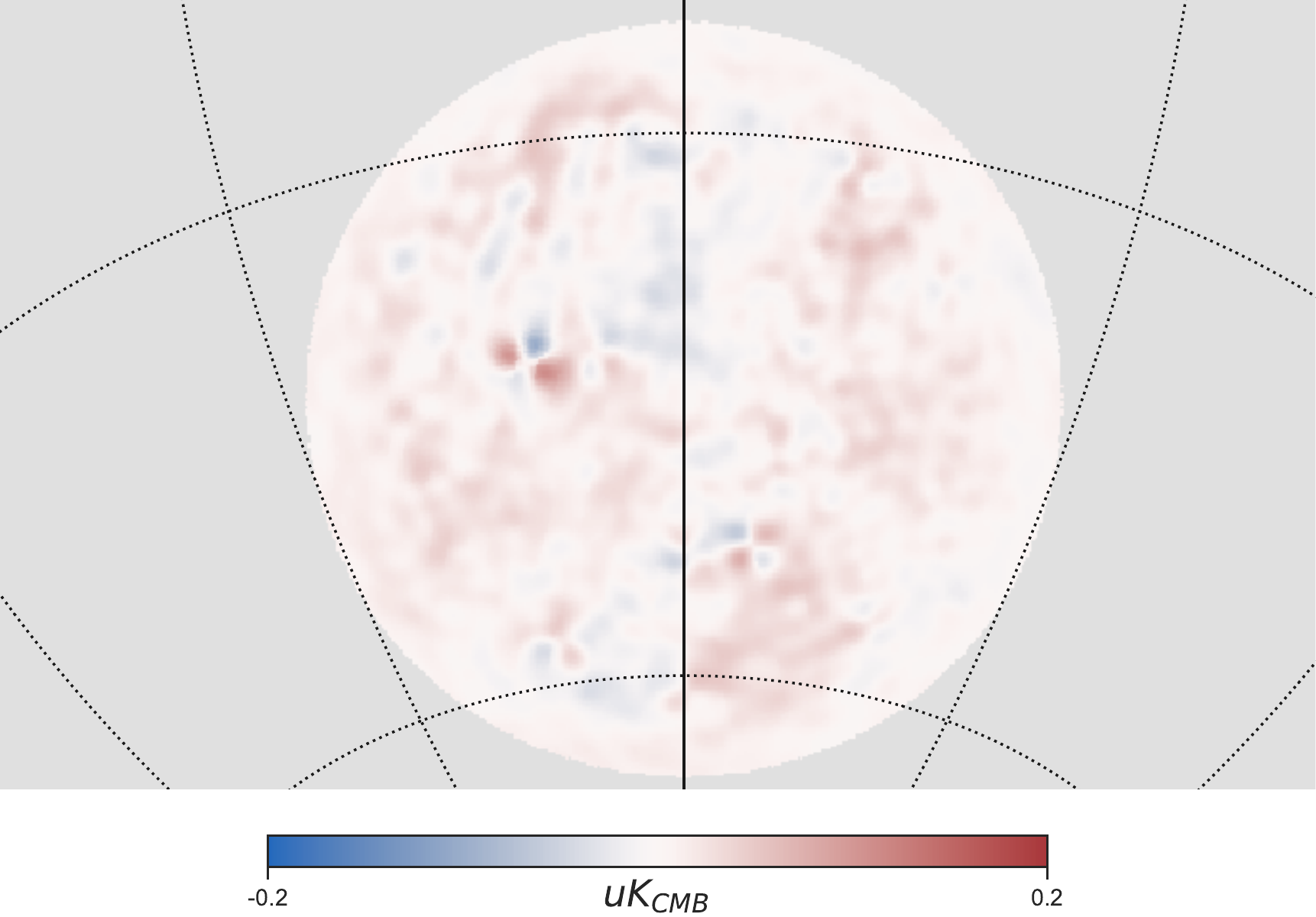}}\hskip 0.5em
        \subfloat{\includegraphics[width=0.2\textwidth]{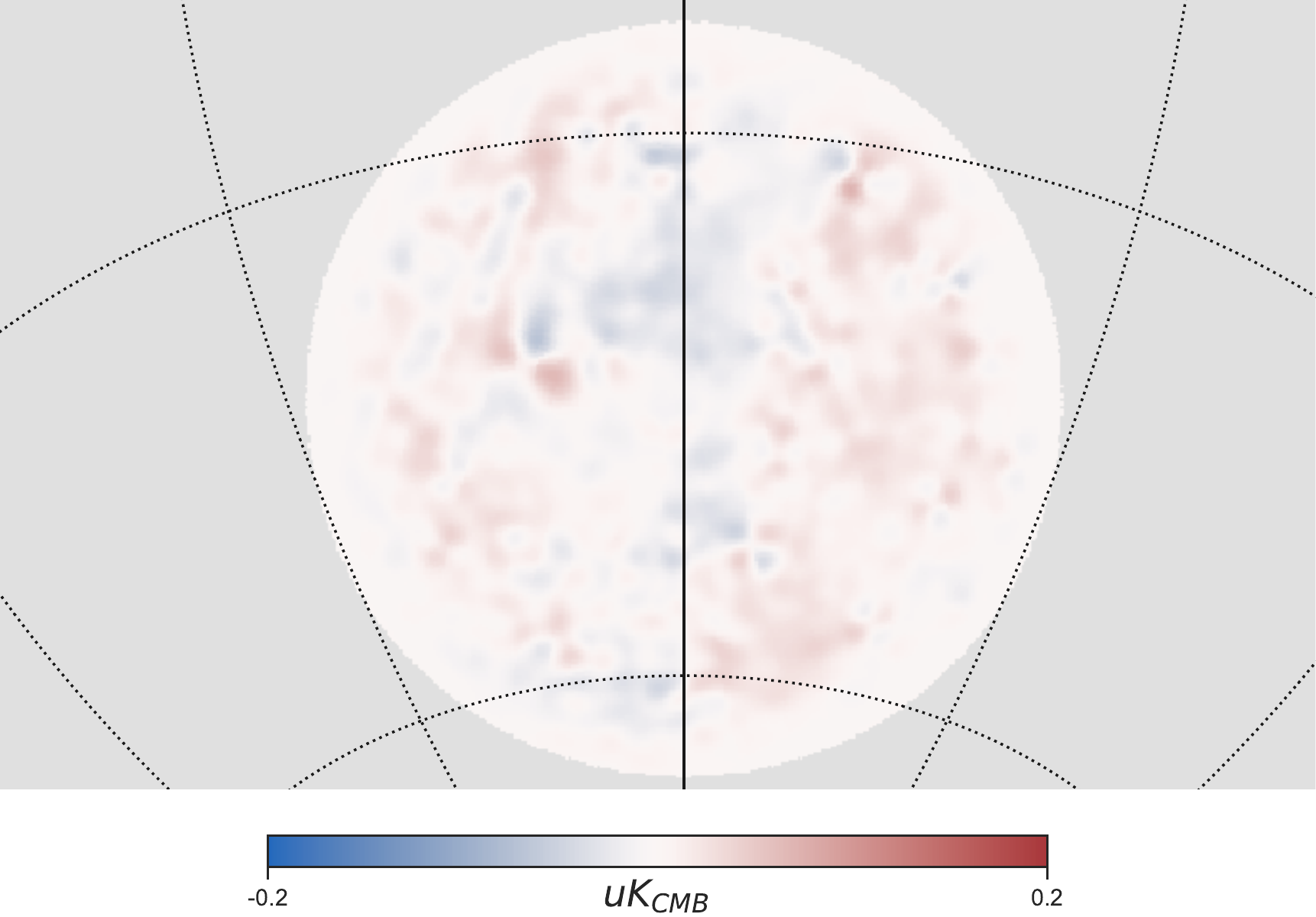}}
    \end{center}
    \caption{The figures in the first and the third row show the $B$ maps after ILC performing for the CMB-S4 sky patch. The figures in the second and the fourth row show their corresponding foreground residual. Each column represents a kind of ILC method. From left to right, They correspond PILC, pixelILC, harmonicILC and NILC respectively. The edges of all the maps are also apodized by 6 degree. }\label{fig:CMBS4Maps}
\end{figure}

\begin{figure}
    \centering
    \includegraphics[width=0.8\textwidth]{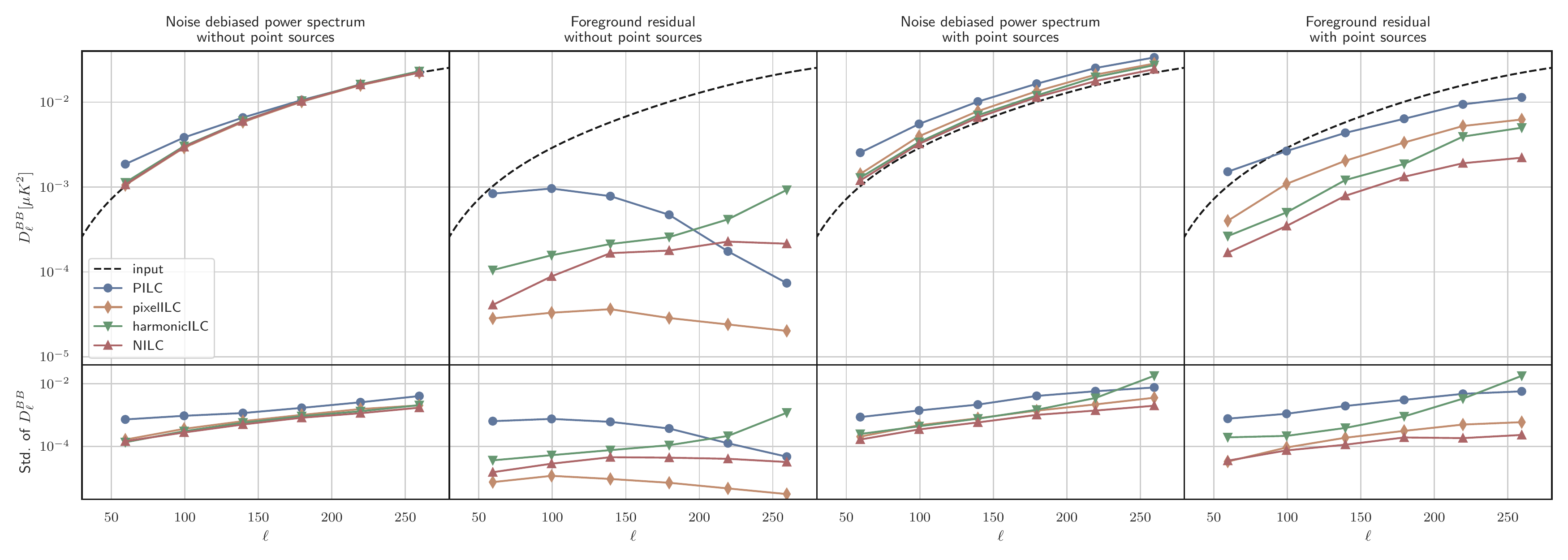}
    \caption{The noise de-biased $B$ mode power spectra and their standard deviation of resulting and residual maps shown in Figure \ref{fig:CMBS4Maps}.}
    \label{fig:CMBS4Power}
\end{figure}

 The same analysis method as in the previous section was also performed on the CMB-S4 patch, which is shown in Figure \ref{fig:skypatch_mask}\subref{subfig:skypatchCMBS4}. According to the CMB-S4 forecasting paper\cite{CMBS4Forecast}, totally 9 channels are used to do the constraints on $r$. The noise level for each channel is shown in Table\ref{tab:setup}. Another difference from the ILC performance in the previous section is that the common resolution is chosen to be $80$ arcmin because the worst-resolution is $77$ arcmin. Figure \ref{fig:CMBS4Maps} shows the sky maps after ILC performance and also their corresponding foreground residual with or without point sources consideration and Figure \ref{fig:CMBS4Power} shows their corresponding $B$ mode power spectra. An obvious and direct conclusion from the power spectra is that the existence of compact sources did have a big impact on the results like we seen in the northern sky observation case shown in Section \ref{subsec:pointsources}. Section \ref{subsec:discussion} elaborates more on the analyses on map and power spectra level, especially the analyses of the residual foreground power spectra.


\subsection{$r$ constraint}\label{sec:parameter}

\begin{figure}
    \centering
    \includegraphics[width=0.8\textwidth]{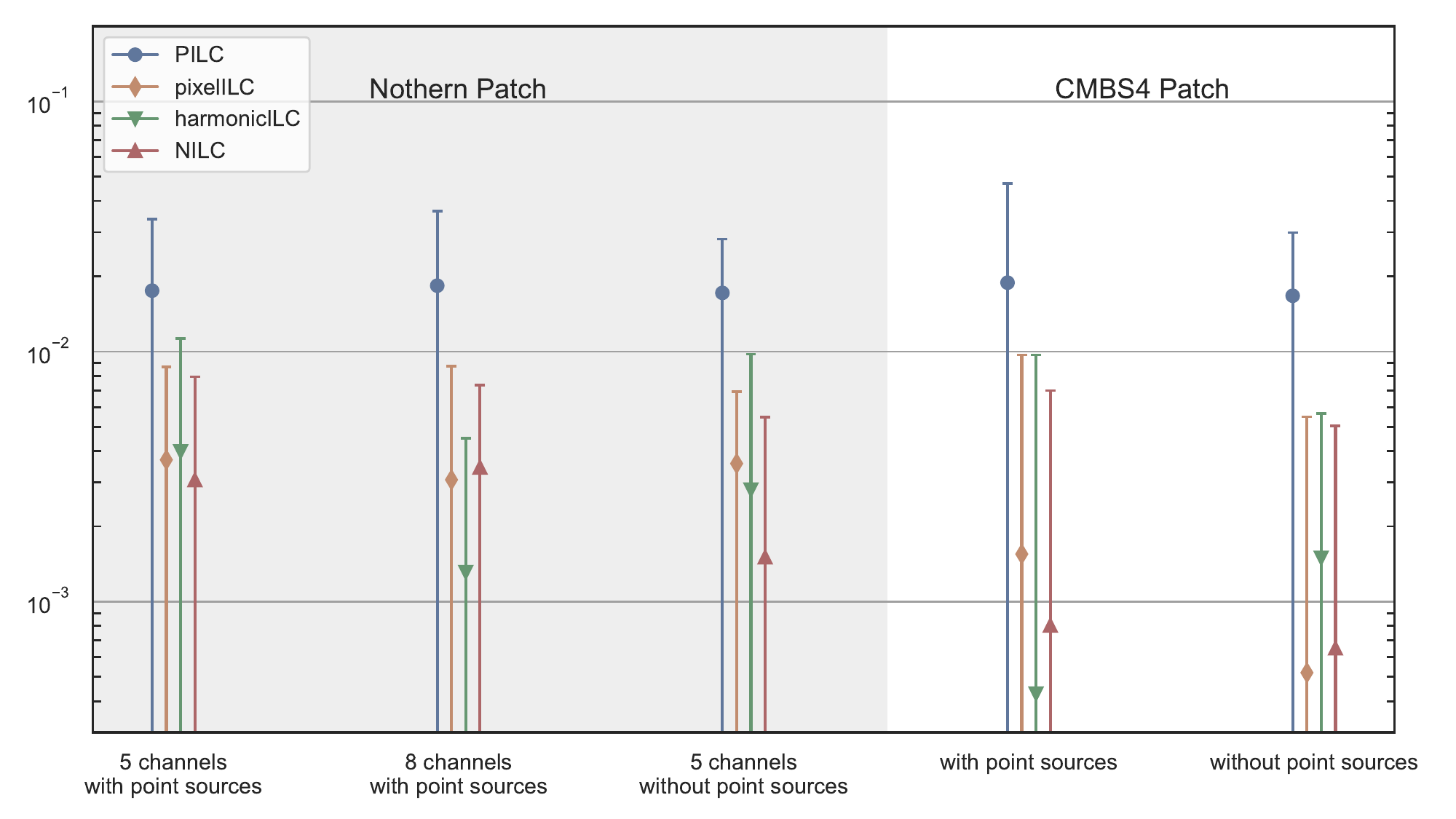}
    \caption{Constraints on $r$, for a model with $r=0$, The results from the 200 Monte Carlo simulations for each ILC method and case in this work.}
    \label{fig:rlimits}
\end{figure}

In order to further explore the impact of foreground residual on $r$ measurement, we give and compare the constraints on r from CMB power spectra provided by different ILCs. The results for different cases are shown in Figure \ref{fig:rlimits}, and the error bars which are plotted in the figure are the $95\%$ confidence interval for $r$. Constraints on $r$ are obtained by sampling the likelihood Eq. \eqref{eq:likelihood} for two parameters $r$ and $A_{L}$. The $r$ constraining is shown in Figure  \ref{fig:rlimits}. For all the ILC methods, the error bars denote the $95\%$ confidence interval and the markers denote the best-fit value on $r$, which reflects the bias to some extent. It can be seen that all the ILC methods based on $B$ mode have the $r$ bias lower than $10^{-2}$. The specific $95\%$ upper limit values are detailed in Table \ref{tab:rLimits}.
\begin{table}[]
    \centering
    \caption{The $r$ constraining results for each ILC method and case. The confidence interval is $95\%$.}\label{tab:rLimits}
    \begin{tabular}{c|ccc|cc}
    \hline\hline
    \multirow{3}{*}{ILC methods}&\multicolumn{3}{c|}{Northern patch}&\multicolumn{2}{|c}{CMB-S4 patch}\\
    &5 channels&8 channels&5 channels&\multirow{2}{*}{with PS}&\multirow{2}{*}{without PS}\\
    &\small{with PS}&\small{with PS}&\small{without PS}\\
    \hline
PILC&$r < 0.0339$&$r < 0.0364$&$r < 0.0282$&$r < 0.0470$&$r < 0.0300$\\
pixelILC&$r < 0.00869$&$r < 0.00875$&$r < 0.00692$&$r < 0.00968$&$r < 0.00549$\\
harmonicILC&$r < 0.0113$&$r < 0.00450$&$r < 0.00978$&$r < 0.00971$&$r < 0.00565$\\
NILC&$r < 0.00794$&$r < 0.00734$&$r < 0.00547$&$r < 0.00699$&$r < 0.00505$\\
\hline\hline
    \end{tabular}
\end{table}

\subsection{Discussion of the results}\label{subsec:discussion}
We show the results of ILC analysis on $B$ maps in a Northern patch as well as a Southern (CMB-S4) patch that are suitable for observation with more advanced ground-based telescopes in the next stage. Since ILC methods are kind of blind foreground removal methods, there will be some residual foreground signal after ILC performing. Such foreground residual can be analyzed on the map, power spectrum and parameter space levels, respectively.

\begin{itemize}
    \item \textbf{Map level:} As all the cleaned maps from different ILC methods are dominated by the CMB signal, it's hard to tell the difference between ILC methods. But the residual foreground maps shown in Figure \ref{fig:5band_wops}, \ref{fig:fg_residual_maps_ps} and \ref{fig:CMBS4Maps} help us analyze the differences of these methods from different aspects. In general, the results by PILC have larger foreground residual for every cases. Qualitatively speaking, PILC works on $QU$ bases, which means the minimization will take care of both $E$ and $B$ mode at the same time. In the standard model of cosmology, $B$ mode is significantly smaller than $E$ mode even considering the weak lensing effects. So the covariance term in Eq. \eqref{eq:mainILC} is dominated by $E$ mode and it will make the $B$ mode cleaning worse. Another thing worth noting is that the butterfly-shaped point sources are clearly visible in Figure \ref{fig:fg_residual_maps_ps} and \ref{fig:CMBS4Maps}, especially for pixel domain ILC. NILC works better because it can provide good localization in both pixel space and harmonic space, which helps NILC track the compact sources.
    
    \item \textbf{Power spectrum level:} More information can be extracted from the foreground residual power spectra shown in Figure \ref{fig:powers_noPs}, \ref{fig:powers_Ps} and \ref{fig:CMBS4Power}. In the case of with point sources, there is no doubt that NILC method gives the minimal foreground residual because of the good elimination of the point sources, and it makes the residuals small. But in the case of without point sources, the best method is pixelILC, and even the PILC has a smaller foreground residual on the multipole range beyond the degree scale. 
    A reasonable explanation is that the polarized foreground has only few diffused components. Ideally, if the weights were already known and the linear combination performed on every pixel, there would be no residual foreground. In the real case, the weights cannot be calculated for each pixel. PILC and pixel ILC can calculate a series of global weights applying on the whole map, which means each pixel shares the same weights. It's a good approximation for a small sky patch. Due to the high signal-to-noise ratio on map level, the weights from pixel domain ILC are quite accurate. However, for harmonic ILC and NILC, in general, there are different weights on different scales. For the large scale where foreground signals dominate, harmonic ILC and NILC will adopt the weights which minimize the foreground. Therefore the large scale foreground residual is small. For the small scale where noise dominates, the weights is aiming to minimize noise rather than the foreground. Therefore the small scale foreground residual is larger than the residuals of PILC and pixel ILC because they both use the whole map to calculate the weights. If the foreground contains point sources, there will no global weights which can minimize point source radiation and diffused foreground simultaneously. In this case, the localized NILC can track the point sources better and give the suitable weights which makes the residual foreground smaller than other ILC methods.

    \item \textbf{Parameters space level:} The most important step is to find the bias when constraining on $r$. Table \ref{tab:rLimits} shows the $95\%$ CL upper limits on $r$. The major contribution for the uncertainty of $r$ come from the noise level and bias. For the northern sky patch with point sources, increasing the number of frequency channels from 5 to 8 doesn't help those ILC methods working on pixel domain too much, but the upper limits of harmonic ILC were reduced to $r\sim4\times10^{-3}$. Compare between the cases with and without point sources, the existence of point sources contribute roughly $2\times 10^{-3}\sim 5\times 10^{-3}$ bias on $r$, depending on different ILC methods. The upper limits on $r$ for the southern patch are consistent with the non-delensing forecasting results presented in the CMB-S4 Science Book\cite{2016arXiv161002743A}.
    
\end{itemize}


%% file: conclusion.tex
We propose a new foreground removal method for ground-based CMB observation, which basically includes two steps: 1) Use template cleaning method to get pure $B$ maps (free from $EB$ leakage) of different frequency bands. 2) Use ILC method at pixel, harmonic and wavelet domains on pseudo-scalar $B$ maps to get clean CMB $B$ map. We carry out ILC analysis based on $B$ map, and study its foreground removal efficiency in the case of different frequency band combinations and foreground emission with or without point source. We choose the clean patch in the northern sky which will be chosen at the very high probability to do a deep survey by the ground-based CMB experiments in the northern hemisphere, the map depth of the simulated maps for each band are reasonable for future experiments. 

We compare the reconstructed CMB signals and foreground residual on the map level, the angular power spectra level as well as the final $r$ constraints level. Figure \ref{fig:5band_wops} and Figure \ref{fig:powers_noPs} show the results of doing ILC on $B$ maps on map level and power spectrum level respectively. According to these two figures, doing ILC on $B$ maps is a feasible way and the foreground residual of such way is smaller than just working on $QU$ bases directly. Figure \ref{fig:resulting_maps_ps}, \ref{fig:fg_residual_maps_ps}, \ref{fig:powers_Ps} show the result of the influences of point sources. The same analyses are also performed on the CMB-S4 sky patch, as shown in Figure \ref{fig:CMBS4Maps} and \ref{fig:CMBS4Power}. It’s clear that the existence of point sources in the foreground makes ILC methods less effective, especially for pixel domain ILC. It will bias the $r$ constraining by roughly $3\times10^{-3}$.  Adding more frequency bands can not improve the results dramatically, which means that in the actual observation, in order to obtain better ILC results, the point sources need to be properly processed either at the map level or at the TOD (time order data) level before using ILC which will be studied in our future paper.

In general, NILC method seems to be the most promising ILC method for cleaned $B$ map reconstruction, it can provide relatively small deviation at map, power spectrum as well as the $r$ constraint level. Then is the harmonic ILC, which can also provide a high-level significance of CMB B modes. PILC is not a good choice for foreground cleaning, for it mixes up all the components in pixel and scale domains. The NILC on $B$ map method can give very low bias on $r$ ($\mathcal{O}(10^{-3})$) which comparable to the statistical uncertainty from instrumental noise shown in Table \ref{tab:setup}. Our study shows that, ILC method on $B$ map potentially could be a kind of feasible way to reduce the foreground contamination for future ground-based primordial gravitational wave experiments.